\documentclass[12pt,a4paper]{article}
\usepackage[latin1]{inputenc}
\usepackage[T1]{fontenc} 
\usepackage{arxiv}
\usepackage{amsmath}
\usepackage{braket}
\usepackage{amsfonts}
\usepackage{amssymb}
\usepackage{graphicx}
\usepackage{amsthm}
\usepackage{listings}
\usepackage{float}
\usepackage{caption,bbold}
\usepackage{subcaption}
\usepackage{color}
\usepackage[colorlinks=true,allcolors=blue]{hyperref}
\usepackage{chngcntr}
\usepackage{cite}
\usepackage{authblk}
\usepackage{tikz}
\usepackage{booktabs}  
\DeclareFontFamily{OMS}{oasy}{\skewchar\font48 }
\DeclareFontShape{OMS}{oasy}{m}{n}{%
	<-5.5> oasy5     <5.5-6.5> oasy6
	<6.5-7.5> oasy7     <7.5-8.5> oasy8
	<8.5-9.5> oasy9     <9.5->  oasy10
}{}
\DeclareFontShape{OMS}{oasy}{b}{n}{%
	<-6> oabsy5
	<6-8> oabsy7
	<8->  oabsy10
}{}
\DeclareSymbolFont{oasy}{OMS}{oasy}{m}{n}
\SetSymbolFont{oasy}{bold}{OMS}{oasy}{b}{n}

\DeclareMathSymbol{\smallleftarrow}     {\mathrel}{oasy}{"20}
\DeclareMathSymbol{\smallrightarrow}    {\mathrel}{oasy}{"21}
\DeclareMathSymbol{\smallleftrightarrow}{\mathrel}{oasy}{"24}

\newcommand{\tensor}[1]{\overset{\scriptscriptstyle\smallleftrightarrow}{#1}}

\definecolor{greennew}{rgb}{0.07, 0.53, 0.37}
\DeclareRobustCommand\full  {\tikz[baseline=-0.6ex]\draw[purple,thick] (0,0)--(0.5,0);}
\DeclareRobustCommand\dotted{\tikz[baseline=-0.6ex]\draw[blue,thick,dotted] (0,0)--(0.54,0);}
\DeclareRobustCommand\chain {\tikz[baseline=-0.6ex]\draw[red,thick,dashdotted] (0,0)--(0.5,0);}
\counterwithout{figure}{section}
\title{Schr\"{o}dinger-type 2D coherent states of magnetized uniaxially strained graphene}
\date{ }
\author{
	{\bf Erik D\'iaz-Bautista} \\
Departamento de Formaci\'on B\'asica Disciplinaria, Unidad Profesional Interdisciplinaria de Ingenier\'ia Campus Hidalgo del Instituto Polit\'ecnico Nacional, Pachuca: Ciudad del Conocimiento y la Cultura, Carretera Pachuca-Actopan km 1+500, San Agust\'in Tlaxiaca, 42162 Hidalgo, Mexico\\
 \texttt{ediazba@ipn.mx}}
\begin{document}
	\maketitle
	\begin{abstract}
		We revisit the uniaxially strained graphene immersed in a uniform homogeneous magnetic field orthogonal to the layer in order to describe the time evolution of coherent states build from a semi-classical model. We consider the symmetric gauge vector potential to render the magnetic field, and we encode the tensile and compression deformations on an anisotropy parameter $\zeta$. After solving the Dirac-like equation with an anisotropic Fermi velocity, we define a set of matrix ladder operators and construct electron coherent states as eigenstates of a matrix annihilation operator with complex eigenvalues. Through the corresponding probability density, we are able to study the anisotropy effects on these states on the $xy$-plane as well as their time evolution. Our results show clearly that the quasi-period of electron coherent states is affected by the uniaxial strain. 
	\end{abstract}


\section{Introduction}\label{sec1:introduction}
The probabilistic interpretation of the wave function $\psi$ has been traditionally used to describe quantum systems and compute observable physical quantities that can be measured in the laboratory. In this way, the quantum theory has been extended to many physics branches allowing to describe the physical phenomena in atomic scales and their subsequent application to the macroscopic world. Underlying this manner of understanding the quantum world, there is an interest in describing quantum systems through a semi-classical model, i.e., to find a way to render the atomic-scale world and its time evolution with analogous classical mechanics tools. Although quantum properties like spin do not have a classical counterpart, certain quantum systems have been described by such semi-classical formalism. Schr\"{o}dinger~\cite{s26} proposed the idea of identifying the most classical states to describe the quantum harmonic oscillator. Later, Glauber~\cite{g63} rediscovered these states for studying light nature, naming them as coherent states (CS). Nowadays, CS have been employed in quantum optics, atomic, nuclear, particle and condensed matter physics (see~\cite{klauder85,gazeau10} and references therein). In this manner, the coherent state formulation has been adopted and generalized with the goal to have the best knowledge about quantum systems.

Following the above ideas, the physical problem of a spinless charged particle moving on the $xy$-plane interacting with a uniform homogeneous orthogonal magnetic field $B_{0}$ has been solved first in the so-called symmetric gauge~\cite{f28}, and then in the well-known Landau gauge~\cite{landau30},
\begin{equation}\label{gauge}
\vec{A}=\frac{B_{0}}{2}\left(-y\hat{i}+x\hat{j}\right), \quad \vec{A}=-B_{0}y\hat{i},
\end{equation}
respectively. The corresponding coherent states have been built in both gauges in order to describe the system dynamics~\cite{landau30,mm69}. Historically, the Landau gauge~\cite{landau30} has been used to address this kind of system because of its simplicity, but more importantly, since it allows us to solve the problem by preserving the translational invariance along a given direction, the $x$-direction for instance. 
Meanwhile, the symmetric gauge~\cite{f28,d31,p30} preserves the rotational invariance and therefore allows us to describe the bidimensional (2D) problem. However, the price to pay is having infinite-fold degenerate energy levels with eigenfunctions that are gauge dependent. Although the energy is gauge invariant, other physical quantities do not, and they must be calculated in each gauge. The issue of the gauge-dependency of the states of charge carriers in magnetic fields, and also of the corresponding CS, is still open~\cite{davighi19}. Nevertheless, the symmetric gauge has been used in several works to study the charged particle dynamics through the coherent states formulation, focusing on different aspects of two-dimensional coherent states~\cite{mm69,fk70,lms89,krp96,sm03,kr05,re08,d17} and the importance of the so-called magnetic translation operators~\cite{z64,b64,l83,wz94,fw99}. For instance, in condensed matter physics, CS have been implemented in the calculation of the partition function for different systems allowing to describe some physical properties as the magnetic susceptibility~\cite{fk70}, to study Landau diamagnetism and de Haas-van Alphen oscillations~\cite{aremua15}, and some properties of fermionic atoms trapped in an optical square lattice subjected to an external and classical non-Abelian gauge field~\cite{goldman09}.

On the other hand, {\it straintronics}~\cite{bukharaev18} is a new research area that explores anisotropy effects on electronic, transport, and optical properties of condensed matter systems by strain engineering methods in order to develop new technologies. A fruitful application of straintronics can be found in graphene~\cite{wallace47,novoselov04,novoselov05,novoselov051}, one of the best known 2D Dirac materials. This material has attracted the scientific interest due to its interesting mechanical, electronic, and optical properties, and also because it allows to connect different areas of research in physics, e.g. condensed matter, high energy, electrodynamics, quantum field theory, theoretical and experimental physics~\cite{novoselov051,pmp06,pc07,ng07,k07,Kuru,hrp10,og13,midya14,jk14,valenzuela15,j15,em17,Diaz3,concha18,cdr19,cf19}. Focusing on the arising phenomena, when a mechanic deformation is applied to graphene, its electrons behave as if they were immersed in a fictitious magnetic field. This effect has provided interesting theoretical~\cite{gbot17} and experimental~\cite{tppj09} results. Another way to study the anisotropy effects consists in considering uniform uniaxial deformations that induce a tensor character to the Fermi velocity in the material in the low-energy regime, namely in the energy range of 0 eV-0.3 eV. Although this modifies the dispersion relation from the pristine case, the generation of any pseudo-magnetic field is prevented~\cite{og13,Betancur,Oliva2017,cdr19,perezpedraza20} and the motion equations are still tractable.

In this work, we are interested in describing the anisotropy effects on electron dynamics in magnetized graphene when either tensile or compression uniaxially deformations are applied. In order to consider such anisotropy effects, the background magnetic field is described by the symmetric vector potential given in Eq.~(\ref{gauge}), and the system is studied through the coherent state formulation. Our aim consists of determining the time evolution of the coherent states associated with this physical system and how it is also affected by the strain affects. Thus, this work is organized as follows. In Sec.~\ref{sec2:model}, we briefly discuss the tight-binding (TB) model used to describe the uniaxially strained graphene, as well as the Dirac-Weyl (DW) Hamiltonian. In Sec.~\ref{sec4:SUstates}, we construct the $SU(2)$ coherent states as a linear combination of the eigenvectors of the Dirac-Weyl Hamiltonian. We also analyze the anisotropy effects on their probability density and the cyclotron motion that they depict. In Sec.~\ref{sec5:SCstates}, we obtain the Schr\"{o}dinger-type 2D coherent states as eigenestates of a matrix annihilation operator with complex eigenvalue. We also analyze the anisotropy effects on their corresponding probability density, the cyclotron motion, and their time evolution. We describe the occupation number distribution of states as well. In Sec.~\ref{sec6:conclusions}, we formulate our conclusions and remarks.

\section{The model}\label{sec2:model}
In condensed matter physics, the tight-binding (TB) model~\cite{slater54,harrison80}, based on linear combination of atomic orbitals, is an approximation, which consists in determining the intensity of the interference between atomic orbitals. It is used for the calculation of electronic band structure of molecules and solids employing a superposition of wave functions for isolated atoms located at each atomic site in the crystal lattice. This model assumes that as the atomic orbitals are further away from each other, they will interfere less. 

\begin{figure}[h!]
	\centering
	\includegraphics[width=0.7\linewidth]{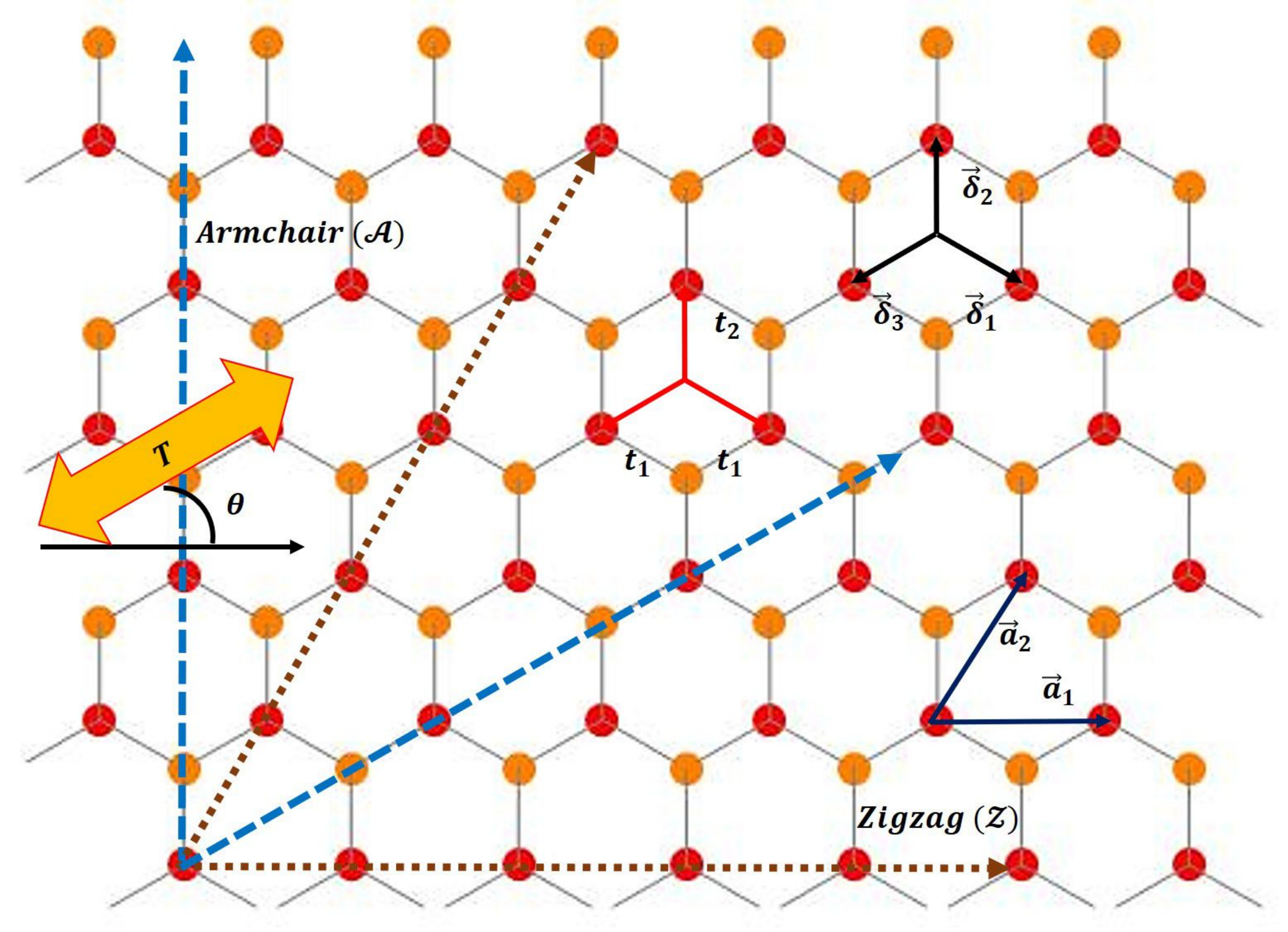}
	\caption{Schematic representation of uniaxially strained graphene where zigzag (armchair) direction is parallel to the $x$-axis ($y$-axis). Both directions repeat alternately every $30^\circ$. Uniaxial strain along the zigzag or armchair direction has two different nearest-neighbor hopping parameters $t_{1}$ and $t_{2}$. The positions of nearest neighbors are $\vec{\delta_1}$, $\vec{\delta_2}$ and $\vec{\delta_3}$. The lattice vectors $\vec{a}_{1}$ and $\vec{a}_{2}$ allow connecting the whole sites in the deformed hexagonal lattice.}
	\label{fig:lattice}
\end{figure}

\subsection{Effective tight-binding model in stained graphene}
In strained graphene~\cite{og13,Pereira,Midtvedt,Betancur2}, as well as in other hexagonal-like lattices, the TB approach can be reduced to an effective model of two energy bands~\cite{Goerbig,Goerbig2,Betancur2}, assuming that there are only two atoms per unit cell in an infinity extended layer and neglecting second nearest-neighbor interactions and  the overlap between $p_z$ orbitals. Thus, the effective model will only depend on two hopping parameters $t_1$ and $t_2$ that quantify the probability amplitude that an electron hops to the nearest atom. In the case in which a uniaxial tension $T$ is applied to the graphene layer along an arbitrary direction $\theta$ respect to the $x$-axis (see Fig.~\ref{fig:lattice}), the uniaxial strain tensor is given by
\begin{equation}
\tensor{\epsilon}=\left(\begin{array}{c c}
\cos^{2}(\theta)-\nu\sin^{2}(\theta) & (1+\nu)\cos(\theta)\sin(\theta) \\
(1+\nu)\cos(\theta)\sin(\theta) & \sin^{2}(\theta)-\nu\cos^{2}(\theta)
\end{array}\right)\epsilon,
\end{equation}
where the tensile strain $\epsilon$ quantifies the percentage of deformation~\cite{Pereira} which is proportional to the magnitude of tension $T$, and $\nu\sim0.178$ is the Poisson ratio of graphene~\cite{Pereira2,Cadelano}. As a consequence, the atomic sites are displaced modifying the hopping parameter values that are related to the bond lengths through an exponential decay rule. For clarifying the latter 
and without loss of generality, let us consider deformations applied along zigzag ($\mathcal{Z}$) direction ($\theta=0$) and armchair ($\mathcal{A}$) direction ($\theta=\pi/2$). Thus, the deformation tensors are, respectively~\cite{Pereira,Colombo,Cadelano,Landau}:
\begin{equation}\label{str}
{\bf \epsilon}_{\mathcal{Z}} = \left(\begin{array}{cc} 
1 & 0\\
0 & -\nu
\end{array}\right)\epsilon, \qquad {\bf \epsilon}_{\mathcal{A}} = \left(\begin{array}{cc} 
-\nu & 0\\
0 & 1
\end{array}\right)\epsilon.
\end{equation}

Furthermore, the atomic positions in uniaxially strained graphene are given by $\vec{r} = (\mathbb{I} + \tensor{\epsilon})\vec{r}_0$, where $\vec{r}_0$ is the vector position of the sites on the pristine sample and $\mathbb{I}$ denotes the $2\times2$ unity matrix. 
Therefore, the deformed lattice vectors for uniaxial strain in the $\mathcal{Z}$ direction are~\cite{diazbetancur20}
	\begin{equation}\label{lattvszz}
	\vec{a}^{\mathcal{Z}}_1 =  \sqrt{3}a_0(1 + \epsilon)\hat{i}, \quad
	\vec{a}^{\mathcal{Z}}_2 =  \frac{\sqrt{3}}{2}a_0[(1 + \epsilon)\hat{i} + \sqrt{3}(1 -\nu\epsilon)\hat{j}],
	\end{equation}
while for the $\mathcal{A}$ direction, they are
	\begin{equation}\label{lattvsac}
	\vec{a}^{\mathcal{A}}_1 =  \sqrt{3}a_0(1 - \nu\epsilon)\hat{i}, \quad
	\vec{a}^{\mathcal{A}}_2 =  \frac{\sqrt{3}}{2}a_0[(1 - \nu\epsilon)\hat{i} + \sqrt{3}(1 +\epsilon)\hat{j}],
	\end{equation}
where $a_0=1.42$ \r{A} is the bond length in pristine graphene \cite{Castro}. Likewise, the nearest-neighbor sites, $\vec{\delta}_1 = 2\vec{a}_1/3 - \vec{a}_2/3$, $\vec{\delta}_2 = 2\vec{a}_2/3 - \vec{a}_1/3$, and $\vec{\delta}_3 = -\vec{\delta}_1 -\vec{\delta}_2$, will be also related to the tensile strain.

Under all above assumptions, let us consider the following Hamiltonian in the Fourier basis from a plane-wave ansatz~\cite{Betancur2,wallace47,Goerbig}:
\begin{equation}
H_{\text{TB}}^{K} = \sum^3_{j = 1}\left[\begin{array}{cc}
0 & t_j\textrm{e}^{i\vec{k}\cdot\vec{\delta}_j}\\
t_j\textrm{e}^{-i\vec{k}\cdot\vec{\delta}_j} & 0
\end{array}\right].
\label{H}
\end{equation} 
We can obtain an explicit expression for the hopping parameters, $t_j = t\,\textrm{exp}[-\beta(\delta_j/a_{0} - 1)]$, where $\beta$ is the Gr\"uneisen constant, $t$ is the hopping in pristine graphene, and $\delta_j$ are the deformed bond lengths~\cite{Pereira,Castro,Betancur2,Papas}. Thus, for $\mathcal{Z}$ deformations, we have that
	\begin{equation}\label{dzz}
	\delta^{\mathcal{Z}}_1 =  \delta^{\mathcal{Z}}_3= a_0\sqrt{\left(1 + \frac{1}{4}(3 - \nu)\epsilon\right)^2 + \frac{3}{16}(1+ \nu)^2\epsilon^2}, \quad
	\delta^{\mathcal{Z}}_2 =  a_0(1 -\nu\epsilon),
	\end{equation}
while for $\mathcal{A}$ deformations, we get
	\begin{equation}\label{dac}
	\delta^{\mathcal{A}}_1 =  \delta^{\mathcal{A}}_3= a_0\sqrt{\left(1 + \frac{1}{4}(1 - 3\nu)\epsilon\right)^2 + \frac{3}{16}(1+ \nu)^2\epsilon^2}, \quad
	\delta^{\mathcal{A}}_2 =  a_0(1 + \epsilon).
	\end{equation}

On the other hand, the electronic band structure of uniaxially strained graphene is directly obtained from the eigenenergies of the Hamiltonian in Eq.~(\ref{H}),
\begin{equation}
E_s(\vec{k}) = s\left|\sum^3_{j = 1}t_j\textrm{e}^{-i\vec{k}\cdot\vec{\delta}_j}\right|,
\end{equation}
where the band index $s = 1\,(-1)$ corresponds to the conduction (valence) band. 
Taking into account that density functional theory and TB calculations agree at the low-energy regime~\cite{Ribeiro}, the TB Hamiltonian (\ref{H}) can be expanded around the Dirac point by taking $\vec{k} = \vec{q} + \vec{K}_D$, such that $\vert\vec{q}\vert\ll\vert\vec{K}_{D}\vert$, where the Dirac point position $\vec{K}_D$ also satisfies
\begin{equation}\label{hopping}
\sum^3_jt_j\exp\left(-i\vec{K}_D\cdot\vec{\delta}_j\right) = 0 \quad \Longrightarrow \quad \cos[\vec{K}_D\cdot(\vec{\delta}_1 - \vec{\delta}_2)] = -\frac{t_2}{2t_1}.
\end{equation}

Thus, the effective Dirac-like Hamiltonian in the continuum approximation turns out to be
\begin{equation}\label{HD}
H=v_{\rm F}\left(a\,p_{x}\sigma_{x}+b\,p_{y}\sigma_{y}\right),
\end{equation}
where $\sigma_{x/y}$ are the Pauli matrices and
	\begin{equation}
	a = \frac{2}{3}\sum^3_{j = 1}\frac{\delta_{jx}}{a_0}\frac{t_j}{t}\sin(\vec{K}_D\cdot\vec{\delta}_j),\quad
	b = \frac{2}{3}\sum^3_{j = 1}\frac{\delta_{jy}}{a_0}\frac{t_j}{t}\cos(\vec{K}_D\cdot\vec{\delta}_j),
	\label{w}
	\end{equation}
which can be expressed as functions of the lattice vectors and hopping parameters by taking into account the relation in Eq.~(\ref{hopping}),
\begin{equation}\label{ayb}
a =  \frac{2}{3a_0t}\sqrt{a^2_{1x}t^2_1 + (a_{2x} - a_{1x})a_{2x}t^2_2}, \quad
b =  \frac{2}{3a_0t}\sqrt{a^2_{1y}t^2_1 + (a_{2y} - a_{1y})a_{2y}t^2_2}.
\end{equation}

\begin{figure}
	\centering
	\begin{tabular}{ccc}
		(a) \qquad \qquad \qquad \qquad \qquad \qquad & (b) \qquad \qquad \qquad \qquad \qquad \qquad & (c) \qquad \qquad \qquad \qquad \qquad \qquad \\
		\includegraphics[trim = 0mm 0mm 0mm 0mm, scale= 0.42, clip]{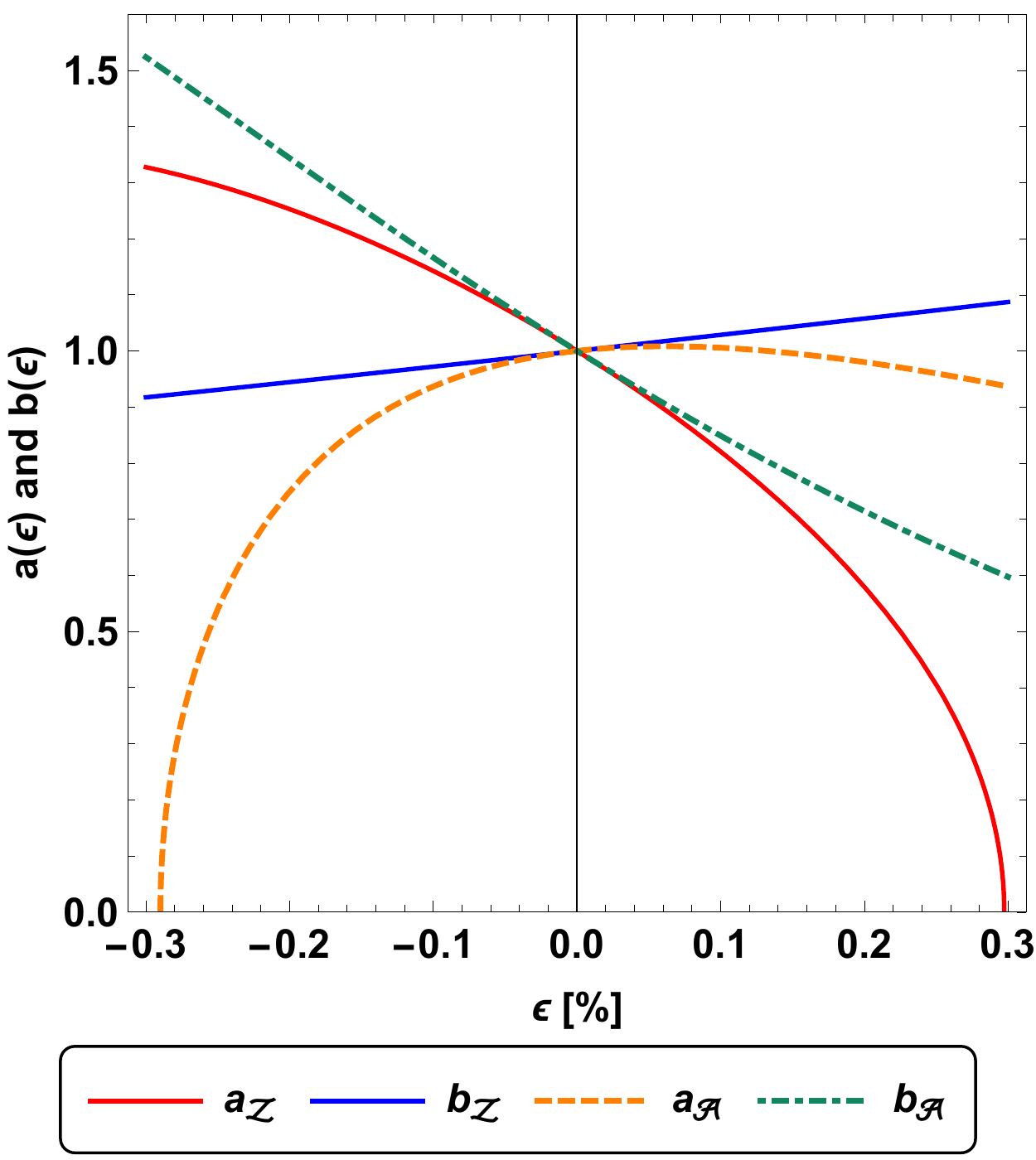} &
		\includegraphics[trim = 0mm 0mm 0mm 0mm, scale= 0.42, clip]{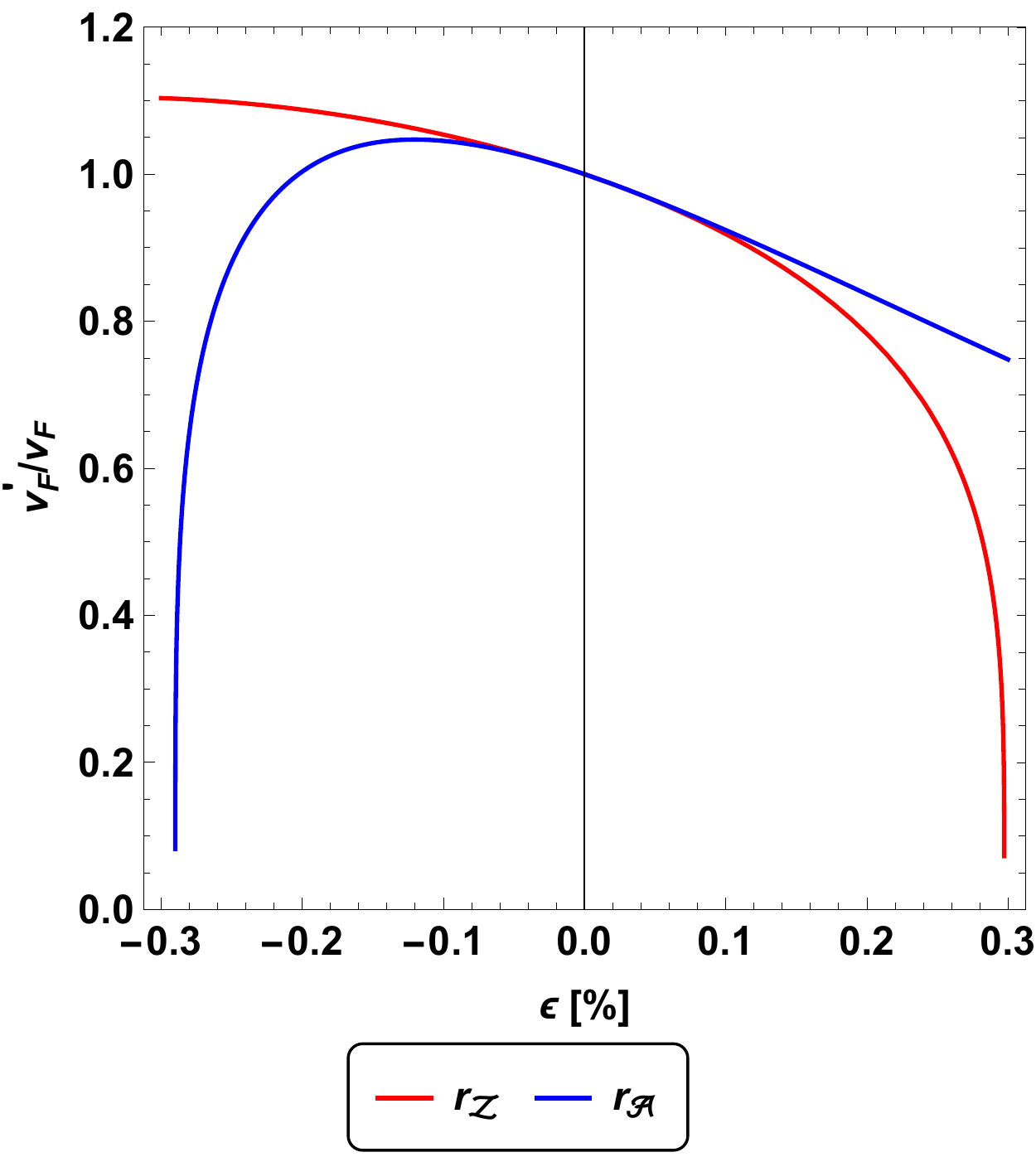}&
		\includegraphics[trim = 0mm 0mm 0mm 0mm, scale= 0.42, clip]{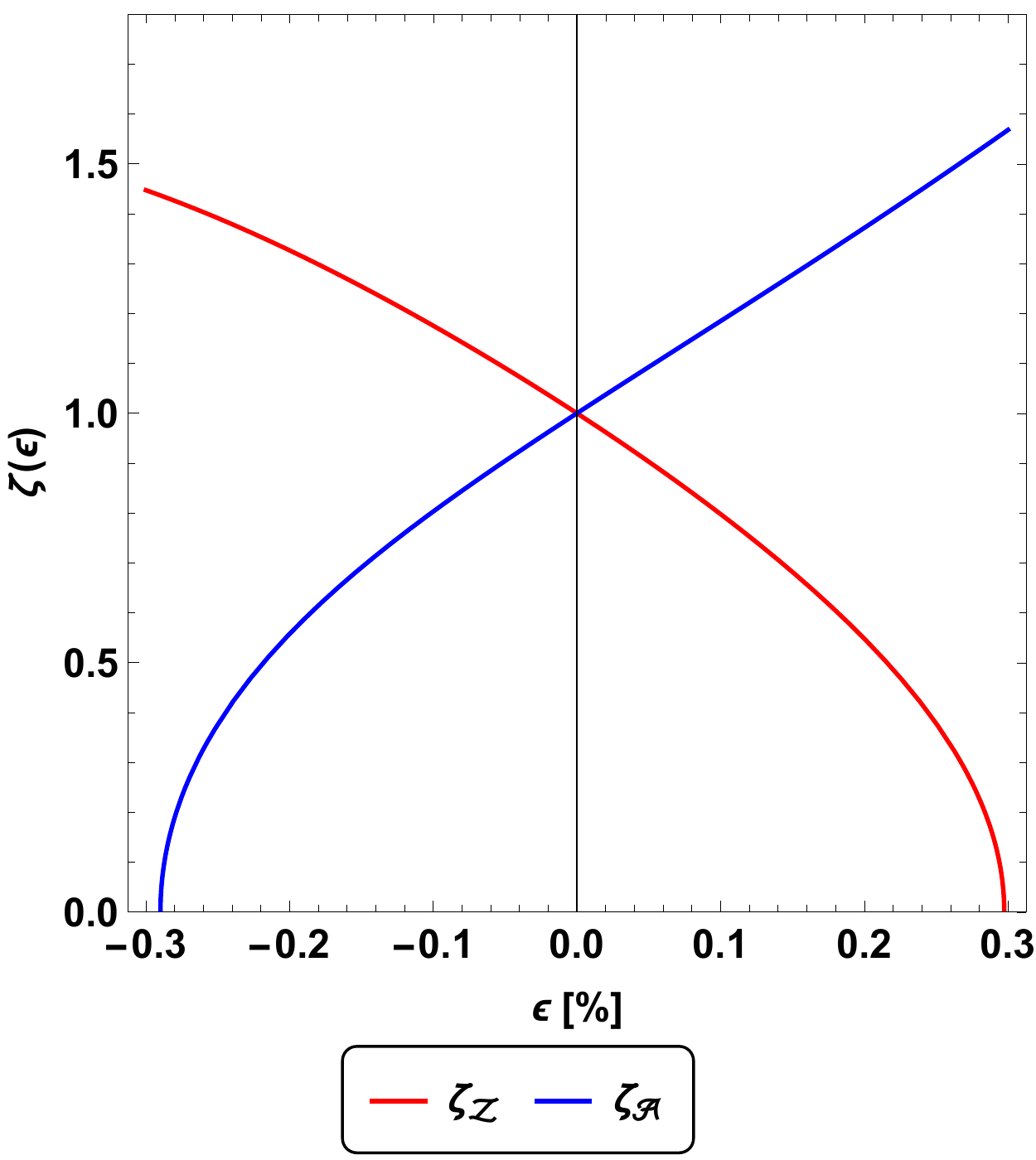}
	\end{tabular}
	\caption{(a) Geometrical parameters $a$ and $b$, (b) effective Fermi velocity $v'_{\rm F} = v_{\rm F}\sqrt{ab}$, and (c) strain parameter $\zeta = a/b$ as functions of the tensile strain $\epsilon$ along the zigzag and armchair directions. The elastic response is isotropic (anisotropic) and linear (nonlinear) for $\epsilon < 10 \%$ ($\epsilon > 10 \%$).}
	\label{GA}
\end{figure}

As is discussed in~\cite{Betancur}, the physical properties of strained graphene are encoded in the geometrical parameters (\ref{ayb}) (see Fig.~\ref{GA}). 

In the forthcoming sections, we describe the strain effects on the electron coherent states through these geometrical parameters.

\subsection{Dirac-Weyl equation under uniaxial strain}\label{sec3:dirac}
Now, let us consider an effective Dirac model around a Dirac point, namely, $K_{\rm D}$, in the first Brillouin zone. In order to investigate the anisotropy effects on a graphene layer interacting with a orthogonal homogeneous magnetic field $\vec{B}_{0}$, we assume
the symmetric gauge vector potential,
\begin{equation}
\vec{A}=\frac{\vec{B}_{0}\times\vec{r}}{2},
\end{equation}
and the Peierls substitution $\vec{p}\rightarrow\vec{\Pi}=\vec{p}+e\vec{A}$ in the Dirac-like Hamiltonian $H$ in Eq.~(\ref{HD}). Thus, the Dirac-Weyl (DW) Hamiltonian in the corresponding eigenvalue equation $H\vert\Psi\rangle=E\vert\Psi\rangle$ is rewritten as follows:
\begin{equation}\label{3}
H=\sqrt{\omega_{\rm B}}\hbar\,v'_{\rm F}\left[\begin{array}{cc}
0 & -iA^{-} \\
iA^{+} & 0
\end{array}\right],
\end{equation}
where the dimensionless operators
\begin{equation}\label{ladderoperators}
A^{\pm}=\frac{\mp\,i}{\sqrt{\omega_{\rm B}}\hbar}\left(\zeta^{1/2}\Pi_{x}\pm \frac{i}{\zeta^{1/2}}\Pi_{y}\right)
\end{equation}
satisfy the commutation relation,
\begin{equation}\label{5}
[A^{-},A^{+}]=\mathbf{1},
\end{equation}
with $\omega_{\rm B}=2eB_{0}/\hbar=(3.038\times10^{-3})B_{0}$ [(nm$^{2}$T)$^{-1}$] being the cyclotron frequency, and $v'_{\rm F}$ and $\zeta$ depend on the anisotropy direction (see Fig.~\ref{GA}).

Then, the action of the Hamiltonian~(\ref{3}) onto a state $\vert\Psi\rangle=\left(\begin{array}{c c}
\vert\psi_1\rangle & \vert\psi_2\rangle
\end{array}\right)^{\rm T}$ gives place to two coupled equations that can be decoupled to obtain the following equations for each component (see Appendix~\ref{appA}),
\begin{subequations}\label{hamiltonians}
	\begin{align}
\mathcal{H}^{-}\vert\psi_1\rangle&=A^{-}A^{+}\vert\psi_1\rangle=\mathcal{E}\vert\psi_1\rangle, \label{8a} \\
\mathcal{H}^{+}\vert\psi_2\rangle&=A^{+}A^{-}\vert\psi_2\rangle=\mathcal{E}\vert\psi_2\rangle \label{8b}
\end{align}
\end{subequations}
with $\mathcal{E}\equiv(E/\hbar\,v'_{\rm F}\sqrt{\omega_{\rm B}})^2$. Thus, we have two Schr\"{o}dinger equations whose eigenvalues are related as follows (see Appendix~\ref{appB}):
\begin{equation}\label{energies}
\mathcal{E}_{1,n-1}=\mathcal{E}_{2,n}=n, \quad n\geq 1, \quad \mathcal{E}_{2,0}=0,
\end{equation}
such that the energy spectrum turns out to be
\begin{equation}\label{LLs}
E_{n} = s v'_{\rm F} \sqrt{2ne\hbar B_{0}},
\end{equation}
where $v'_{\rm F}=v_{\rm F}\sqrt{ab}$ is the effective Fermi velocity and the index $s$ denotes the positive (negative) energy corresponding to electrons in the conduction (valence) band. We will only focus on electrons in the conduction band ($s=1$). 
According to Fig. \ref{GA}, the quantity $\sqrt{ab}$ depends on negative and positive deformations, affecting then the Landau level (LL) spectrum spacing~\cite{Betancur,Sahalianov}. 
This fact will be addressed in detail when we discuss the time evolution of the electron coherent states.

\begin{figure}[h!]
	\centering
	\includegraphics[width=0.5\linewidth]{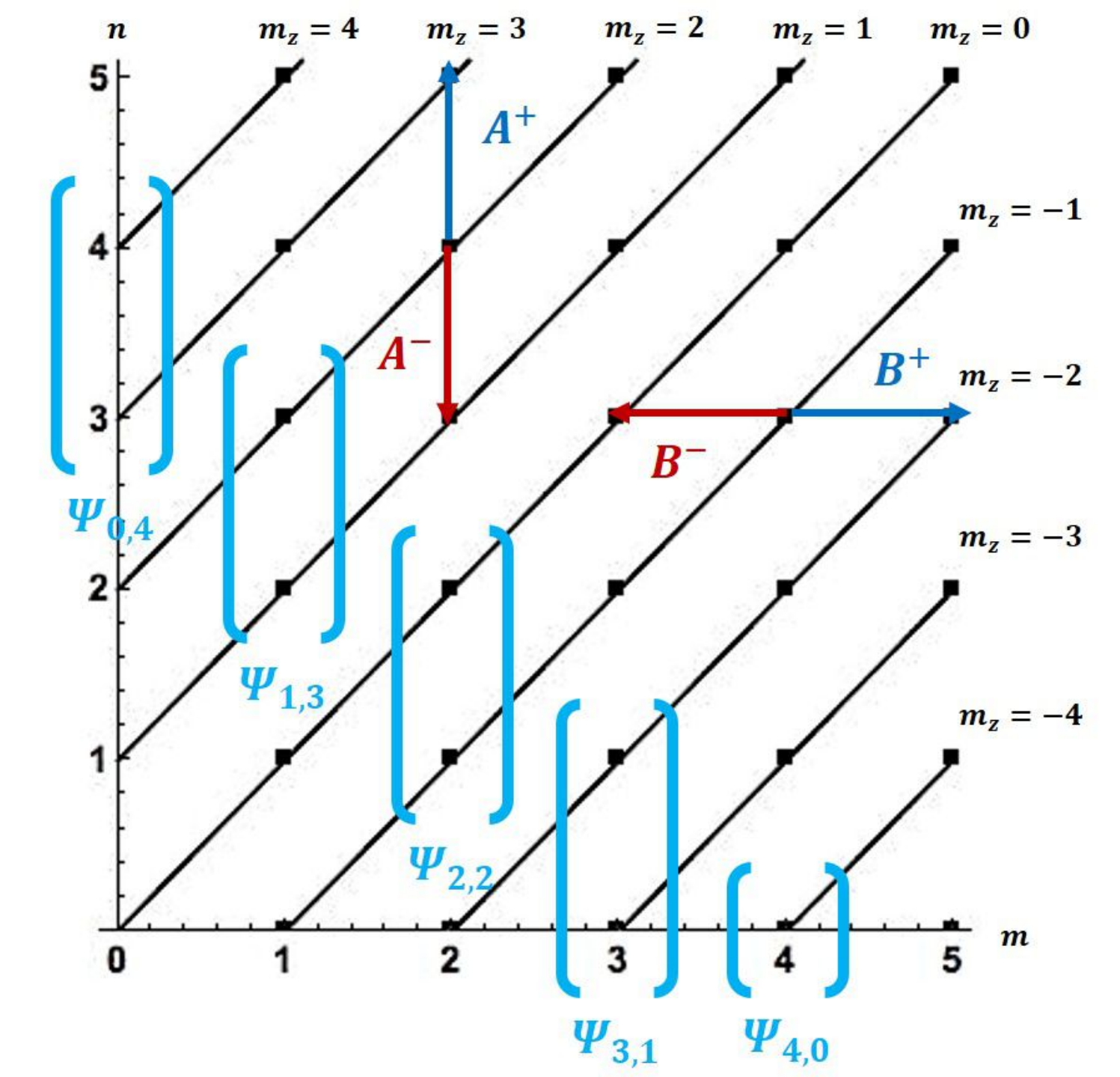}
	\caption{Space of scalar states $\psi_{m,n}$ is represented univocally by coordinates $(m,n)$. Inclined lines connect states with the same angular momentum $m_{z}=n-m$. The spinor states $\Psi_{m,n}$, $0\leq m,n\leq4$, that contribute to the linear combination of $\Psi_{\nu=4}$ are shown.}
	\label{fig:fig1}
\end{figure}

Now, proceeding as in Appendix~\ref{appB}, the wave functions of the normalized eigenstates of the Hamiltonian $\mathcal{H}^+$ turn out to be \cite{f28}
\begin{align}\label{34}
\nonumber\psi_2(x,y)\equiv\psi_{m,n}(\rho,\theta)&=(-1)^{\min(m,n)}\sqrt{\frac{\omega_{\rm B}}{4\pi}\frac{\min(m,n)!}{\max(m,n)!}}\left(\frac{\sqrt{\omega_{\rm B}}}{2}\rho\right)^{\vert n-m\vert}\exp\left(-\frac{\omega_{\rm B}}{8}\rho^2+i(n-m)\theta\right)\\
&\quad\times L_{\min(m,n)}^{\vert n-m\vert}\left(\frac{\omega_{\rm B}}{4}\rho^2\right)
\end{align}
with $n,m=0,1,2,\dots$ and $L_a^b(x)$ denoting the associated Laguerre polynomials. The normalized eigenstates of the Hamiltonian $\mathcal{H}^-$ are obtained by acting $A^{-}$ onto $\vert\psi_1\rangle$, as $\vert\psi_1\rangle\equiv\vert\psi_{m,n-1}\rangle=A^{-}\vert\psi_{m,n}\rangle/\sqrt{n}$. In addition, according to Appendix~\ref{appB}, the eigenstates of the Hamiltonians $\mathcal{H}^\pm$ are labeled by two positive integers $m,n$, namely, (see Fig.~\ref{fig:fig1})
\begin{equation}\label{28}
\vert\psi_1\rangle\equiv\vert\psi_{m,n-1}\rangle, \quad \vert\psi_2\rangle\equiv\vert\psi_{m,n}\rangle,
\end{equation}
that correspond to the eigenvalues of the number operators $N=A^{+}A^{-}$ and $M=B^ {+}B^{-}$, respectively. These scalar ladder operators are described in detail in Appendix~\ref{appC}. 
Thus, the eigenstates $\vert\Psi_{m,n}\rangle$ that describe the wave function of electrons in strained graphene in a uniform magnetic field are
\begin{equation}\label{eigenstates}
\vert\Psi_{m,0}\rangle=\left(\begin{array}{c}
0 \\
i\vert\psi_{m,0}\rangle
\end{array}\right) \quad \vert\Psi_{m,n}\rangle=\left(\begin{array}{c}
\vert\psi_{m,n-1}\rangle \\
i\vert\psi_{m,n}\rangle
\end{array}\right),
\end{equation}
where $\langle\psi_{m',n'}\vert\psi_{m,n}\rangle=\delta_{m'm}\delta_{n'n}$.

Moreover, the states $\vert\psi_{m,n}\rangle$ are also eigenstates of the $z$-component angular momentum-like operator given by $L_z=N-M$ with eigenvalue $m_z=n-m$. By defining the $z$-component of the total angular momentum operator as $\mathbb{L}_z=L_z\otimes\mathbb{I}+\sigma_z/2$ such that
\begin{align}\label{total_angular}
\mathbb{L}_z\vert\Psi_{m,n}\rangle&=\left(m_z-\frac12\right)\vert\Psi_{m,n}\rangle=j\,\vert\Psi_{m,n}\rangle,
\end{align}
we can conclude that the eigenstates $\vert\Psi_{m,n}\rangle$ of the Hamiltonian~(\ref{3}) are also eigenstates of $\mathbb{L}_z$ with rational eigenvalue $j\equiv m_z-1/2$. This fact was developed in~\cite{diaz2019coherent}.  

\subsection{Annihilation operators}
In order to apply the coherent state formulation, let us consider the matrix operators defined in~\cite{Diaz4,diaz2019coherent} for anisotropic 2D Dirac materials,
\begin{subequations}\label{annihop1}
	\begin{align}
	\mathbb{A}^-=\left[\begin{array}{c c}
	\frac{\sqrt{N+2}}{\sqrt{N+1}}A^- & 0 \\
	0 & A^-
	\end{array}\right], &\quad \mathbb{A}^+=\left[\begin{array}{c c}
	A^+\frac{\sqrt{N+2}}{\sqrt{N+1}} & 0 \\
	0 & A^+
	\end{array}\right], \label{41a}\\
	\mathbb{B}^-=\left[\begin{array}{c c}
	B^- & 0 \\
	0 & B^-
	\end{array}\right], &\quad \mathbb{B}^+=\left[\begin{array}{c c}
	B^+ & 0 \\
	0 & B^+
	\end{array}\right], \label{41b}
	\end{align}
\end{subequations}
whose actions onto the states $\vert\Psi_{m,n}\rangle$ are
\begin{equation}
\mathbb{A}^-\vert\Psi_{m,n}\rangle=\sqrt{n}\vert\Psi_{m,n-1}\rangle, \quad \mathbb{B}^-\vert\Psi_{m,n}\rangle=\sqrt{m}\vert\Psi_{m-1,n}\rangle.
\end{equation}

For our purposes, we also define the following matrix operator in terms of the above ones:
\begin{equation}\label{annihop2}
\mathbb{J}^-=\alpha^\ast\mathbb{A}^{-}+\beta^\ast\mathbb{B}^{-},
\end{equation}
where $^\ast$ denotes complex conjugate and $\alpha,\beta\in\mathbb{C}$ with $\vert\alpha\vert^2+\vert\beta\vert^2=1$.

In Appendix~\ref{appD}, we discuss the commutation relations that the above matrix operators fulfill.

\begin{figure}[h!]
	\centering
	\begin{tabular}{ccc}
		(a) $\alpha=\sqrt{3}/2$ \qquad \qquad \quad & (b) $\alpha=\sqrt{3}\exp\left(i\pi/4\right)/2$ \qquad \qquad & (c) $\alpha=-\sqrt{3}/2$ \qquad \qquad \\
		\includegraphics[trim = 0mm 0mm 0mm 0mm, scale= 0.34, clip]{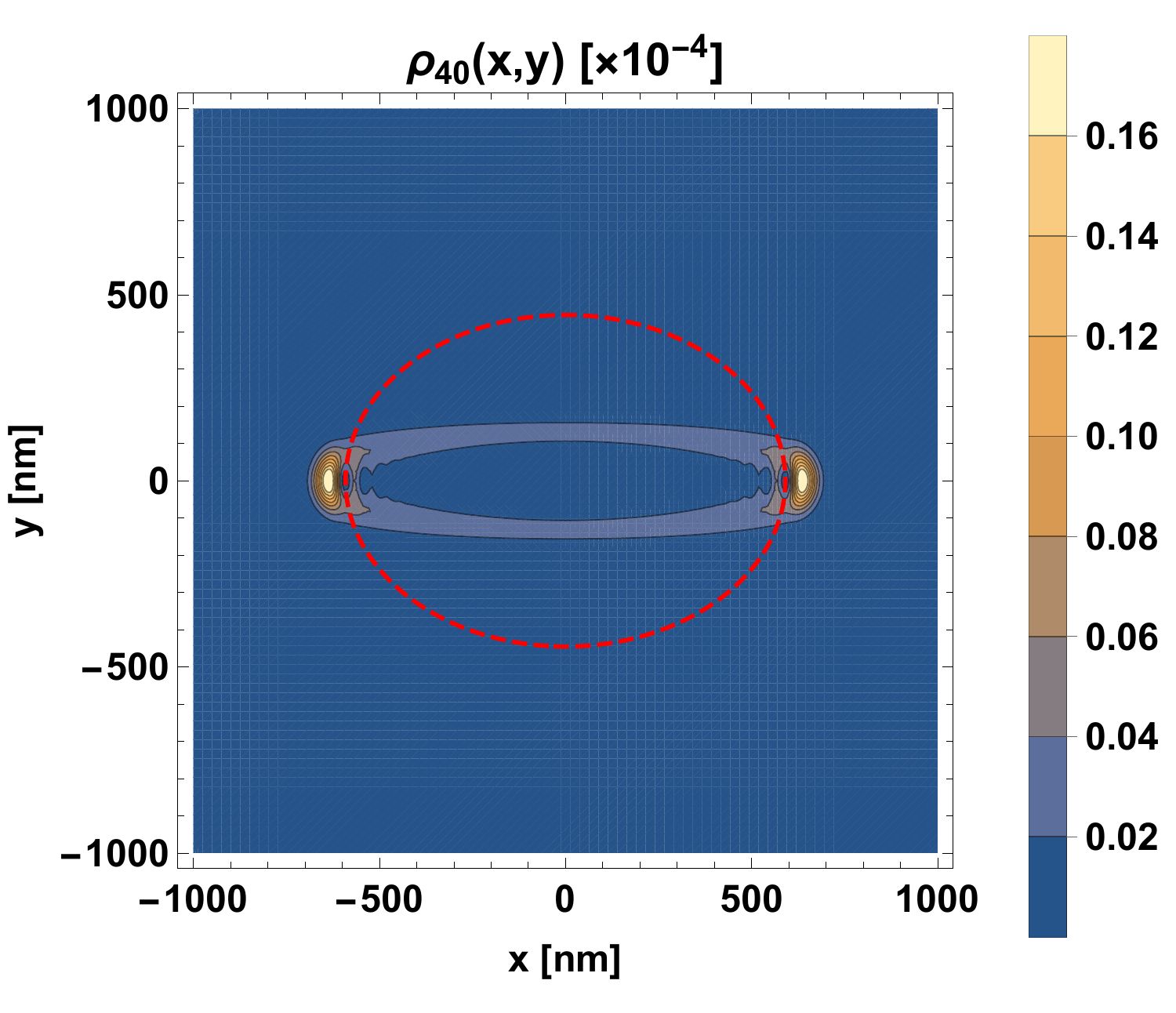} &
		\includegraphics[trim = 0mm 0mm 0mm 0mm, scale= 0.34, clip]{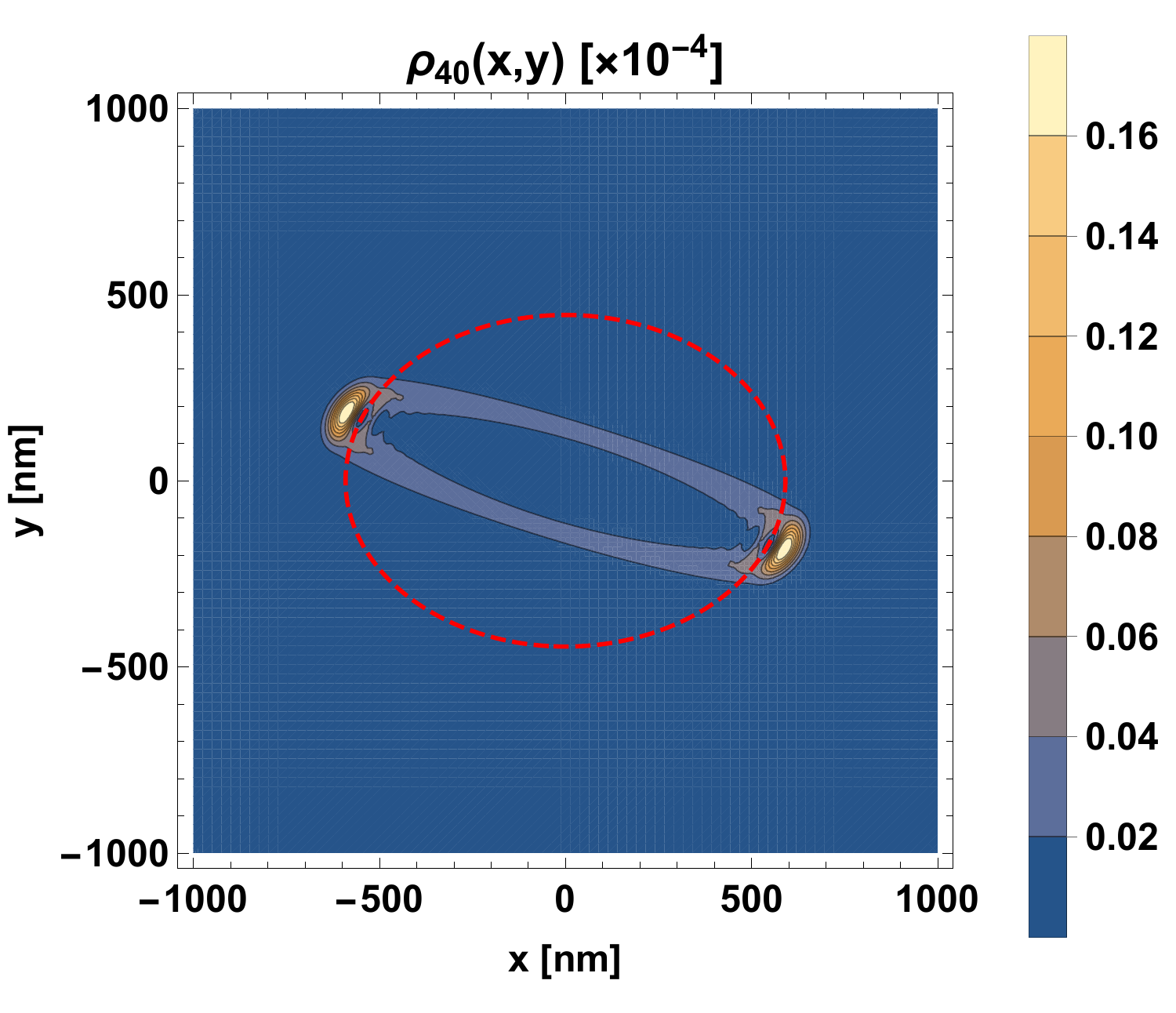}&
		\includegraphics[trim = 0mm 0mm 0mm 0mm, scale= 0.34, clip]{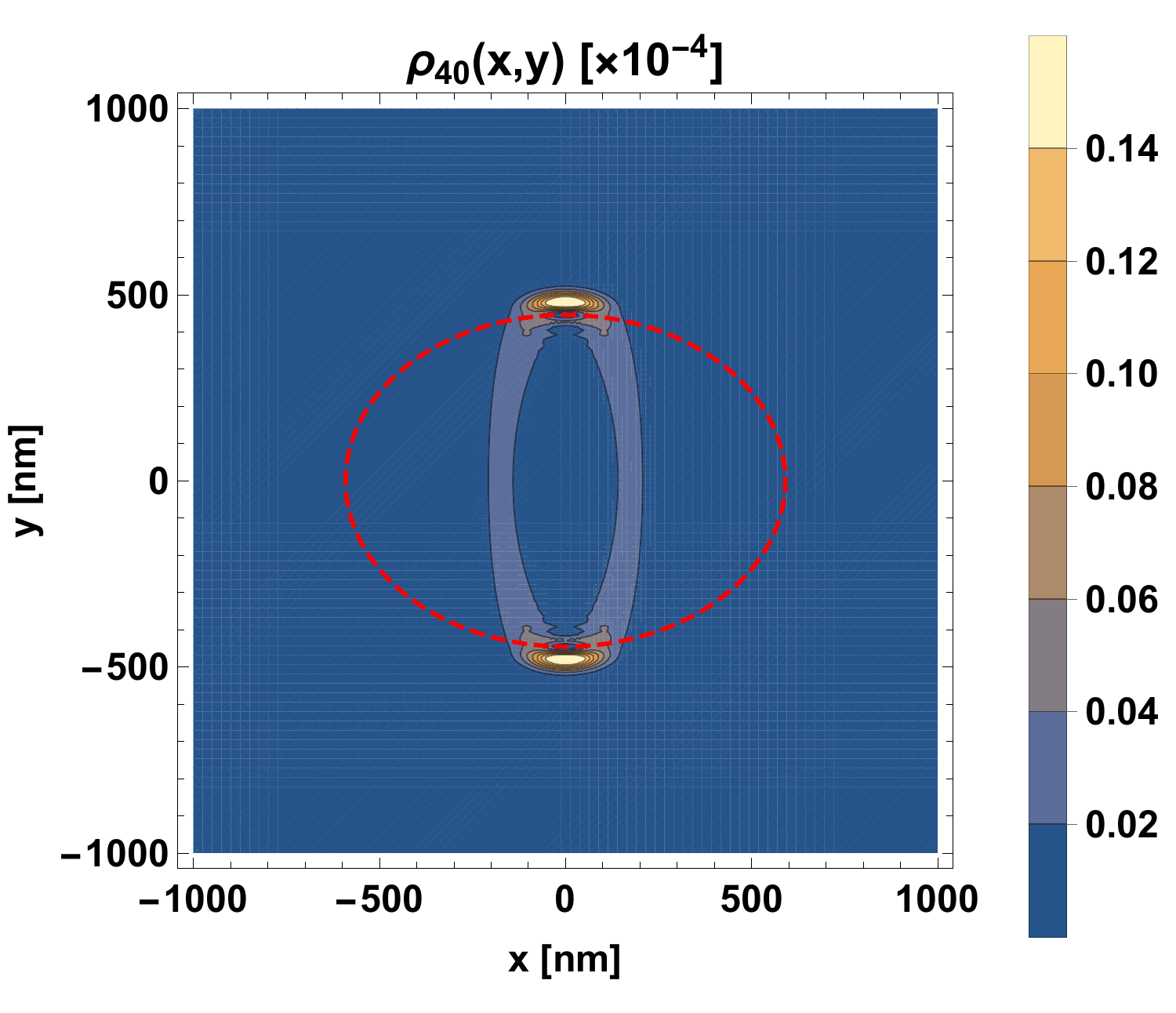}
	\end{tabular}
	\begin{tabular}{ccc}
		(d) {\color{white}$\alpha=\sqrt{3}/2$} \qquad \qquad \quad & (e) {\color{white}$\alpha=\sqrt{3}\exp\left(i\pi/4\right)/2$} \qquad \qquad & (f) {\color{white}$\alpha=-\sqrt{3}/2$} \qquad \qquad \\
		\includegraphics[trim = 0mm 0mm 0mm 0mm, scale= 0.34, clip]{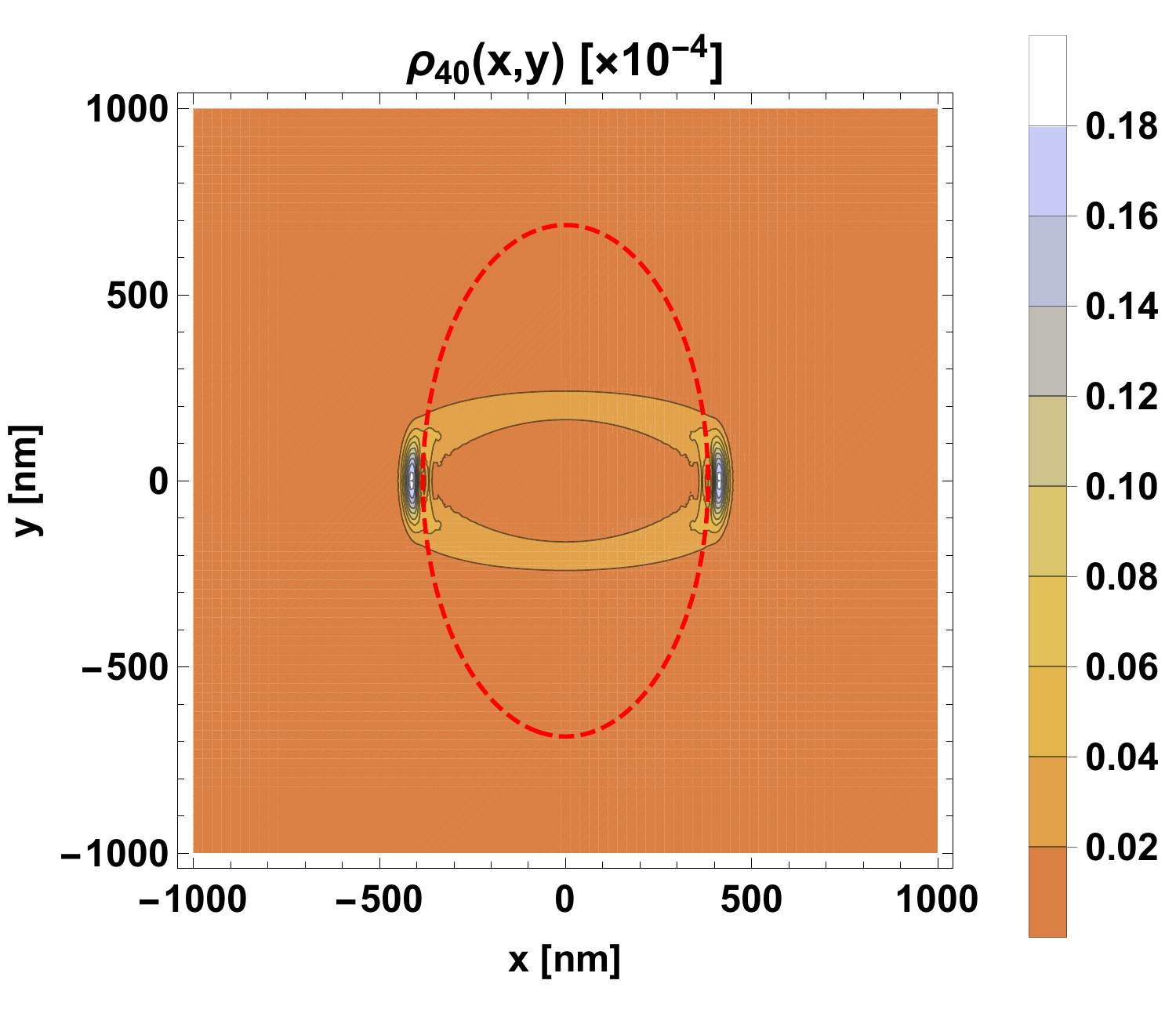} &
		\includegraphics[trim = 0mm 0mm 0mm 0mm, scale= 0.34, clip]{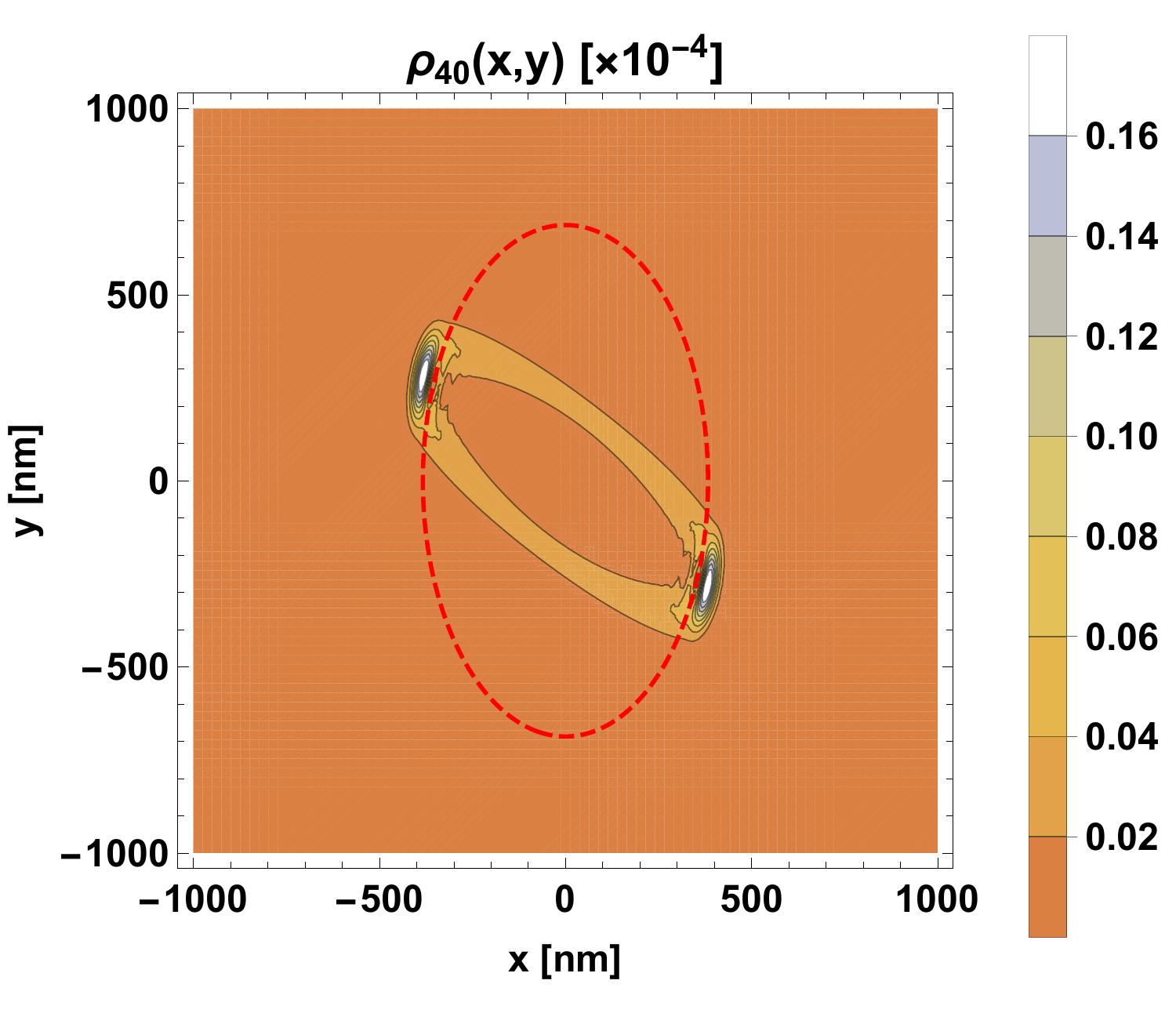}&
		\includegraphics[trim = 0mm 0mm 0mm 0mm, scale= 0.34, clip]{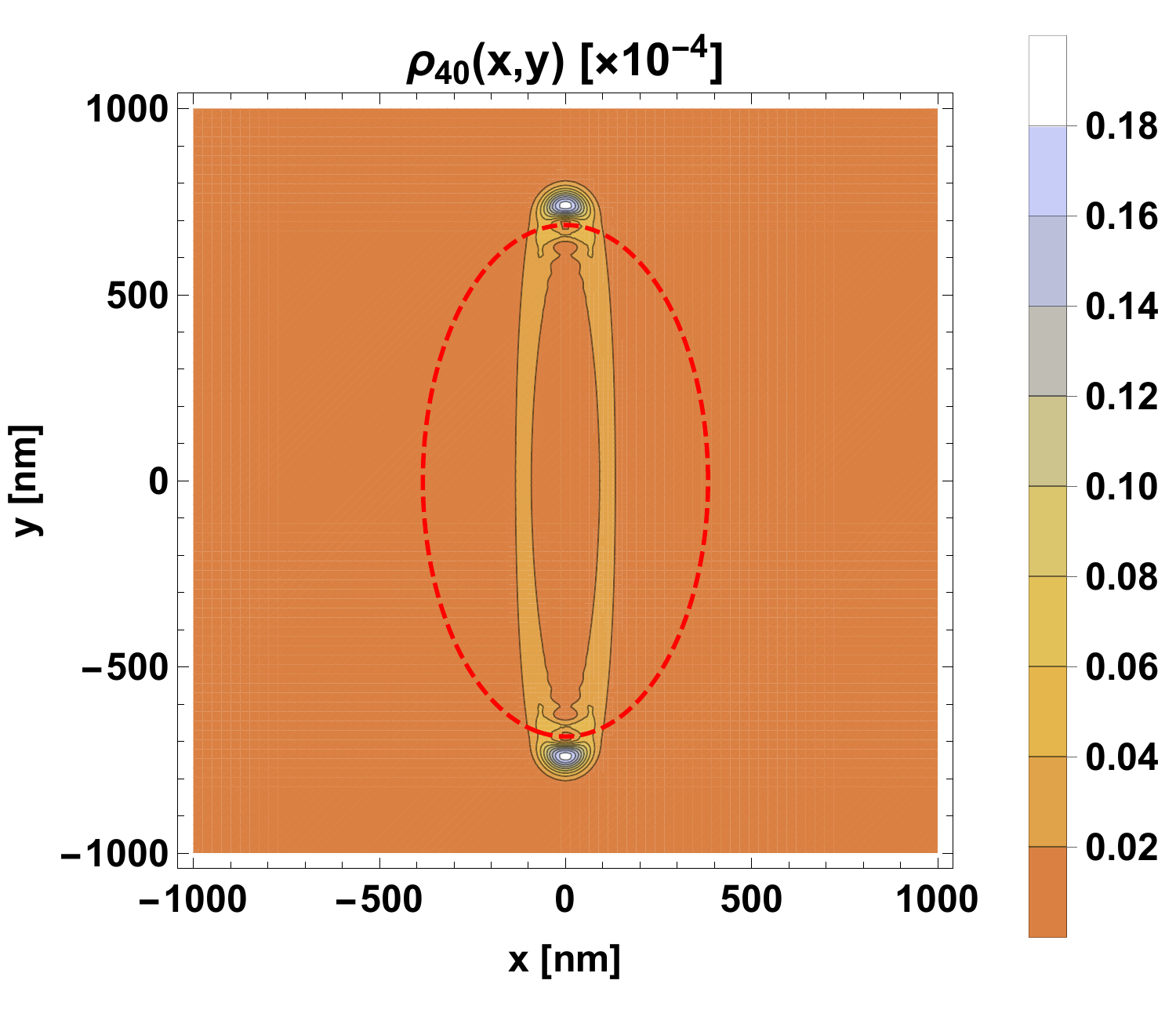}
	\end{tabular}
	\caption{\label{fig:density3}Probability density $\rho_{\nu}(x,y)$ in Eq.~(\ref{density}) for different values of $\alpha$ (from left to right) and $\epsilon=-20\%$ along (a-c) the $\mathcal{Z}$ direction and (d-f) the $\mathcal{A}$ direction. The red dashed line shows the classical trajectory in~(\ref{path1}). $\nu=40$, $B_{0}=0.3$ T, and $\beta=1/2$.}
\end{figure}

\section{$SU(2)$ coherent states}\label{sec4:SUstates}
\begin{figure}[h!]
	\centering
	\begin{tabular}{ccc}
	(a) $\alpha=1/\sqrt{2}$ \qquad \qquad \quad & (b) $\alpha=\exp\left(i\pi/3\right)/\sqrt{2}$ \qquad \qquad & (c) $\alpha=-1/\sqrt{2}$ \qquad \qquad \\
	\includegraphics[trim = 0mm 0mm 0mm 0mm, scale= 0.34, clip]{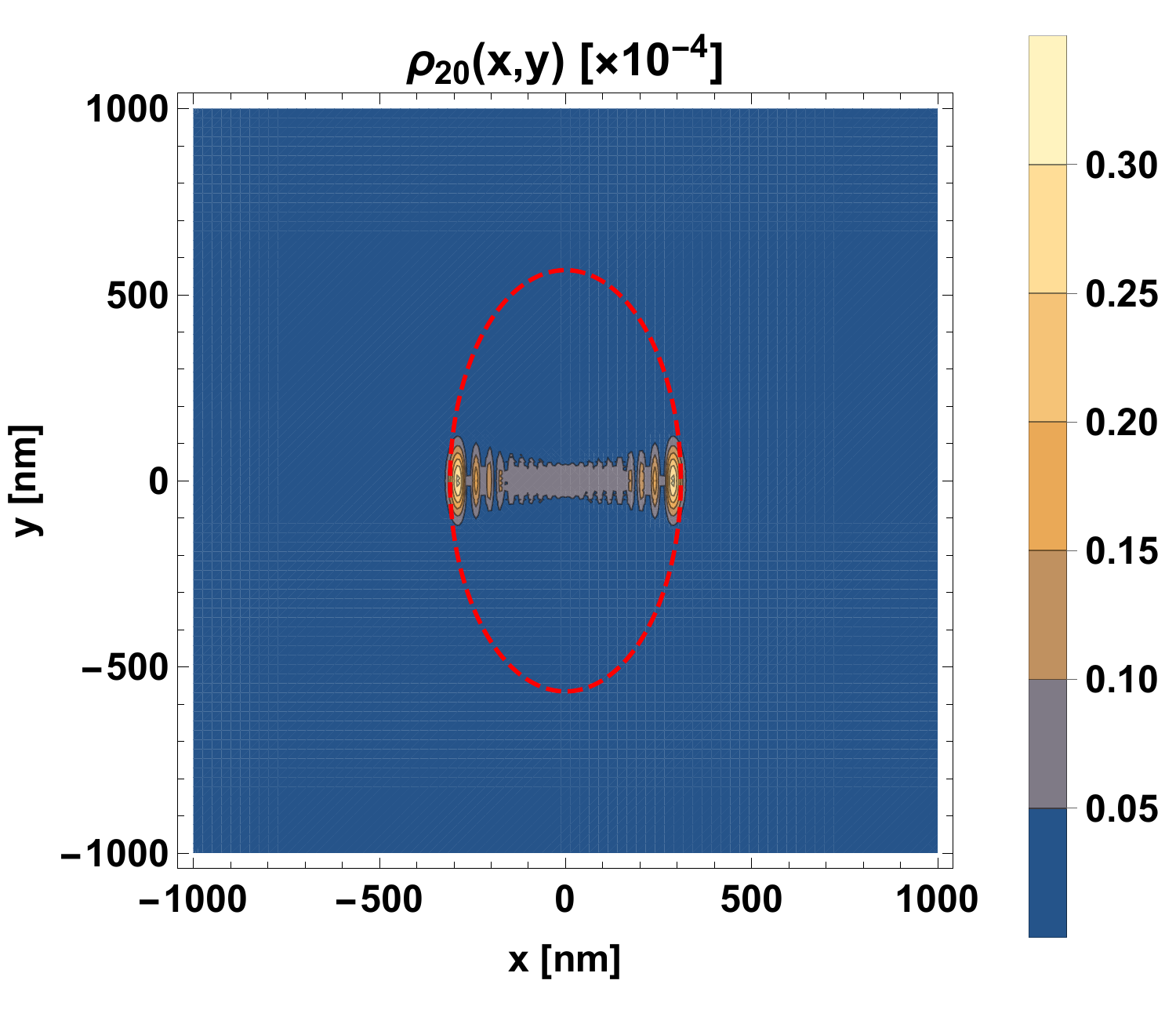} &
	\includegraphics[trim = 0mm 0mm 0mm 0mm, scale= 0.34, clip]{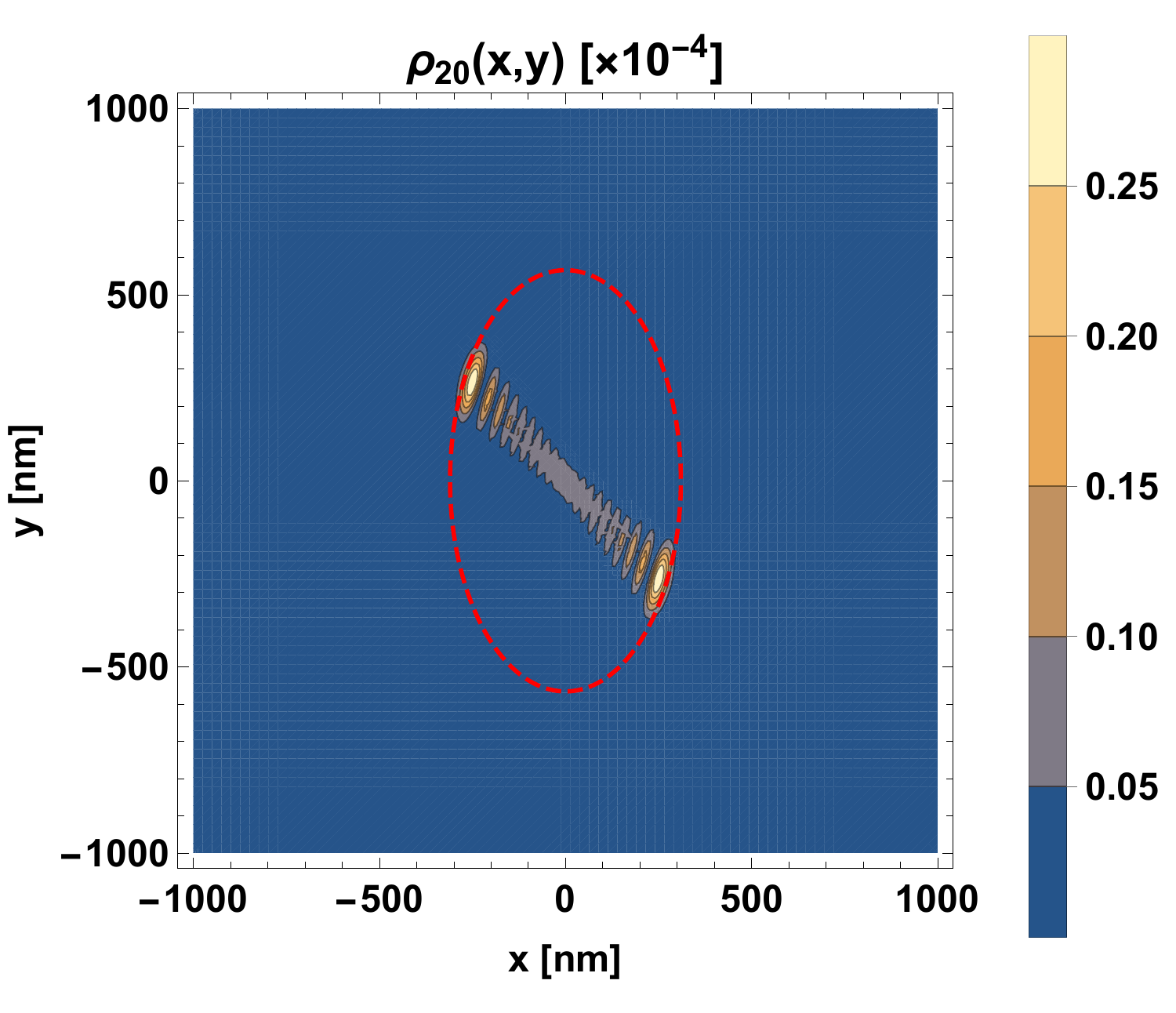}&
	\includegraphics[trim = 0mm 0mm 0mm 0mm, scale= 0.34, clip]{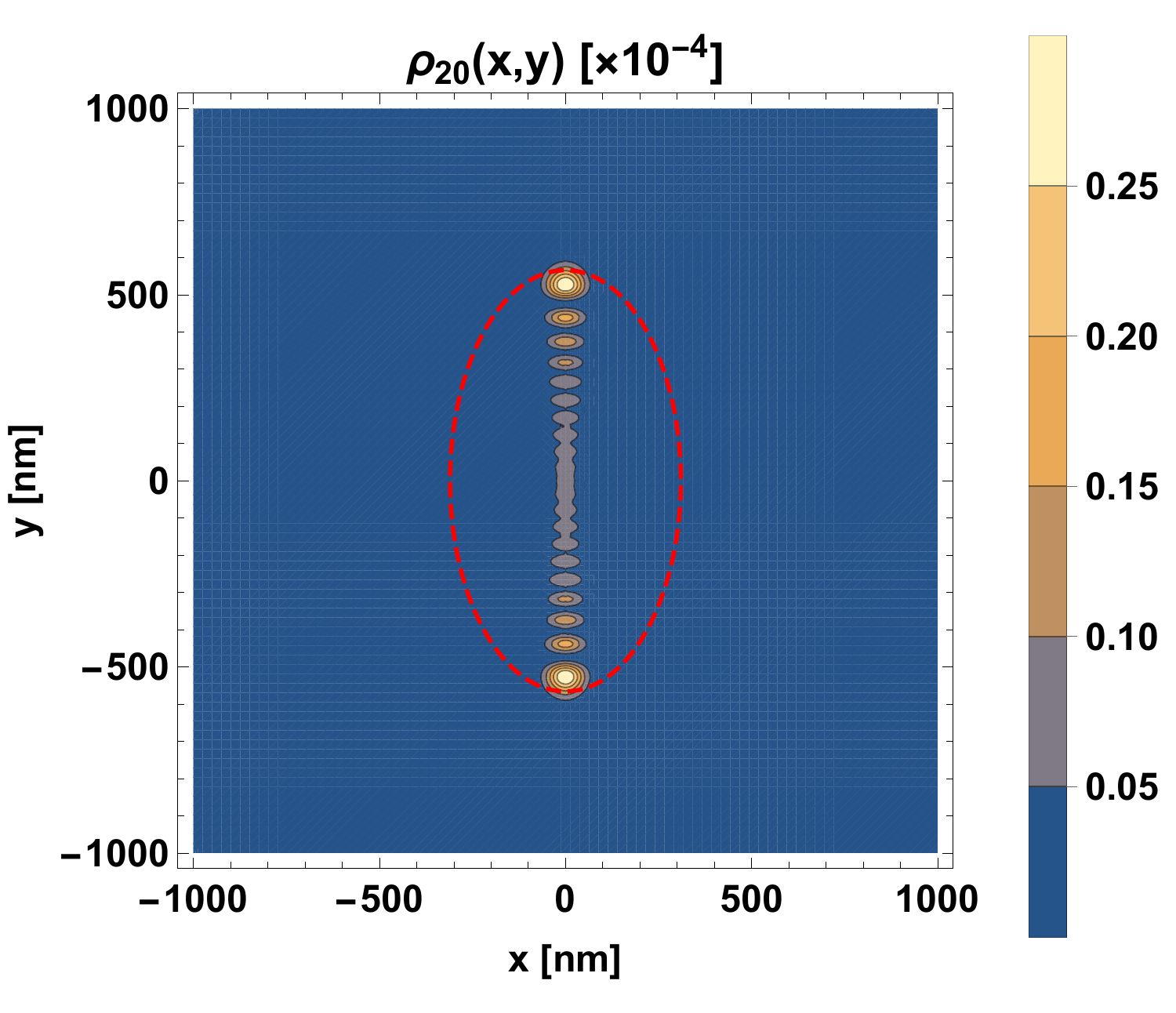}
\end{tabular}
	\begin{tabular}{ccc}
	(d) {\color{white}$\alpha=1/\sqrt{2}$} \qquad \qquad \quad & (e) {\color{white}$\alpha=\exp\left(i\pi/3\right)/\sqrt{2}$} \qquad \qquad & (f) {\color{white}$\alpha=-1/\sqrt{2}$} \qquad \qquad \\
	\includegraphics[trim = 0mm 0mm 0mm 0mm, scale= 0.34, clip]{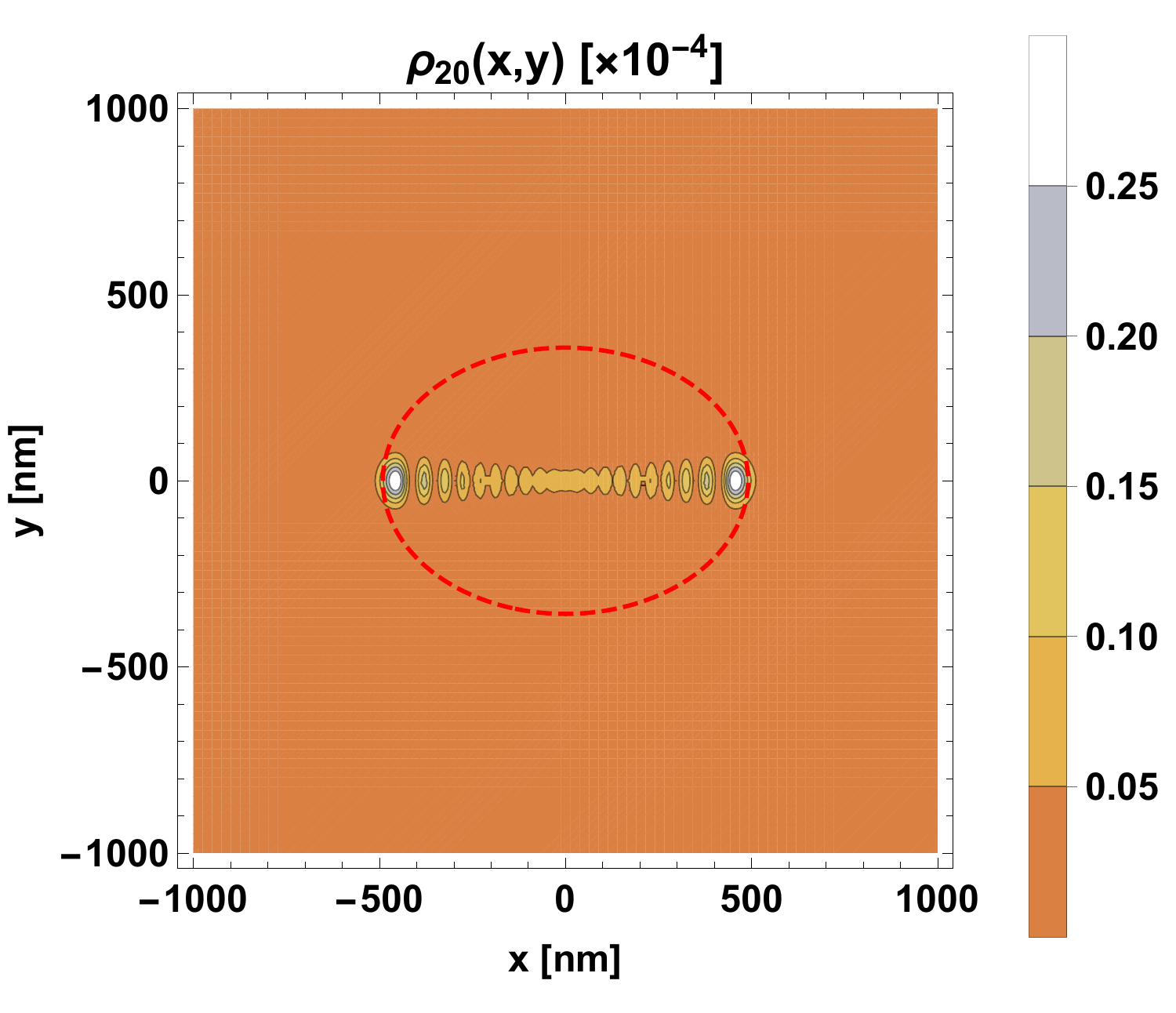} &
	\includegraphics[trim = 0mm 0mm 0mm 0mm, scale= 0.34, clip]{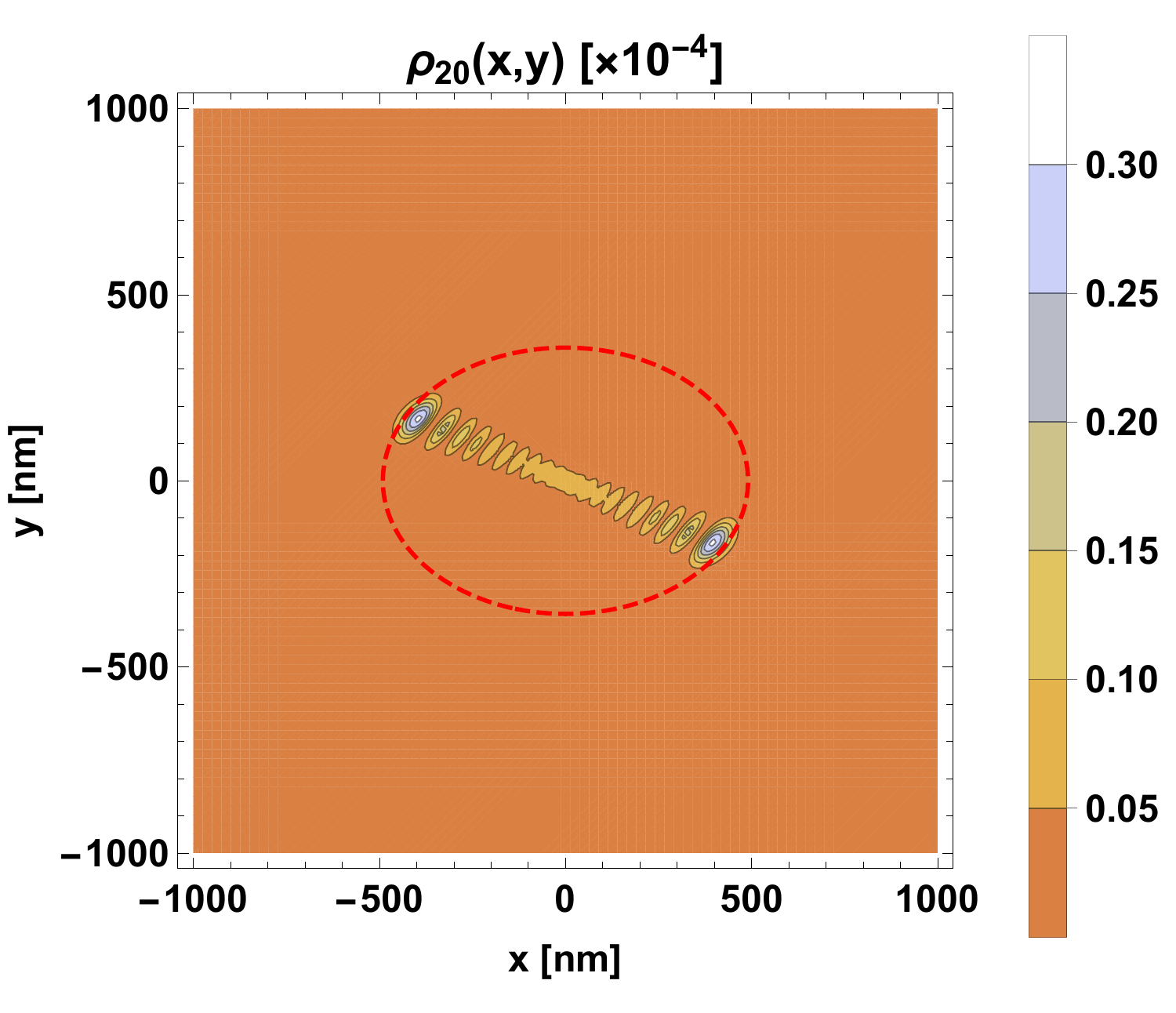}&
	\includegraphics[trim = 0mm 0mm 0mm 0mm, scale= 0.34, clip]{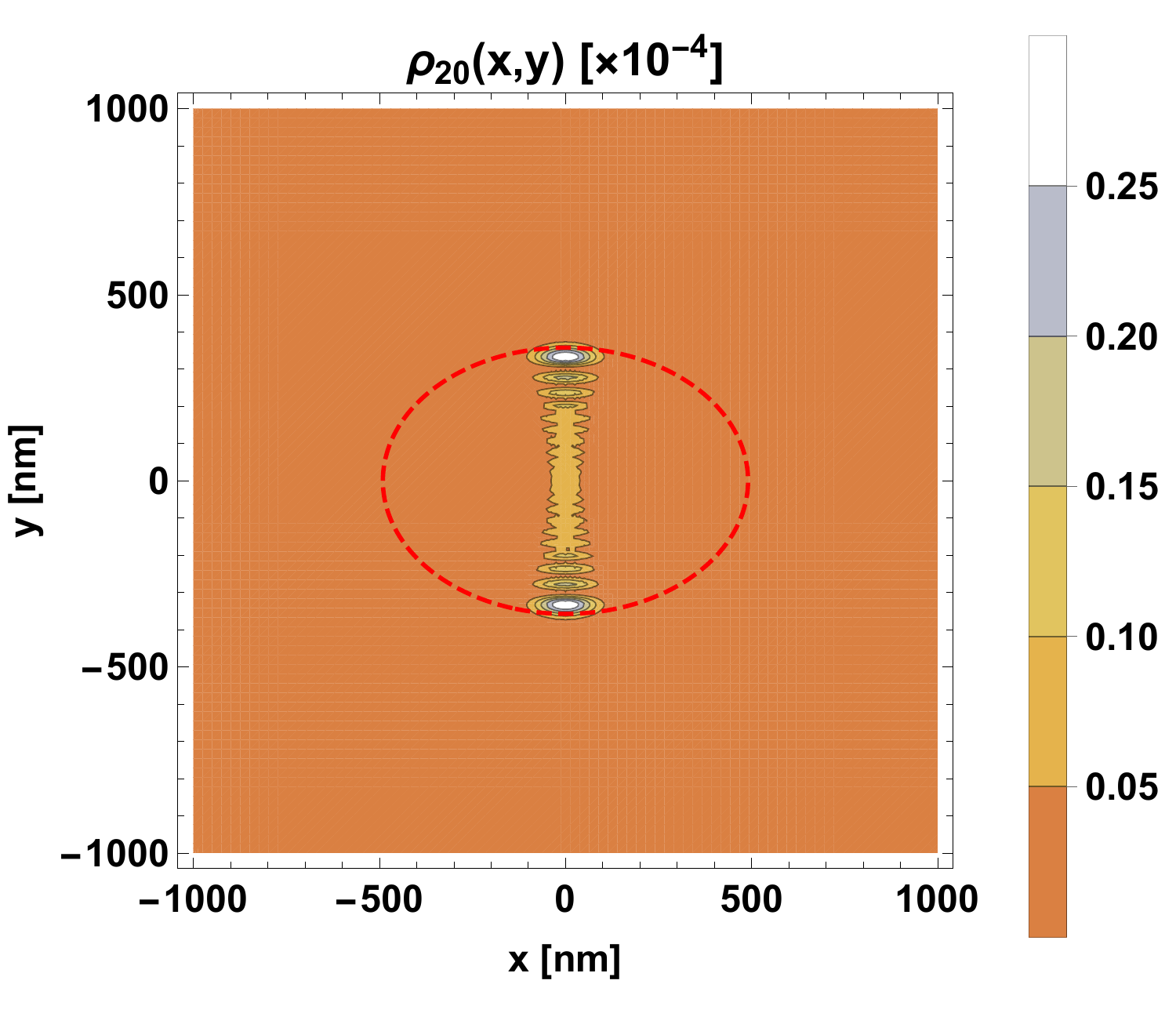}
\end{tabular}
	\caption{\label{fig:density2}Probability density $\rho_{\nu}(x,y)$ in Eq.~(\ref{density}) for different values of $\alpha$ (from left to right) and $\epsilon=20\%$ along (a-c) the $\mathcal{Z}$ direction and (d-f) the $\mathcal{A}$ direction. The red dashed line shows the classical trajectory in~(\ref{path1}). $\nu=20$, $B_{0}=0.3$ T, and $\beta=1/\sqrt{2}$.}
\end{figure}

By adopting the terminology in~\cite{moran2019}, let us define a type of states labeled by the occupation number $\nu=m+n$, constructed as the following linear combination (see Fig.~\ref{fig:fig1}):
\begin{equation}\label{SU2}
\vert\Psi_{\nu}\rangle=\sum_{n=0}^{\nu}\alpha^{n}\beta^{\nu-n}\sqrt{\left(\begin{array}{c}
	\nu \\ n
	\end{array}\right)}\vert\Psi_{\nu-n,n}\rangle, \quad \left(\begin{array}{c}
\nu \\ n
\end{array}\right)=\frac{\nu!}{(\nu-n)!\,n!},
\end{equation}
whose normalized probability density $\rho_{\nu}(x,y)$ is written as (see Figs.~\ref{fig:density3} and \ref{fig:density2})
\begin{equation}\label{density}
\rho_{\nu}(x,y)=\mathcal{N}_{\nu}^{2}\left(\left\vert\sum_{n=1}^{\nu}\alpha^{n}\beta^{\nu-n}\sqrt{\left(\begin{array}{c}
	\nu \\ n
	\end{array}\right)}\psi_{\nu-n,n-1}(x,y)\right\vert^2+\left\vert\sum_{n=0}^{\nu}\alpha^{n}\beta^{\nu-n}\sqrt{\left(\begin{array}{c}
	\nu \\ n
	\end{array}\right)}\psi_{\nu-n,n}(x,y)\right\vert^2\right),
\end{equation}
where $\mathcal{N}_{\nu}^{-1}=\sqrt{2-\vert\beta\vert^{2\nu}}$ and the wave functions in Cartesian coordinates are given by
\begin{equation}
\psi_{m,n}(x,y)=\left\{\begin{array}{c c}
(-1)^{m}\sqrt{\frac{\omega_{\rm B}}{4\pi}\frac{m!}{n!}}\xi^{n-m}\exp\left(-\frac{\vert\xi\vert^2}{2}\right)L_{m}^{n-m}(\vert\xi\vert^2), & n\geq m,\\
(-1)^{n}\sqrt{\frac{\omega_{\rm B}}{4\pi}\frac{n!}{m!}}\xi^{\ast m-n}\exp\left(-\frac{\vert\xi\vert^2}{2}\right)L_{n}^{m-n}(\vert\xi\vert^2), & m\geq n,
\end{array}\right.
\end{equation}
with $\xi=\frac{\sqrt{\omega_{\rm B}}}{2}(\zeta^{-1}x+i\zeta y)$ and $L_{k}^{\alpha}(x)$ denotes the generalized Laguerre polynomials as before. Hence, the whole Hilbert space $\mathcal{H}$, spanned by the spinors $\vert\Psi_{m,n}\rangle$, is now decomposed into $(\nu+1)$-dimensional subspaces $\mathcal{H}_{\nu}$,
\begin{equation}
\mathcal{H}_{\nu}={\rm span}\{\vert\Psi_{\nu}\rangle\,\vert\,\nu=0,1,\dots\} \quad \Longrightarrow\quad \mathcal{H}=\bigoplus_{\nu=0}^{\infty}\mathcal{H}_{\nu}.
\end{equation}
Thus, the states $\vert\Psi_{\nu}\rangle$ constitute a kind of $SU(2)$ coherent states in the Schwinger boson representation~\cite{ichihashi78,Novaes02,auerbach11}.

\begin{figure}[h!]
	\centering
	\includegraphics[width=0.45\linewidth]{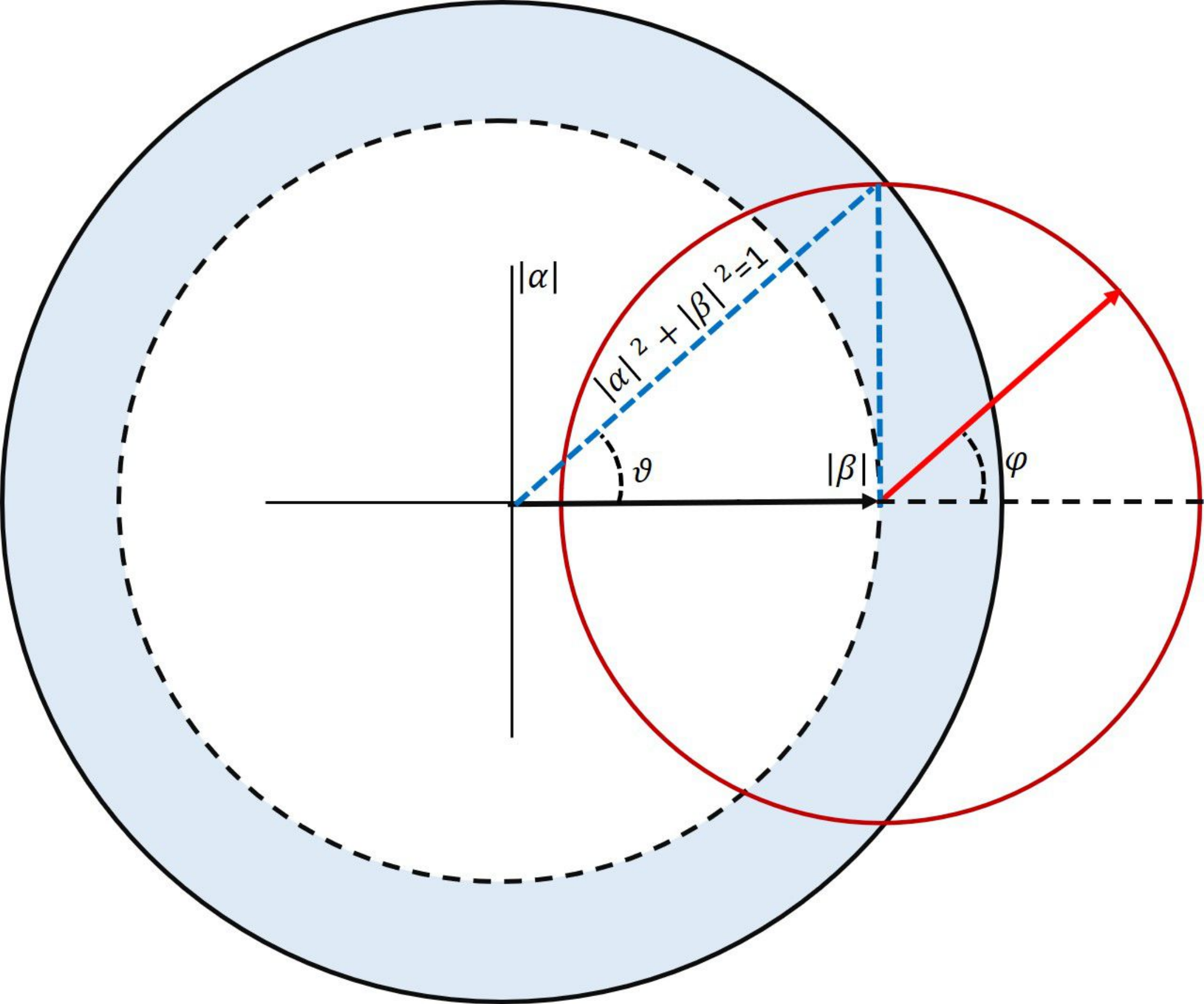}
	\caption{Unit circle $\vert\alpha\vert^{2}+\vert\beta\vert^{2}=1$ shown in a black continuous line. The horizontal (vertical) axis corresponds to $\vert\beta\vert$ ($\vert\alpha\vert$) and the natural relative phase $\vartheta$ is also indicated (blue dotted triangle). The relative phase $\varphi$ between $\alpha$ and $\beta$ is defined in the counterclockwise direction in the red circle of radius $\vert\alpha\vert$.}
	\label{fig:fig2}
\end{figure}

Figures~\ref{fig:density3} and~\ref{fig:density2} show how the probability density $\rho_{\nu}(x,y)$ is affected by the deformations applied along the zigzag and armchair directions, respectively, with strain values close to $20\%$. As we can see, when either a compression or a tensile deformation is applied along $\mathcal{Z}$ and $\mathcal{A}$ directions, two peaks appear in opposite positions over the classical trajectory. In addition, a relative phase $\varphi$ between $\alpha$ and $\beta$ causes that the probability density peaks are concentrated on a line whose angle with the $x$-axis turns out to be $\theta_{x}=\varphi/2$. Besides, as the relative phase $\varphi$ changes, the maximum values of $\rho_{\nu}(x,y)$ follow an elliptical path whose semi-major axis is aligned to either the $x$-axis or the $y$-axis, according to which direction the deformation is applied (see Table~\ref{tab:table1}). Moreover, there is a natural relative phase $\vartheta$ between $\vert\alpha\vert$ and $\vert\beta\vert$ that is determined as $\tan(\vartheta)=\vert\alpha\vert/\vert\beta\vert$ in the unit circle $\vert\alpha\vert^{2}+\vert\beta\vert^{2}=1$ (see Fig.~\ref{fig:fig2}).


\begin{table}[h!]
	\begin{center}
		\caption{Semi-major axis direction according to the type of deformation applied in the graphene layer.}
		\label{tab:table1}
		\renewcommand{\arraystretch}{1.1}
		\begin{tabular}{c|c|c} 
			\textbf{Lattice} & \textbf{Compression} & \textbf{Tensile}\\
			\textbf{direction} & \textbf{deformation ($\epsilon<0$)} & \textbf{deformation ($\epsilon>0$)} \\
			\hline
			Zigzag ($\mathcal{Z}$) & $x$-aligned & $y$-aligned\\
			Armchair ($\mathcal{A}$) & $y$-aligned & $x$-aligned
		\end{tabular}
	\end{center}
\end{table}

\subsection{Orthogonal condition and completeness relation}
The $SU(2)$ coherent states in Eq.~(\ref{SU2}) satisfy the orthogonal condition,
\begin{align}
\nonumber\langle\Psi_{\mu}\vert\Psi_{\nu}\rangle&=\left(\sum_{m=0}^{\mu}\delta^{\ast m}\eta^{\ast \mu-m}\sqrt{\left(\begin{array}{c}
	\mu \\ m
	\end{array}\right)}\langle\Psi_{\mu-m,m}\vert\right)\left(\sum_{n=0}^{\nu}\alpha^{n}\beta^{\nu-n}\sqrt{\left(\begin{array}{c}
	\nu \\ n
	\end{array}\right)}\vert\Psi_{\nu-n,n}\rangle\right) \\
&=\left[2(\alpha\delta^{\ast}+\beta\eta^{\ast})^{\nu}-(\beta\eta^\ast)^{\nu}\right]\delta_{\mu\nu},
\end{align}
which reduces to the standard orthonormal condition up to the normalization constant, when $\delta=\alpha$ and $\eta=\beta$:
\begin{equation}
\langle\Psi_{\mu}\vert\Psi_{\nu}\rangle=\left(2-\vert\beta\vert^{2\nu}\right)\delta_{\mu\nu}.
\end{equation}

On the other hand, by considering the measure
\begin{equation}
d\mu(\alpha,\beta)=\frac{\exp\left(-\vert\alpha\vert^{2}-\vert\beta\vert^{2}\right)}{\Gamma(\nu+1)}\,\delta\left(\vert\alpha\vert^{2}+\vert\beta\vert^{2}-1\right)d^{2}\alpha\,d^{2}\beta,
\end{equation}
the $SU(2)$ coherent states resolve the identity in the following way:
\begin{align}
\nonumber\frac{1}{\pi^{2}}\int_{\mathbb{C}^{3}}d\mu(\alpha,\beta)\vert\Psi_{\nu}\rangle\langle\Psi_{\nu}\vert&=\frac{1}{\pi^{2}}\int_{\mathbb{C}^{3}}d^{2}\alpha\,d^{2}\beta\frac{\exp\left(-\vert\alpha\vert^{2}-\vert\beta\vert^{2}\right)}{\Gamma(\nu+1)}\,\delta\left(\vert\alpha\vert^{2}+\vert\beta\vert^{2}-1\right)\vert\Psi_{\nu}\rangle\langle\Psi_{\nu}\vert \\
&=\sum_{n=0}^{\nu}\vert\Psi_{\nu-n,n}\rangle\langle\Psi_{\nu-n,n}\vert=\mathbb{I}_{\nu},
\end{align}
where $\mathbb{I}_{\nu}$ is the identity operator for the states $\vert\Psi_{\nu}\rangle$ in each subspace $\mathcal{H}_{\nu}$ and the $\delta$-distribution guarantees that the integration is performed in the circle $\vert\alpha\vert^{2}+\vert\beta\vert^{2}=1$ in the complex plane. Hence, the identity operator in the Hilbert space $\mathcal{H}$ can be expressed as
\begin{equation}\label{identity}
\mathbb{I}=\bigoplus_{\nu=0}^{\infty}\mathbb{I}_{\nu}.
\end{equation}

\subsection{Cyclotron motion}
The operators $B^\pm$ are related with the so-called magnetic translation operators, which clasically correspond to the coordinates of the center of a circular orbit \cite{z64,b64,l83}, 
\begin{equation}\label{occo}
x_{0}=x-\frac{2\Pi_{y}}{\omega_{\rm B}\hbar}, \quad y_{0}=y+\frac{2\Pi_{x}}{\omega_{\rm B}\hbar}.
\end{equation}
In terms of the orbit center-coordinate operators, we have that
\begin{equation}
B^{-}=\frac{\sqrt{\omega_{\rm B}}}{2}\left(\zeta^{-1/2}x_{0}+i\zeta^{1/2}y_{0}\right), \quad B^{+}=\frac{\sqrt{\omega_{\rm B}}}{2}\left(\zeta^{-1/2}x_{0}-i\zeta^{1/2}y_{0}\right),
\end{equation}
or reciprocally,
\begin{equation}
x_{0}=\frac{\zeta^{1/2}}{\sqrt{\omega_{\rm B}}}\left(B^{-}+B^{+}\right), \quad  y_{0}=\frac{\zeta^{-1/2}}{i\sqrt{\omega_{\rm B}}}\left(B^{-}-B^{+}\right).
\end{equation}
In this system, the so-called magnetic translation operator $\vec{K}=\vec{\Pi}+e\,\vec{r}\times\vec{B}_{0}$, where $\vec{K}=\omega_{\rm B}\hbar(y_{0},-x_{0})/2$, is the generator of translations. From a classical approach, the position of the center of circular motion is a constant of motion leading to the degeneracy of LLs since the electron energy does not depend on the orbit center~\cite{Goerbig}.

\begin{figure}[h!]
	\centering
	\includegraphics[width=0.45\linewidth]{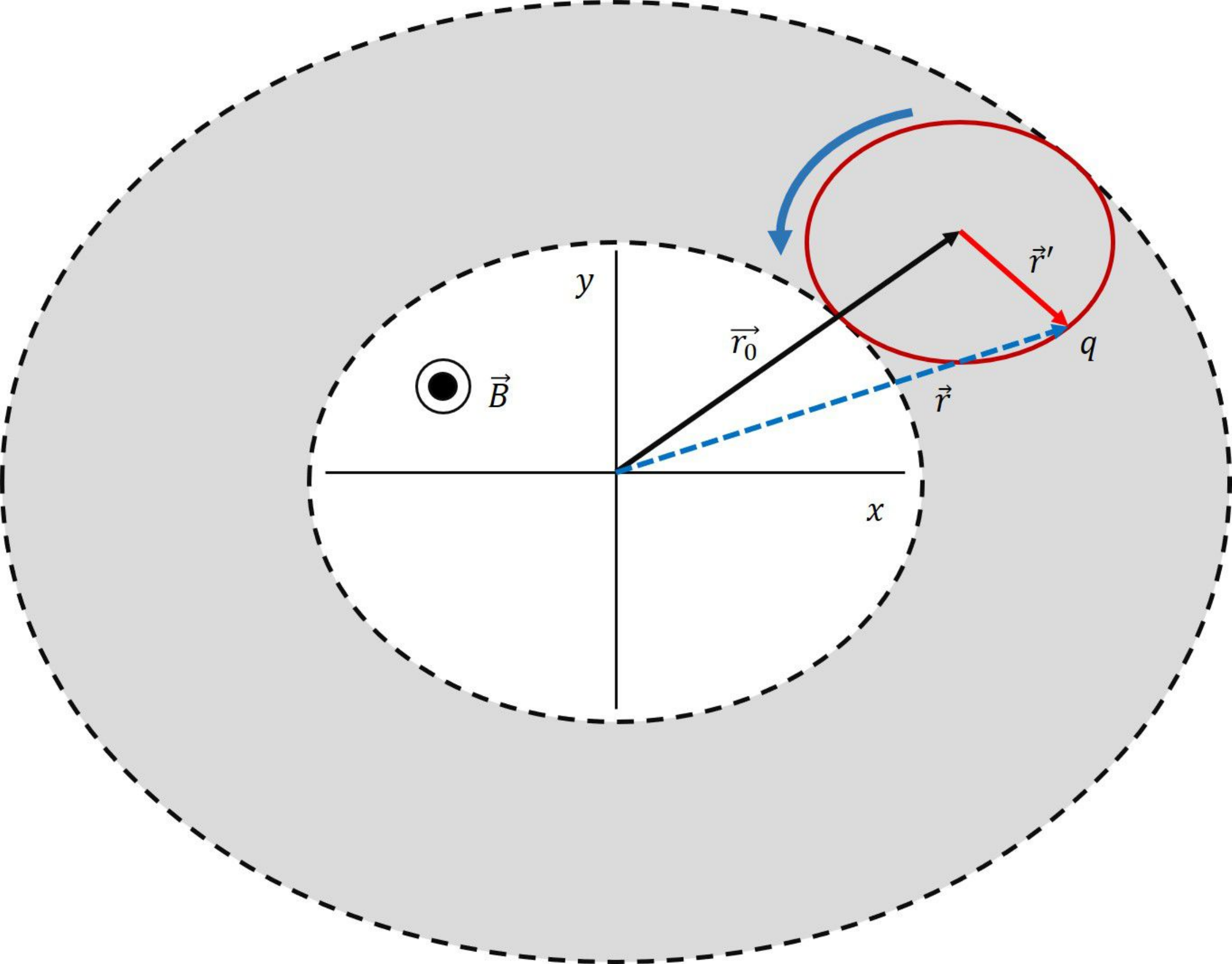}
	\caption{Classical elliptic trajectory for an electron ($q=-e$) in a homogeneous magnetic field. The vector $\vec{r}_{0}=(x_{0},y_{0})$ (in black) locates the center of the elliptic path around which the charge particle moves, while $\vec{r}'$ (in red) denotes the vector position of the particle respect to the point $(x_{0},y_{0})$.}
	\label{fig:fig3}
\end{figure}

By defining the matrix orbit center-coordinate operators as
\begin{equation}
\mathbb{X}_{0}=x_{0}\otimes\mathbb{I}, \quad \mathbb{Y}_{0}=y_{0}\otimes\mathbb{I},
\end{equation}
it follows that for the $SU(2)$ coherent states,
\begin{subequations}\label{meanvalues1}
	\begin{align}
\langle\mathbb{X}_{0}\rangle_{\nu}^{\alpha,\beta}&=0, \quad \langle\mathbb{Y}_{0}\rangle_{\nu}^{\alpha,\beta}=0, \\
\zeta^{-1}\langle\mathbb{X}_{0}^{2}\rangle_{\nu}^{\alpha,\beta}=\zeta\langle\mathbb{Y}_{0}^{2}\rangle_{\nu}^{\alpha,\beta}&=
\frac{\left(4\nu\vert\beta\vert^{2}-\left(2\nu+1\right)\vert\beta\vert^{2\nu}+2\right)l_{\rm B}^{2}}{2\left(2-\vert\beta\vert^{2\nu}\right)},
\end{align}
\end{subequations}
where $l_{\rm B}^{2}=2/\omega_{\rm B}$ is the magnetic length. 

Now, we consider the position observable of a particle in a circle as
\begin{equation}
r_{x}=x-x_{0}=\frac{2\Pi_{y}}{\omega_{\rm B}\hbar}, \quad r_{y}=y-y_{0}=-\frac{2\Pi_{x}}{\omega_{\rm B}\hbar},
\end{equation}
which correspond to the coordinates of the radius vector of a charge particle moving in a closed path with center at the point $(x_{0},y_{0})$ (see Fig.~\ref{fig:fig3}). These operators can be expressed in terms of the ladder operators $A^{\pm}$ as follows:
\begin{equation}
r_{x}=\frac{\zeta^{1/2}}{\sqrt{\omega_{\rm B}}}\left(A^{-}+A^{+}\right), \quad r_{y}=-\frac{\,\zeta^{-1/2}}{i\sqrt{\omega_{\rm B}}}\left(A^{-}-A^{+}\right).
\end{equation}

For the $SU(2)$ coherent states, we define the matrix operators,
\begin{equation}
\mathbb{R}_{x}=r_{x}\otimes\mathbb{I}, \quad \mathbb{R}_{y}=r_{y}\otimes\mathbb{I},
\end{equation}
and thus,
\begin{subequations}\label{meanvalues2}
	\begin{align}
	\langle \mathbb{R}_{x}\rangle_{\nu}^{\alpha,\beta}&=0, \quad \langle \mathbb{R}_{y}\rangle_{\nu}^{\alpha,\beta}=0, \\
	\zeta^{-1}\langle\mathbb{R}_{x}^{2}\rangle_{\nu}^{\alpha,\beta}=\zeta\langle\mathbb{R}_{y}^{2}\rangle_{\nu}^{\alpha,\beta}&=
	\frac{\left(4\nu\vert\alpha\vert^{2}+\left(1-\vert\alpha\vert^{2}\right)^{\nu}\right)l_{\rm B}^{2}}{2\left(2-(1-\vert\alpha\vert^{2})^{\nu}\right)}.
	\end{align}
\end{subequations}

\begin{figure}[h!]
	\centering
	\includegraphics[width=0.6\linewidth]{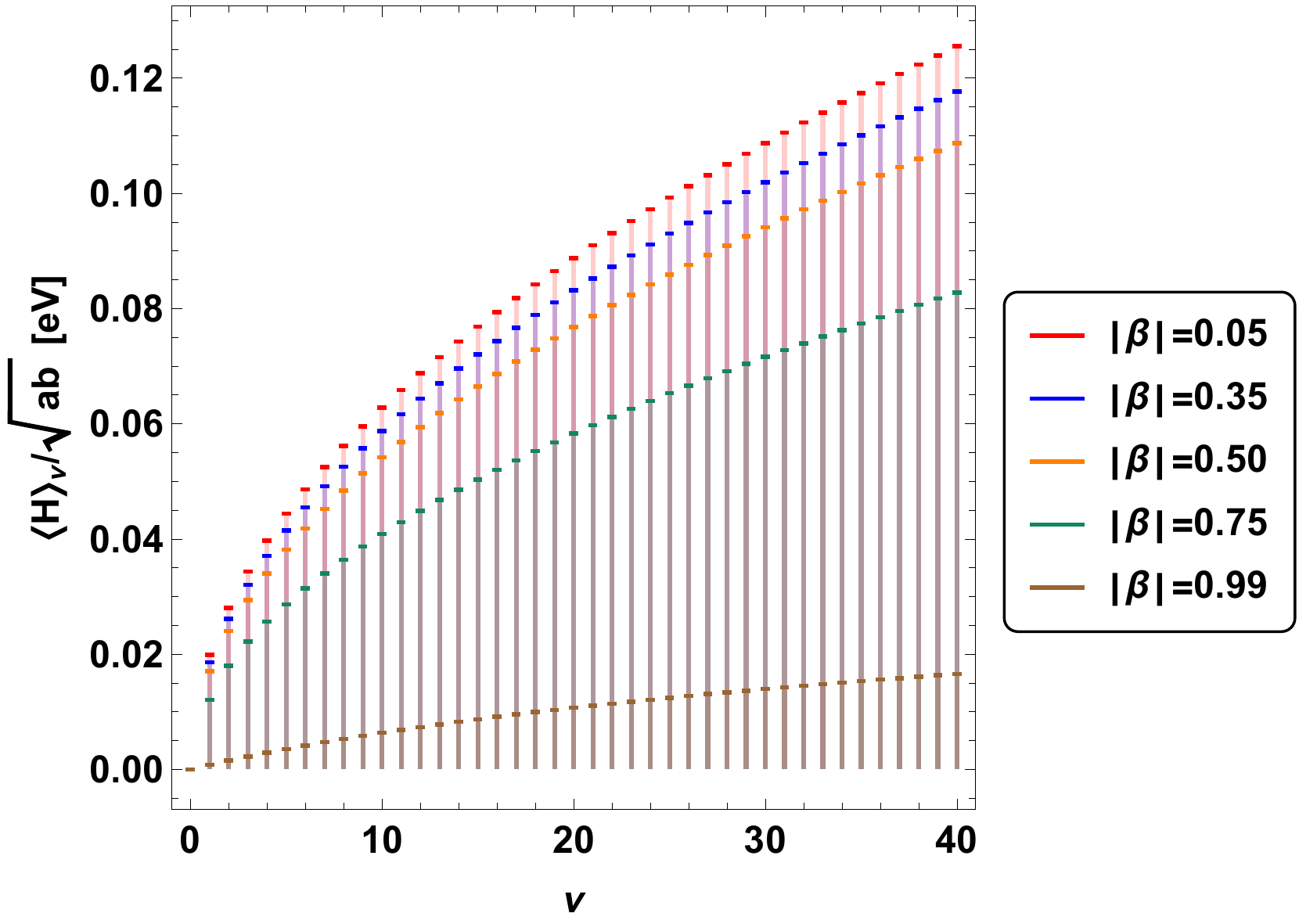}
	\caption{Mean energy value $\langle H\rangle_{\nu}/\sqrt{ab}$ is shown for different values of $\vert\beta\vert$ and $B_{0}=0.3$ T.}
	\label{fig:Hnu}
\end{figure}

Let us now consider the operator of the square of the distance of the center of the circle from the coordinate origin,
\begin{equation}
r_{0}^2=\zeta^{-1}x_{0}^{2}+\zeta y_{0}^{2}=\frac{2}{\omega_{\rm B}}\left(2B^{+}B^{-}+1\right),
\end{equation}
and the expression for the classical elliptical trajectory, which is given by
\begin{equation}
r^{2}=\zeta^{-1}r_{x}^{2}+\zeta\,r_{y}^{2}=\zeta^{-1}\left(x-x_{0}\right)^{2}+\zeta\left(y-y_{0}\right)^{2}=\frac{2}{\omega_{\rm B}}\left(2A^{+}A^{-}+1\right).
\end{equation}
Setting the following matrix operators:
\begin{equation}
\mathbb{R}_{0}^{2}=r_{0}^{2}\otimes\mathbb{I}, \quad \mathbb{R}^{2}=r^{2}\otimes\mathbb{I},
\end{equation}
it is straightforward to show that the expected value of $\mathbb{R}_{0}^{2}$ and $\mathbb{R}^{2}$ in the basis of the eigenstates~(\ref{eigenstates}) is
\begin{equation}
\langle\mathbb{R}_{0}^{2}\rangle=\frac{2\left(2m+1\right)}{\omega_{\rm B}}=(2m+1)\,l_{\rm B}^{2}, \quad \langle\mathbb{R}^{2}\rangle=\begin{cases}
l_{\rm B}^{2}, & n=0, \\
2n\,l_{\rm B}^{2}, & n>0,
\end{cases}
\end{equation}
while for the $SU(2)$ coherent states, we have
\begin{subequations}
	\begin{align}
\langle\mathbb{R}_{0}^{2}\rangle_{\nu}^{\alpha,\beta}&=\zeta^{-1}\langle\mathbb{X}_{0}^{2}\rangle_{\nu}^{\alpha,\beta}+\zeta\langle\mathbb{Y}_{0}^{2}\rangle_{\nu}^{\alpha,\beta}=
\frac{\left(4\,\nu\vert\beta\vert^{2}-(2\nu+1)\vert\beta\vert^{2\nu}+2\right)l_{\rm B}^{2}}{2-\vert\beta\vert^{2\nu}}, \\
\langle\mathbb{R}^{2}\rangle_{\nu}^{\alpha,\beta}&=\zeta^{-1}\langle\mathbb{R}_{x}^{2}\rangle_{\nu}^{\alpha,\beta}+\zeta\langle\mathbb{R}_{y}^{2}\rangle_{\nu}^{\alpha,\beta}=
\frac{\left(4\nu\vert\alpha\vert^{2}+\left(1-\vert\alpha\vert^{2}\right)^{\nu}\right)l_{\rm B}^{2}}{2-(1-\vert\alpha\vert^{2})^{\nu}}.
\end{align}
\end{subequations}

Coming back to Figs.~\ref{fig:density3} and \ref{fig:density2}, they show the classical trajectory
\begin{equation}\label{path1}
\frac{x^{2}}{\varepsilon\,\zeta\langle\mathbb{R}^{2}\rangle_{\nu}^{\alpha,\beta}}+\frac{y^{2}}{\varepsilon\,\zeta^{-1}\langle\mathbb{R}^{2}\rangle_{\nu}^{\alpha,\beta}}=1,
\end{equation}
where $\varepsilon=2$ ($\varepsilon=4$) for $\epsilon<0$ ($\epsilon>0$) has been set by hand.

\subsection{Mean energy value}
Considering the Hamiltonian in Eq.~(\ref{3}), we compute the mean energy value for the $SU(2)$ coherent states (see Fig.~\ref{fig:Hnu}),
\begin{equation}\label{Hnu}
\frac{\langle H\rangle_{\nu}}{\hbar\,v'_{F}\sqrt{\omega_{\rm B}}}=\frac{2}{2-\vert\beta\vert^{2\nu}}\sum_{n=1}^{\nu}\left(\begin{array}{c}
\nu \\ n
\end{array}\right)(1-\vert\beta\vert^{2})^{n}\vert\beta\vert^{2(\nu-n)}\sqrt{n}.
\end{equation}

\section{Schr\"{o}dinger-type 2D coherent states}\label{sec5:SCstates}

Let us now consider the action of the matrix operator $\mathbb{J}^-$ in Eq.~(\ref{annihop2}) onto the un-normalized states (\ref{SU2}),
\begin{equation}
\mathbb{J}^-\vert\Psi_{\nu}\rangle=\sqrt{\nu}\vert\Psi_{\nu-1}\rangle.
\end{equation}
Thus, we can define the Schr\"{o}dinger-type 2D coherent states $\vert\Psi_{z}\rangle$ as eigenstates of the annihilation operator $\mathbb{J}^-$ with complex eigenvalue $z\in\mathbb{C}$, i.e.,
\begin{equation}\label{eigeneq}
\mathbb{J}^-\vert\Psi_{z}\rangle=z\vert\Psi_{z}\rangle, \quad \vert\Psi_{z}\rangle=\sum_{\nu=0}^{\infty}c_{\nu}\vert\Psi_{\nu}\rangle.
\end{equation}

\begin{figure}[h!]
	\centering
	\begin{tabular}{ccc}
		(a) $\alpha=\sqrt{3}/2$ \qquad \qquad \quad & (b) $\alpha=\sqrt{3}\exp\left(i\pi/4\right)/2$ \qquad \qquad & (c) $\alpha=-\sqrt{3}/2$ \qquad \qquad \\
		\includegraphics[trim = 0mm 0mm 0mm 0mm, scale= 0.34, clip]{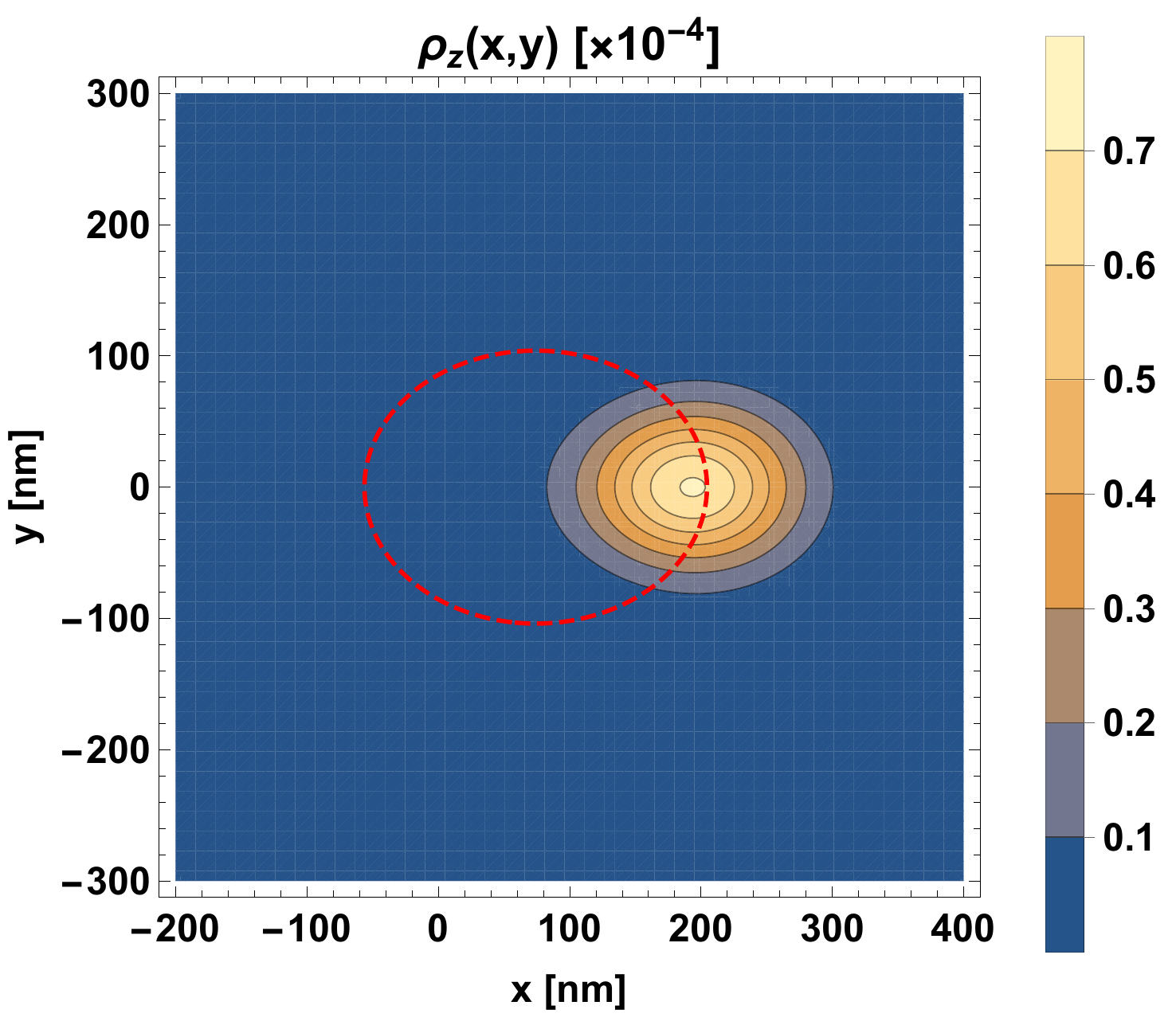} &
		\includegraphics[trim = 0mm 0mm 0mm 0mm, scale= 0.34, clip]{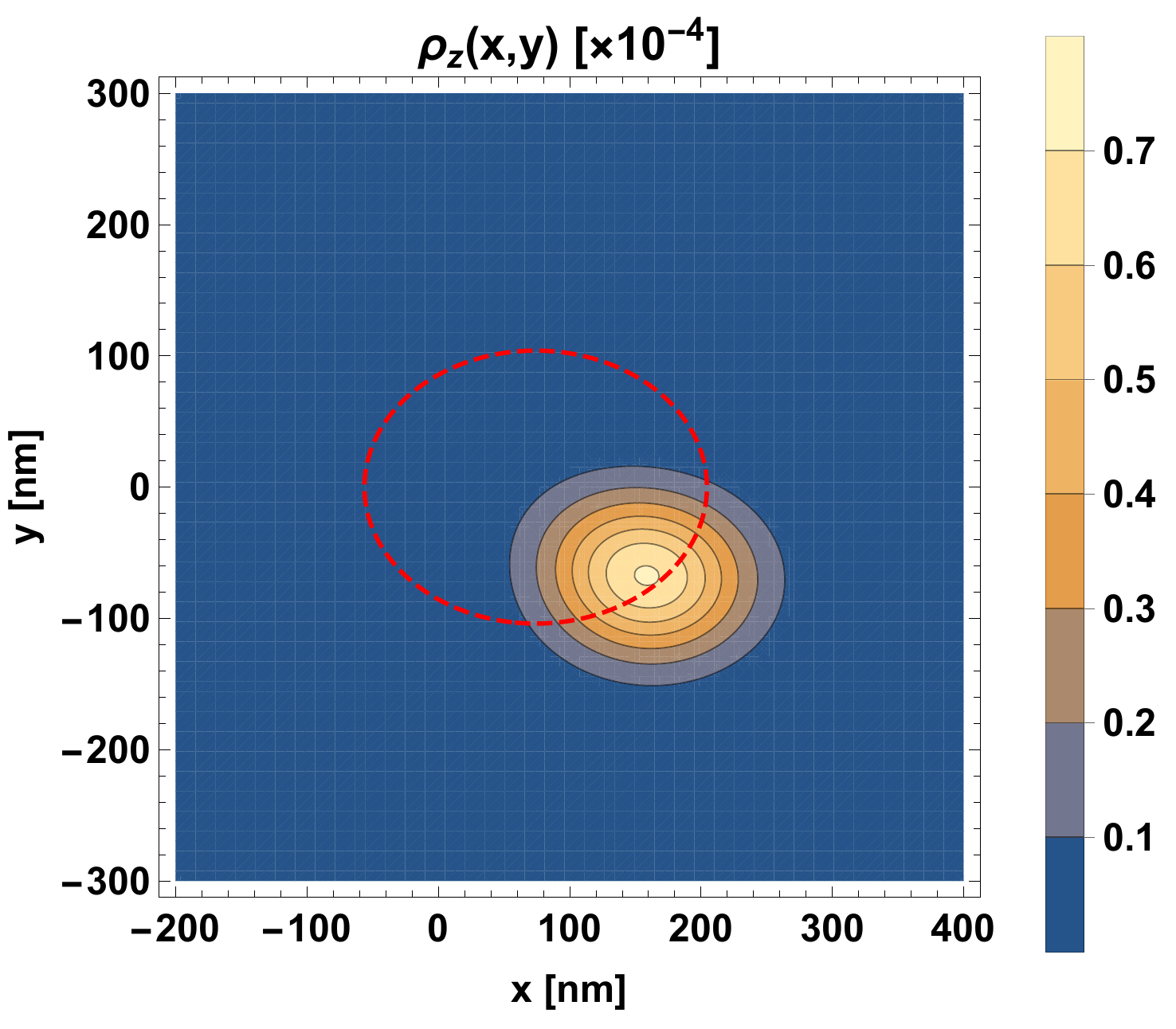}&
		\includegraphics[trim = 0mm 0mm 0mm 0mm, scale= 0.34, clip]{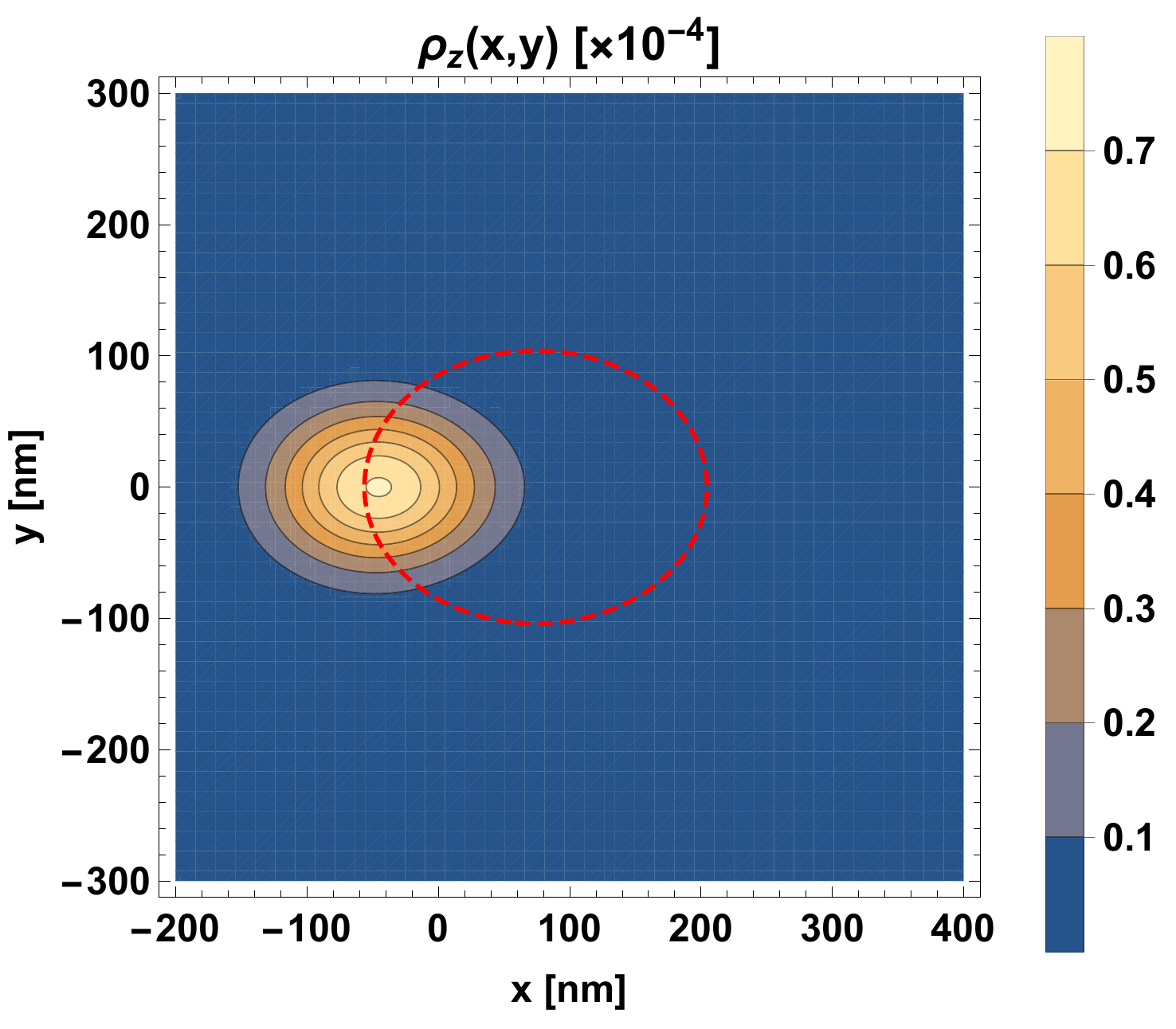}
	\end{tabular}
	\begin{tabular}{ccc}
		(d) {\color{white}$\alpha=\sqrt{3}/2$} \qquad \qquad \quad & (e) {\color{white}$\alpha=\sqrt{3}\exp\left(i\pi/4\right)/2$} \qquad \qquad & (f) {\color{white}$\alpha=-\sqrt{3}/2$} \qquad \qquad \\
		\includegraphics[trim = 0mm 0mm 0mm 0mm, scale= 0.34, clip]{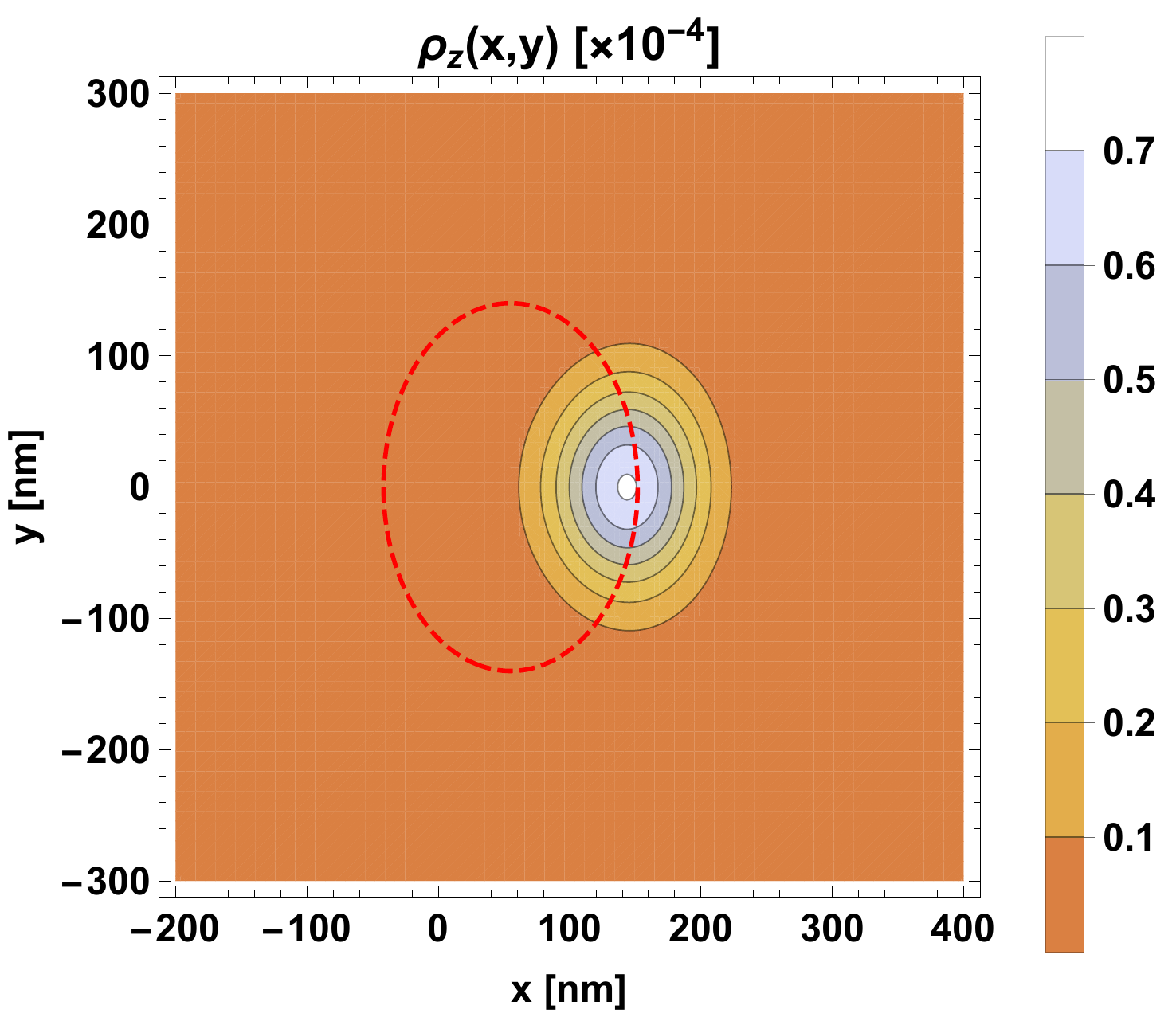} &
		\includegraphics[trim = 0mm 0mm 0mm 0mm, scale= 0.34, clip]{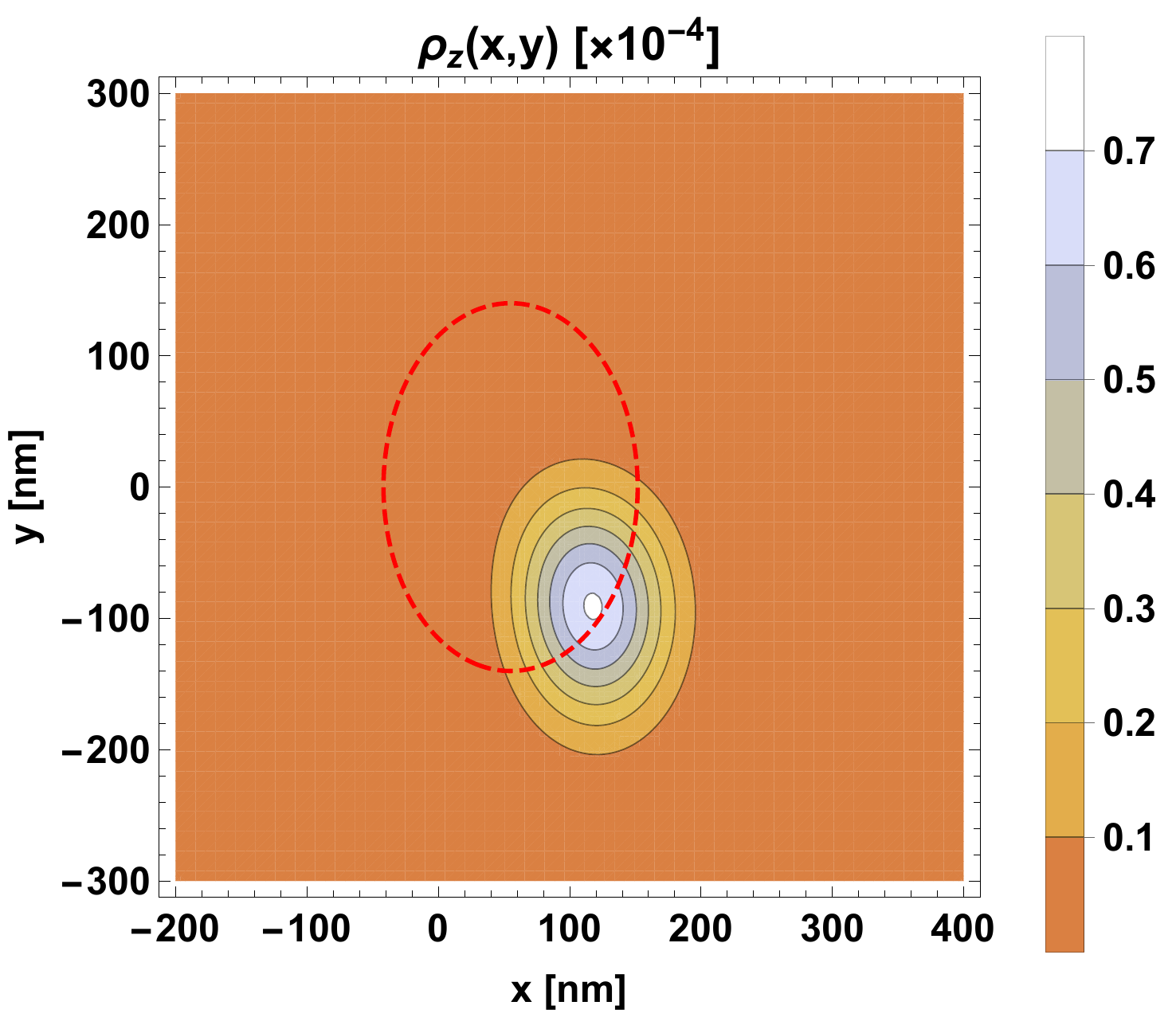}&
		\includegraphics[trim = 0mm 0mm 0mm 0mm, scale= 0.34, clip]{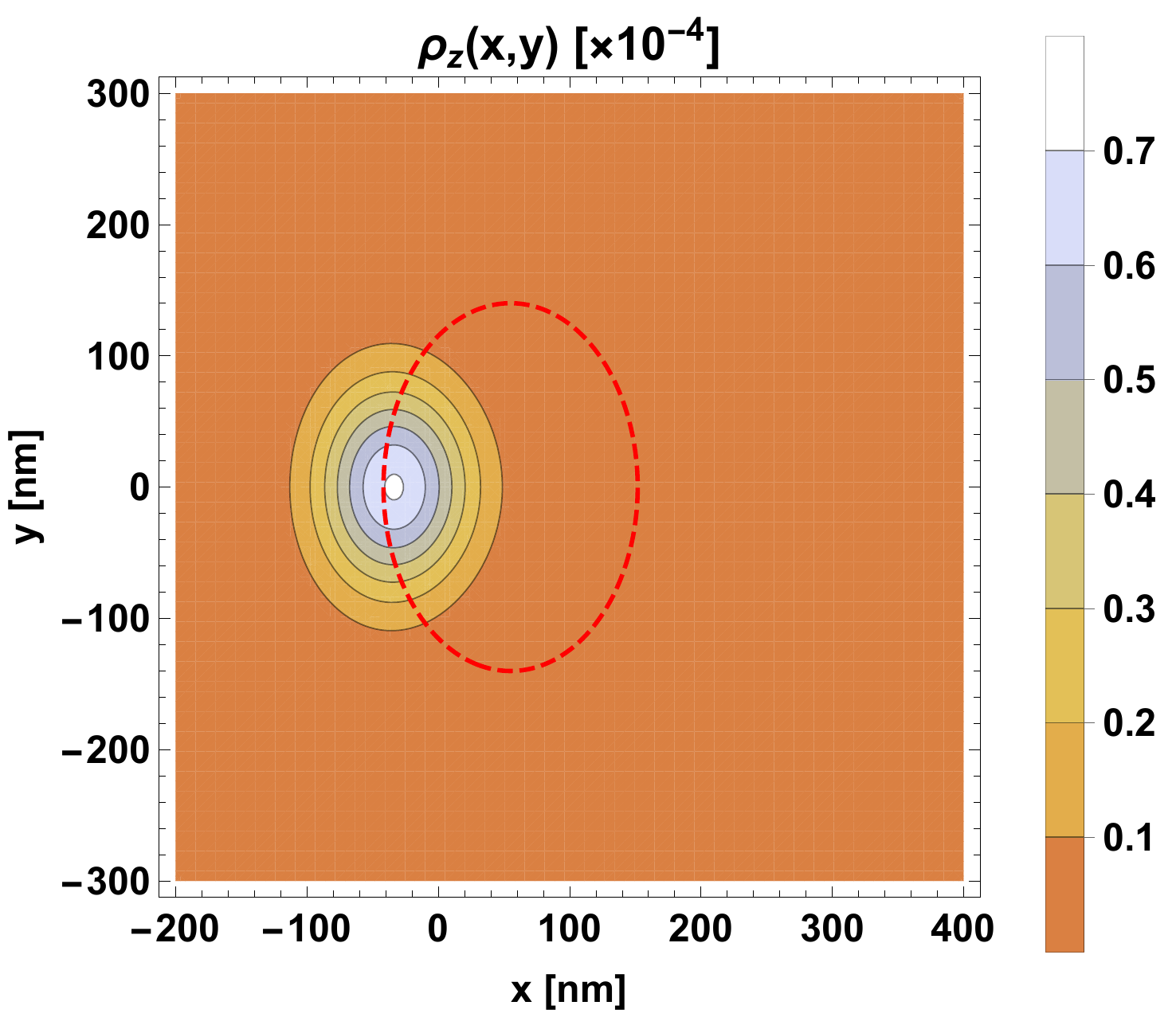}
	\end{tabular}
	\caption{\label{fig:density5}Probability density $\rho_{z}(x,y)$ in Eq.~(\ref{densityz}) for different values of $\alpha$ (from left to right) and $\epsilon=-15\%$ along (a-c) the $\mathcal{Z}$ direction and (d-f) the $\mathcal{A}$ direction. The red dashed line shows the classical trajectory in~(\ref{path2}). $z=2$, $B_{0}=0.3$ T, and $\beta=1/2$.}
\end{figure}

\begin{figure}[h!]
	\centering
	\begin{tabular}{ccc}
		(a) $\alpha=1/\sqrt{2}$ \qquad \qquad \quad & (b) $\alpha=\exp\left(i\pi/3\right)/\sqrt{2}$ \qquad \qquad & (c) $\alpha=-1/\sqrt{2}$ \qquad \qquad \\
		\includegraphics[trim = 0mm 0mm 0mm 0mm, scale= 0.34, clip]{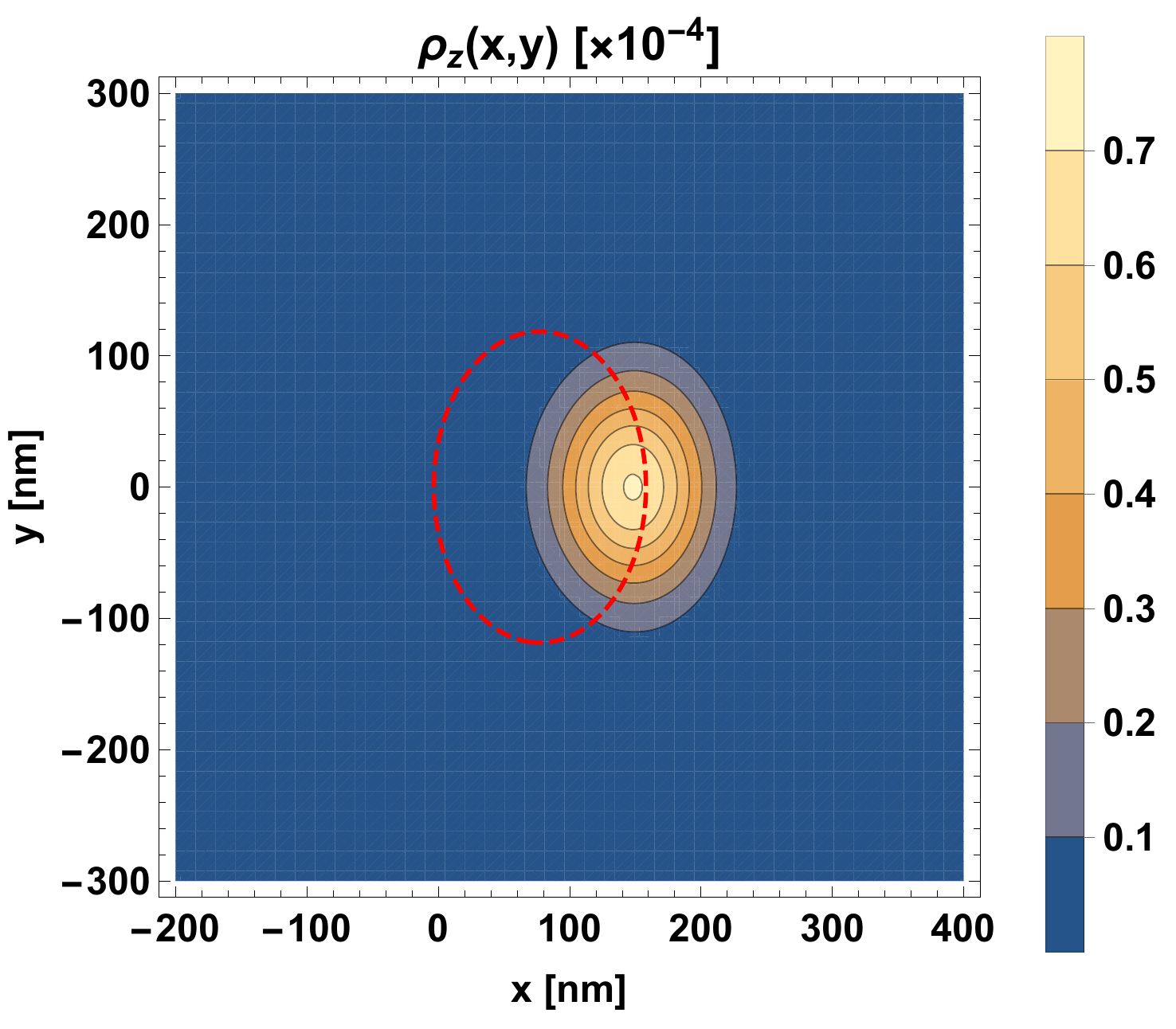} &
		\includegraphics[trim = 0mm 0mm 0mm 0mm, scale= 0.34, clip]{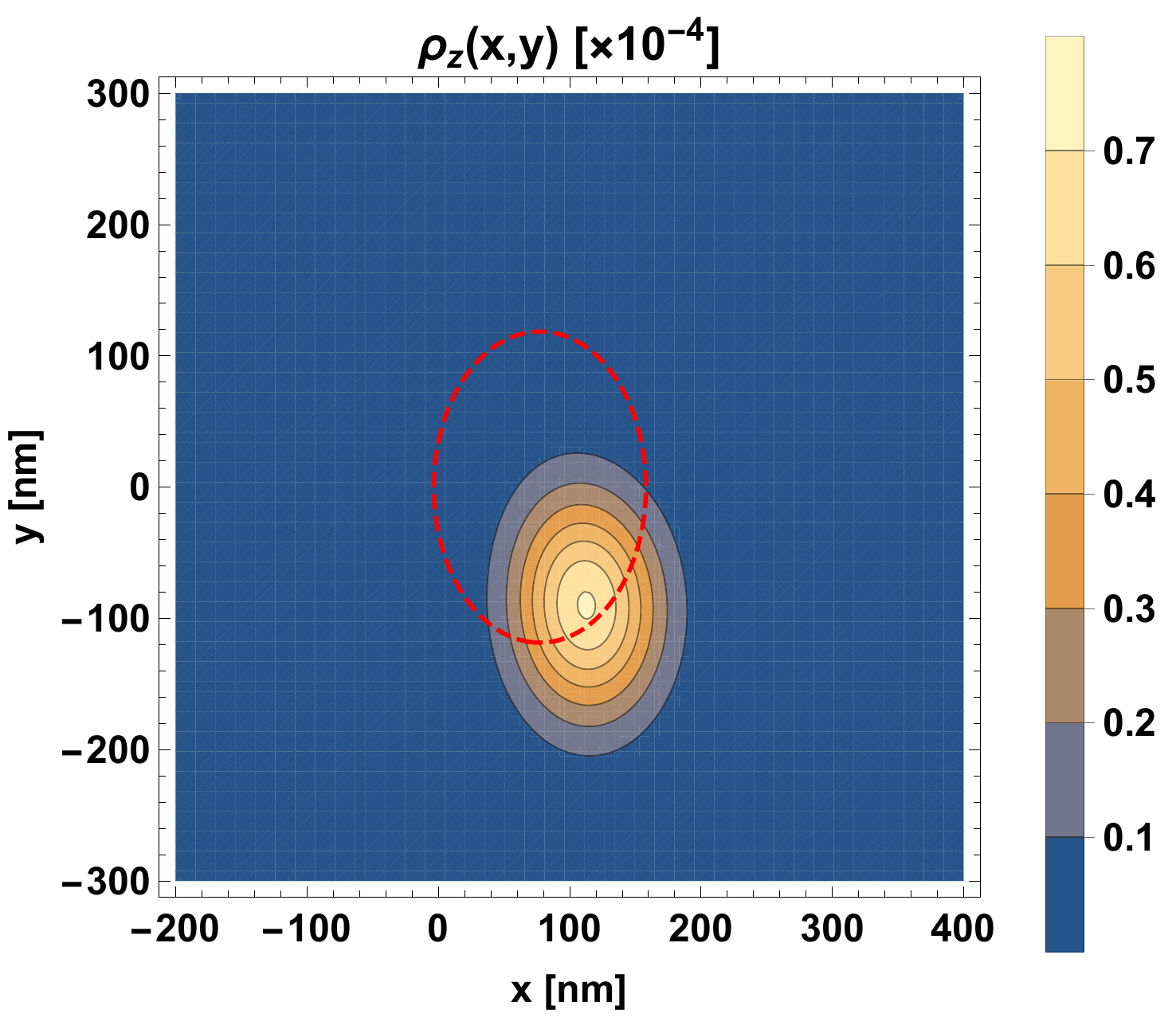}&
		\includegraphics[trim = 0mm 0mm 0mm 0mm, scale= 0.34, clip]{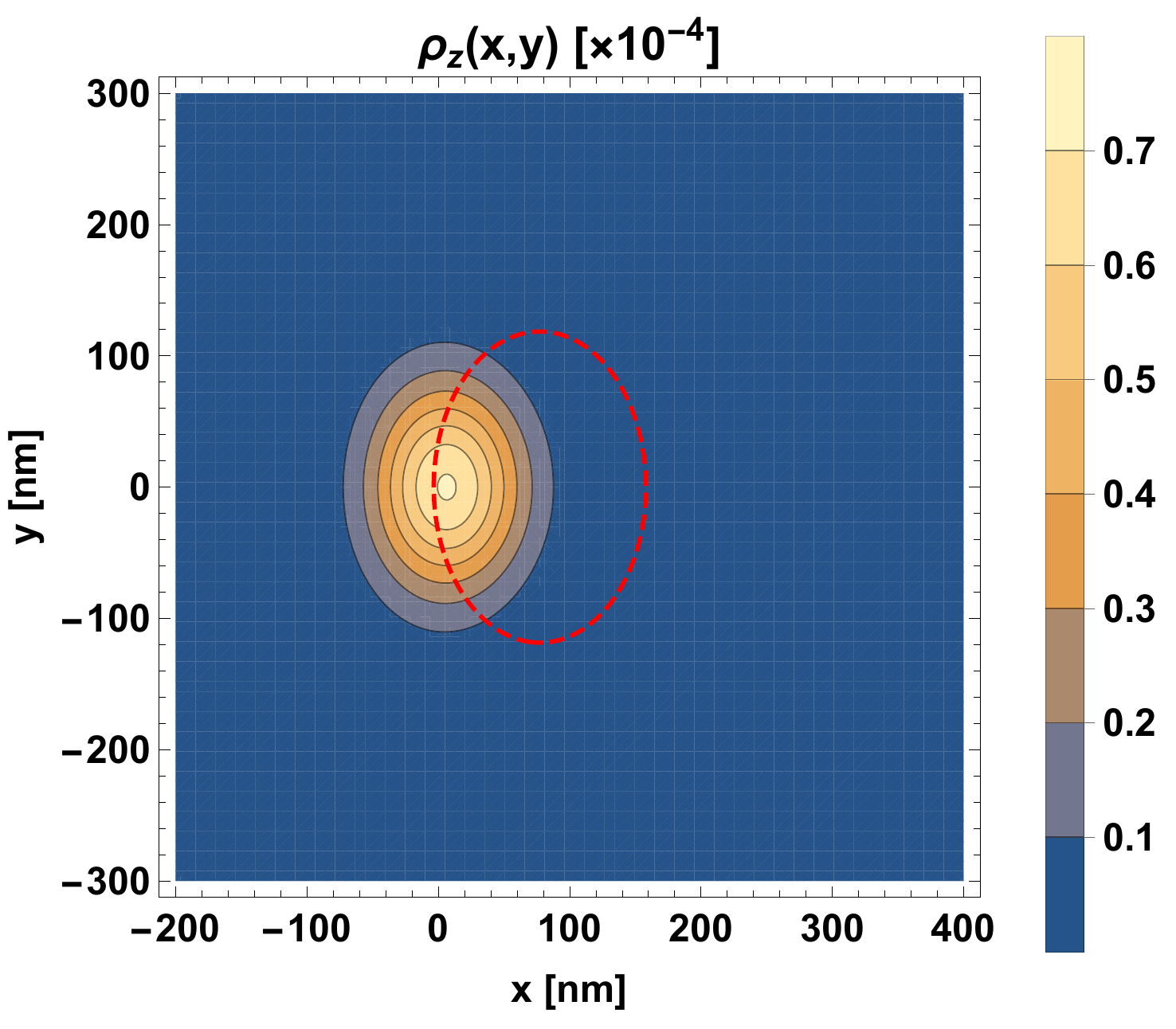}
	\end{tabular}
	\begin{tabular}{ccc}
		(d) {\color{white}$\alpha=1/\sqrt{2}$} \qquad \qquad \quad & (e) {\color{white}$\alpha=\exp\left(i\pi/3\right)/\sqrt{2}$} \qquad \qquad & (f) {\color{white}$\alpha=-1/\sqrt{2}$} \qquad \qquad \\
		\includegraphics[trim = 0mm 0mm 0mm 0mm, scale= 0.34, clip]{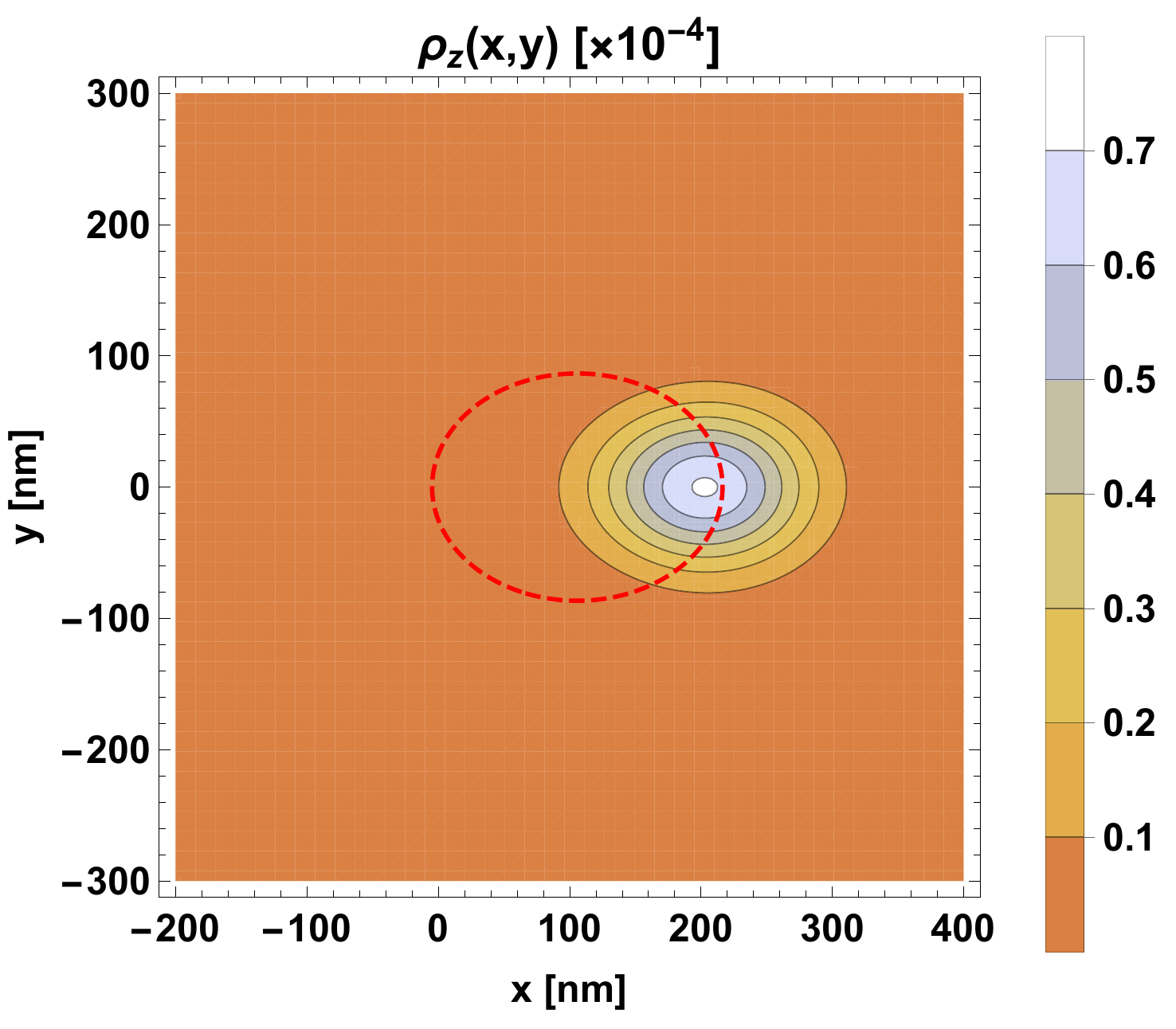} &
		\includegraphics[trim = 0mm 0mm 0mm 0mm, scale= 0.34, clip]{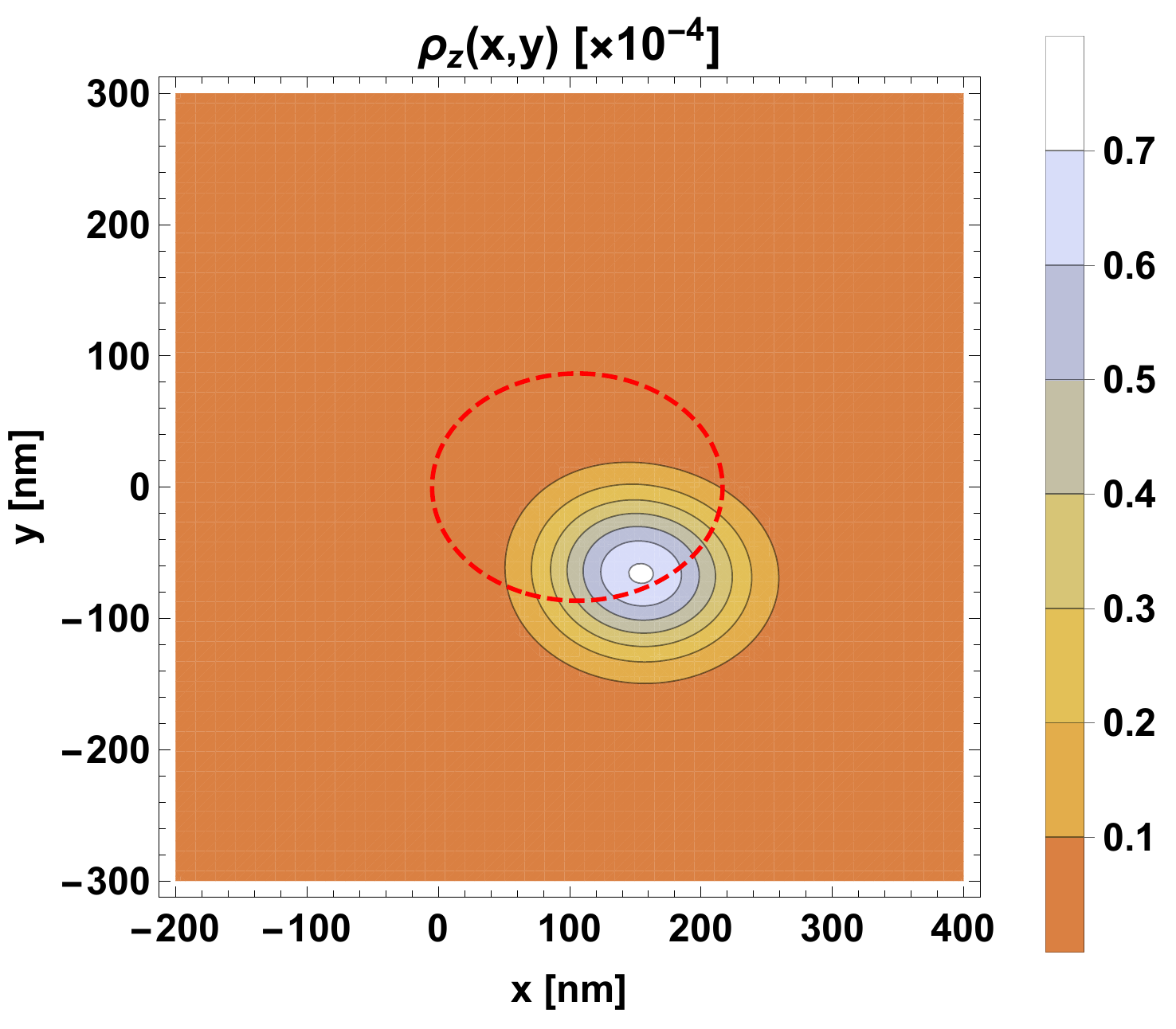}&
		\includegraphics[trim = 0mm 0mm 0mm 0mm, scale= 0.34, clip]{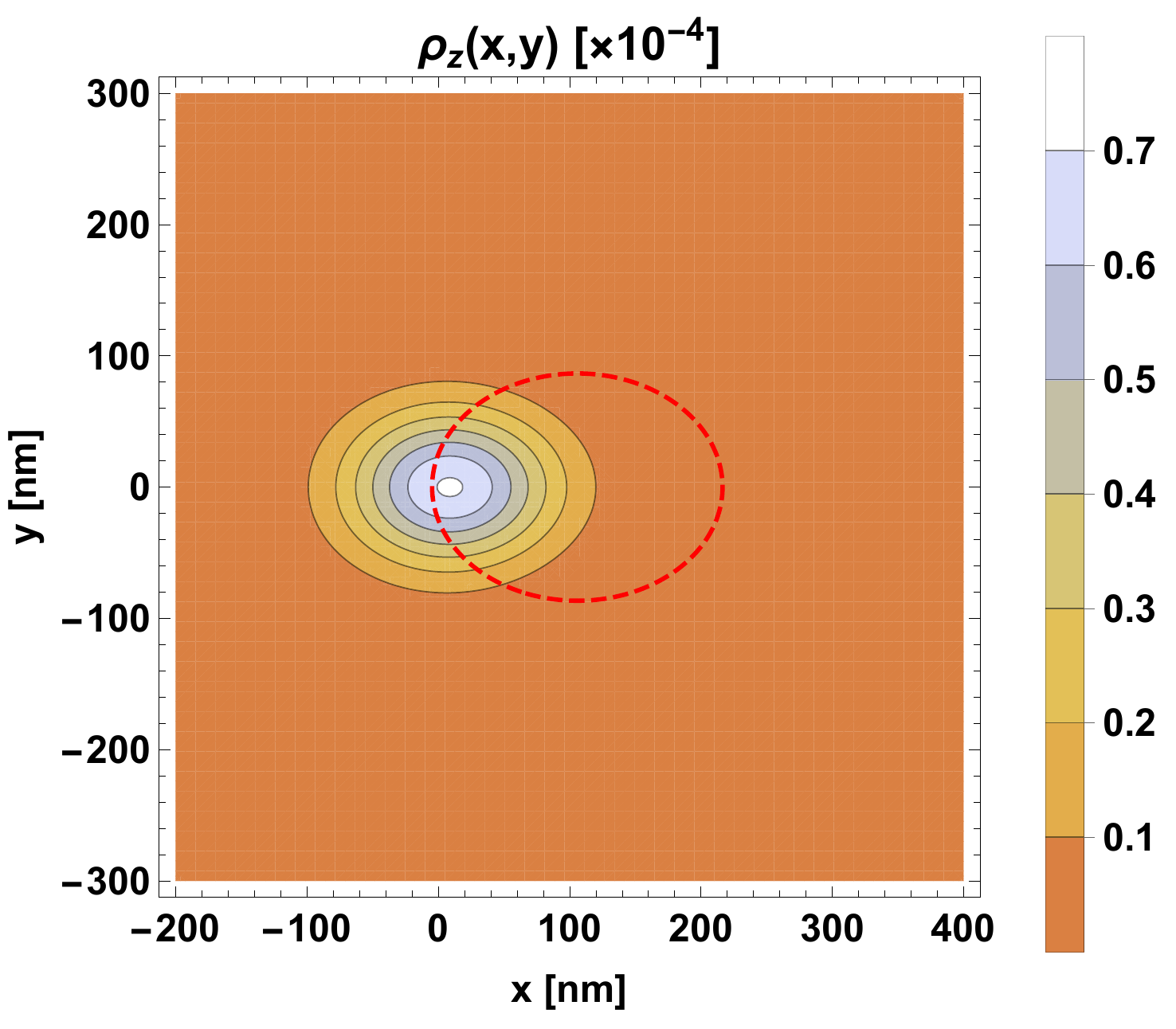}
	\end{tabular}
	\caption{\label{fig:density4}Probability density $\rho_{z}(x,y)$ in Eq.~(\ref{densityz}) for different values of $\alpha$ (from left to right) and $\epsilon=15\%$ along (a-c) the $\mathcal{Z}$ direction and (d-f) the $\mathcal{A}$ direction. The red dashed line shows the classical trajectory in~(\ref{path2}). $z=2$, $B_{0}=0.3$ T, and $\beta=1/\sqrt{2}$.}
\end{figure}

By applying the eigenvalue equation in~(\ref{eigeneq}), we have that
\begin{equation}\label{ScrCS}
\vert\Psi_{z}\rangle=c_{0}\sum_{\nu=0}^{\infty}\frac{z^{\nu}}{\sqrt{\nu!}}\vert\Psi_{\nu}\rangle,
\end{equation}
where $\vert c_{0}\vert^{-1}\equiv\mathcal{N}_{z}^{-1}=\sqrt{2\exp\left(\vert z\vert^2\right)-\exp\left(\vert z\beta\vert^2\right)}$ is determined by the normalization condition $\langle\Psi_{z}\vert\Psi_{z}\rangle=1$. The corresponding probability density reads as (see Figs.~\ref{fig:density5} and~\ref{fig:density4})
\begin{equation}\label{densityz}
\rho_{z}(x,y)=\mathcal{N}_{z}^{2}\left(\left\vert\sum_{\nu=0}^{\infty}\sum_{n=1}^{\nu}\frac{z^{\nu}\alpha^{n}\beta^{\nu-n}}{\sqrt{(\nu-n)!n!}}\psi_{\nu-n,n-1}(x,y)\right\vert^2+\left\vert\sum_{\nu=0}^{\infty}\sum_{n=0}^{\nu}\frac{z^{\nu}\alpha^{n}\beta^{\nu-n}}{\sqrt{(\nu-n)!n!}}\psi_{\nu-n,n}(x,y)\right\vert^2\right).
\end{equation}

\subsection{Orthogonal condition and completeness relation}
The Schr\"{o}dinger-type 2D coherent states in Eq.~(\ref{ScrCS}) satisfy the following relation:
\begin{equation}
\langle\Psi_{z'}\vert\Psi_{z}\rangle=2\exp\left(z'^{\ast}z\right)-\exp\left(z'^{\ast}z\vert\beta\vert^{2}\right).
\end{equation}
Therefore, these states are not orthogonal for $z'\neq z$.

Now, considering the following measure:
\begin{equation}\label{measurez}
d\mu(z,\alpha,\beta)=\frac{\left(2\exp(\vert z\vert^2)-\exp(\vert z\beta\vert^2)\right)\exp\left(-\vert z\vert^{2}-\vert\alpha\vert^{2}-\vert\beta\vert^{2}\right)}{\Gamma(\nu+1)}\,\delta\left(\vert\alpha\vert^{2}+\vert\beta\vert^{2}-1\right)d^{2}z\,d^{2}\alpha\,d^{2}\beta,
\end{equation}
the $SU(2)$ coherent states resolve the identity as follows:
\begin{align}
\nonumber\frac{1}{\pi^{3}}\int_{\mathbb{C}^{4}}d\mu(z,\alpha,\beta)\vert\Psi_{z}\rangle\langle\Psi_{z}\vert&=\frac{1}{\pi^{3}}\int_{\mathbb{C}^{4}}d^{2}z\,d^{2}\alpha\,d^{2}\beta\frac{\left(2\exp(\vert z\vert^2)-\exp(\vert z\beta\vert^2)\right)}{\Gamma(\nu+1)}\\
\nonumber&\quad\times\exp\left(-\vert z\vert^{2}-\vert\alpha\vert^{2}-\vert\beta\vert^{2}\right)\delta\left(\vert\alpha\vert^{2}+\vert\beta\vert^{2}-1\right)\vert\Psi_{z}\rangle\langle\Psi_{z}\vert \\
&=\sum_{\nu=0}^{\infty}\sum_{n=0}^{\nu}\vert\Psi_{\nu-n,n}\rangle\langle\Psi_{\nu-n,n}\vert=\sum_{\nu=0}^{\infty}\mathbb{I}_{\nu}\equiv\mathbb{I},
\end{align}
where the last equality is defined according to Eq.~(\ref{identity}). Thus, the Schr\"{o}dinger-type 2D coherent states represent an over-complete basis for
the whole Hilbert space $\mathcal{H}$. The resolution of the identity suggests these coherent states could have some application in 2D quantization in (anisotropic) 2D Dirac materials.

\subsection{Cyclotron motion}
In order to compute the expected values of the matrix operators $\mathbb{X}$, $\mathbb{Y}_{0}$ and $\mathbb{R}_{x,y}$ for the Schr\"{o}dinger-type 2D coherent states, we define first the dimensionless operators $\mathbb{R}_{q}=r_{q}\otimes\mathbb{I}$ and $\mathbb{S}_{q}=s_{q}\otimes\mathbb{I}$, where
\begin{equation}
r_{q}=\frac{1}{\sqrt{2}i^{q}}\left(B^{-}+(-1)^{q}B^{+}\right), \quad s_{q}=\frac{1}{\sqrt{2}i^{q}}\left(A^{-}+(-1)^{q}A^{+}\right), \quad q=0,1,
\end{equation}
and then, we calculate the mean values of the operators $\mathbb{S}_{q}$ and $\mathbb{R}_{q}$ and their squares,
\begin{subequations}\label{62}
	\begin{align}
	\langle\mathbb{R}_{q}\rangle_{z}&=\mathcal{N}_{z}^{2}\frac{\left(z\beta+(-1)^{q}z^{\ast}\beta^{\ast}\right)}{\sqrt{2}i^{q}}, \label{62a} \\
	\langle\mathbb{R}_{q}^{2}\rangle_{z}&=\frac{\mathcal{N}_{z}^{2}}{2}\left(1+2\vert z\beta\vert^{2}+(-1)^{q}\left(z^{2}\beta^{2}+z^{\ast 2}\beta^{\ast 2}\right)\right), \label{62b} \\
	\langle \mathbb{S}_{q}\rangle_{z}&=\mathcal{N}_{z}^{2}\frac{\left(z\alpha+(-1)^{q}z^{\ast}\alpha^{\ast}\right)}{\sqrt{2}i^{q}}\left(\exp\left(\vert z\vert^{2}\right)+\sum_{\nu=0}^{\infty}\sum_{n=1}^{\nu}\frac{\vert z\vert^{2\nu}\vert\alpha\vert^{2n}\vert\beta\vert^{2(\nu-n)}}{(\nu-n)!\sqrt{(n-1)!(n+1)!}}\right), \label{62c} \\
	\nonumber\langle \mathbb{S}_{q}^{2}\rangle_{z}&=\frac{\mathcal{N}_{z}^{2}}{2}\Bigg[4\vert z\alpha\vert^{2}\exp\left(\vert z\vert^{2}\right)+\exp\left(\vert z\beta\vert^{2}\right)+(-1)^{q}(z^{2}\alpha^{2}+z^{\ast 2}\alpha^{\ast 2}) \\
	&\quad\times\left(\exp\left(\vert z\vert^{2}\right)+\sum_{\nu=0}^{\infty}\sum_{n=1}^{\nu}\frac{\vert z\vert^{2\nu}\vert\alpha\vert^{2n}\vert\beta\vert^{2(\nu-n)}}{(\nu-n)!\sqrt{(n-1)!(n+2)!}}\sqrt{n+1}\right)\Bigg]. \label{62d}
	\end{align}
\end{subequations}

From Eqs.~(\ref{62a}) and~(\ref{62b}) with $q=0,1$, the mean values of the matrix operators $\mathbb{X}_{0}$ and $\mathbb{Y}_{0}$ are given by
\begin{subequations}
	\begin{align}
	\langle\mathbb{X}_{0}\rangle_{z}&=\frac{2\,\zeta^{1/2}\Re\left(z\beta\right)}{\sqrt{\omega_{\rm B}}\left(2\exp\left(\vert z\vert^2\right)-\exp\left(\vert z\beta\vert^2\right)\right)}, \\ \langle\mathbb{X}_{0}^{2}\rangle_{z}&=\frac{\zeta\left(1+4\Re^{2}\left(z\beta\right)\right)}{\omega_{\rm B}\left(2\exp\left(\vert z\vert^2\right)-\exp\left(\vert z\beta\vert^2\right)\right)}, \\
	\langle\mathbb{Y}_{0}\rangle_{z}&=\frac{2\,\zeta^{-1/2}\Im\left(z\beta\right)}{\sqrt{\omega_{\rm B}}\left(2\exp\left(\vert z\vert^2\right)-\exp\left(\vert z\beta\vert^2\right)\right)}, \\ \langle\mathbb{Y}_{0}^{2}\rangle_{z}&=\frac{\zeta^{-1}\left(1+4\Im^{2}\left(z\beta\right)\right)}{\omega_{\rm B}\left(2\exp\left(\vert z\vert^2\right)-\exp\left(\vert z\beta\vert^2\right)\right)}.
	\end{align}
\end{subequations}
Hence, the mean square distance value $\langle\mathbb{R}_{0}^{2}\rangle_{z}$ is given by
\begin{equation}
\langle\mathbb{R}_{0}^{2}\rangle_{z}=\zeta^{-1}\langle\mathbb{X}_{0}^{2}\rangle_{z}+\zeta\langle\mathbb{Y}_{0}^{2}\rangle_{z}=\frac{\left(2\vert z\beta\vert^{2}+1\right)l_{\rm B}^{2}}{2\exp\left(\vert z\vert^2\right)-\exp\left(\vert z\beta\vert^2\right)}.
\end{equation}

On the other hand, from Eqs.~(\ref{62c}) and~(\ref{62d}) with $q=0,1$, the mean values of the matrix operators $\mathbb{R}_{x}$ and $\mathbb{R}_{y}$ read as
\begin{subequations}
	\begin{align}
	\langle\mathbb{R}_{x}\rangle_{z}&=\frac{2\,\zeta^{1/2}\Re\left(z\alpha\right)}{\sqrt{\omega_{\rm B}}\left(2\exp\left(\vert z\vert^2\right)-\exp\left(\vert z\beta\vert^2\right)\right)}\left(\exp\left(\vert z\vert^{2}\right)+\sum_{\nu=0}^{\infty}\sum_{n=1}^{\nu}\frac{\vert z\vert^{2\nu}\vert\alpha\vert^{2n}\vert\beta\vert^{2(\nu-n)}}{(\nu-n)!\sqrt{(n-1)!(n+1)!}}\right), \\
	\nonumber\langle\mathbb{R}_{x}^{2}\rangle_{z}&=\frac{\zeta}{\omega_{\rm B}\left(2\exp\left(\vert z\vert^2\right)-\exp\left(\vert z\beta\vert^2\right)\right)}\Bigg[4\vert z\alpha\vert^{2}\exp\left(\vert z\vert^{2}\right)+\exp\left(\vert z\beta\vert^{2}\right)+2(\Re^{2}\left(z\alpha\right)-\Im^{2}\left(z\alpha\right)) \\
	&\quad\times\left(\exp\left(\vert z\vert^{2}\right)\right)+\sum_{\nu=0}^{\infty}\sum_{n=1}^{\nu}\frac{\vert z\vert^{2\nu}\vert\alpha\vert^{2n}\vert\beta\vert^{2(\nu-n)}}{(\nu-n)!\sqrt{(n-1)!(n+2)!}}\sqrt{n+1}\Bigg], \\
	\langle\mathbb{R}_{y}\rangle_{z}&=-\frac{2\,\zeta^{-1/2}\Im\left(z\alpha\right)}{\sqrt{\omega_{\rm B}}\left(2\exp\left(\vert z\vert^2\right)-\exp\left(\vert z\beta\vert^2\right)\right)}\left(\exp\left(\vert z\vert^{2}\right)+\sum_{\nu=0}^{\infty}\sum_{n=1}^{\nu}\frac{\vert z\vert^{2\nu}\vert\alpha\vert^{2n}\vert\beta\vert^{2(\nu-n)}}{(\nu-n)!\sqrt{(n-1)!(n+1)!}}\right), \\
	\nonumber\langle\mathbb{R}_{y}^{2}\rangle_{z}&=\frac{\zeta^{-1}}{\omega_{\rm B}\left(2\exp\left(\vert z\vert^2\right)-\exp\left(\vert z\beta\vert^2\right)\right)}\Bigg[4\vert z\alpha\vert^{2}\exp\left(\vert z\vert^{2}\right)+\exp\left(\vert z\beta\vert^{2}\right)+2(\Im^{2}\left(z\alpha\right)-\Re^{2}\left(z\alpha\right)) \\
	&\quad\times\left(\exp\left(\vert z\vert^{2}\right)+\sum_{\nu=0}^{\infty}\sum_{n=1}^{\nu}\frac{\vert z\vert^{2\nu}\vert\alpha\vert^{2n}\vert\beta\vert^{2(\nu-n)}}{(\nu-n)!\sqrt{(n-1)!(n+2)!}}\sqrt{n+1}\right)\Bigg].
	\end{align}
\end{subequations}
Thus, we have that
\begin{equation}
\langle\mathbb{R}^{2}\rangle_{z}=\zeta^{-1}\langle\mathbb{R}_{x}^{2}\rangle_{z}+\zeta\langle\mathbb{R}_{y}^{2}\rangle_{z}=\frac{\left(4\vert z\alpha\vert^{2}\exp\left(\vert z\vert^{2}\right)+\exp\left(\vert z\beta\vert^{2}\right)\right)l_{\rm B}^{2}}{\left(2\exp\left(\vert z\vert^{2}\right)-\exp\left(\vert z\beta\vert^{2}\right)\right)}.
\end{equation}

\begin{figure}[h!]
	\centering
	\includegraphics[width=0.6\linewidth]{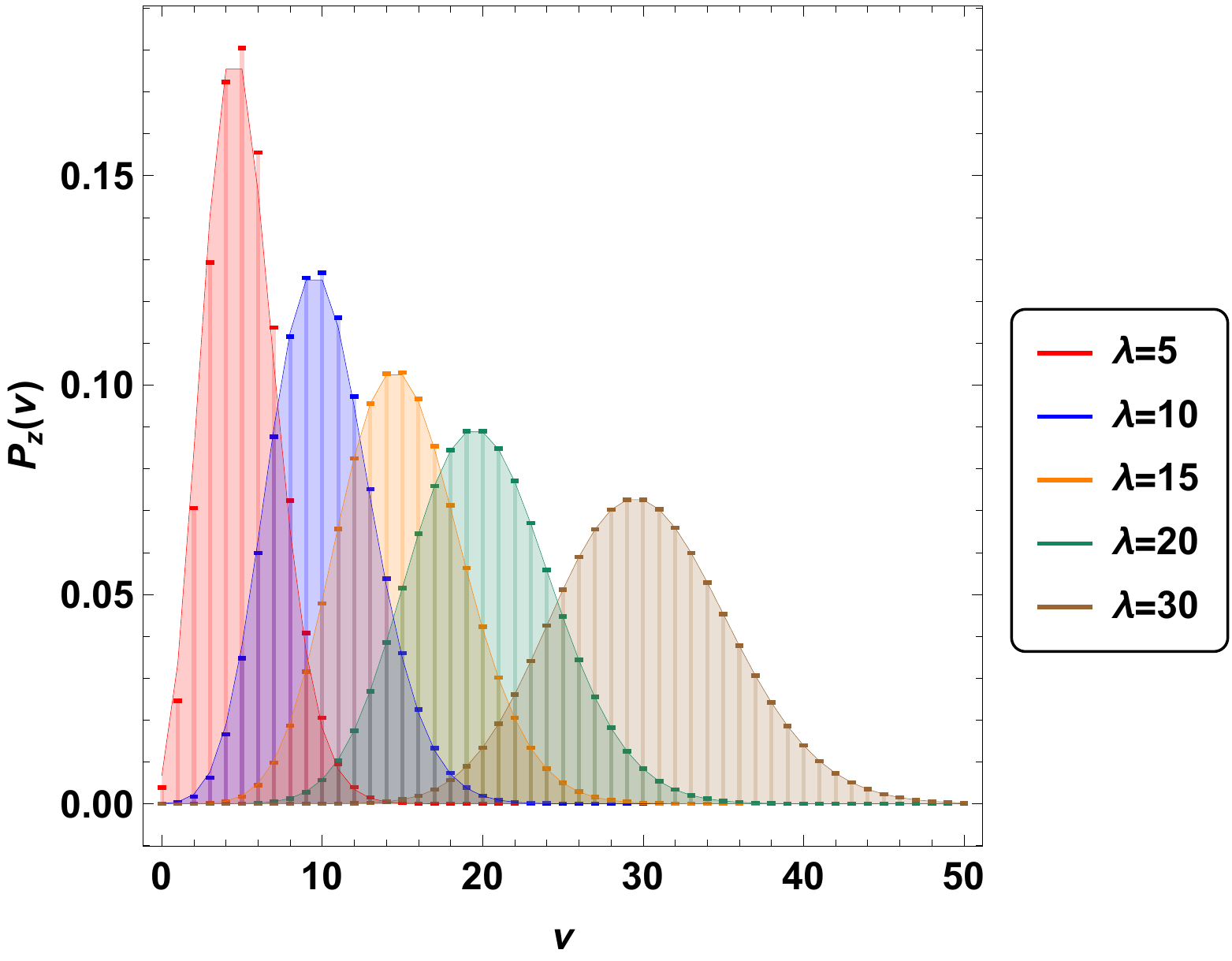}
	\caption{Occupation number distribution $P_{z}(\nu)$ in Eq.~(\ref{distribution}) for the coherent states $\vert\Psi_{z}\rangle$ with $\vert\beta\vert=\sqrt{3}/2$ and different values of $\lambda=\vert z\vert^{2}$. $P_{z}(\nu)$ adjusts to Poisson distribution (solid curves) as $\vert z\vert$ grows.}
	\label{fig:occupation}
\end{figure}

For the Schr\"{o}dinger-type 2D coherent states, Figs.~\ref{fig:density5} and \ref{fig:density4} also show the classical trajectory
\begin{equation}\label{path2}
\frac{(x-\mathcal{N}_{z}^{2}\langle\mathbb{X}_{0}\rangle_{z})^{2}}{\zeta\langle\mathbb{R}^{2}\rangle_{z}}+\frac{(y-\mathcal{N}_{z}^{2}\langle\mathbb{Y}_{0}\rangle_{z})^{2}}{\zeta^{-1}\langle\mathbb{R}^{2}\rangle_{z}}=1.
\end{equation}

\subsection{Occupation number distribution}
The Schr\"{o}dinger-type 2D coherent states are constructed as an infinite linear combination of the $SU(2)$ coherent states established previously, with a Poissonian-like probability of being in a state $\vert\Psi_{\nu}\rangle$,
\begin{equation}\label{distribution}
P_{z}(\nu)=\vert\langle\Psi_{\nu}\vert\Psi_{z}\rangle\vert^{2}=\frac{2-\vert\beta\vert^{2\nu}}{2\exp\left(\lambda\right)-\exp\left( \lambda\vert\beta\vert^{2}\right)}\frac{\lambda^{\nu}}{\nu!}, \quad \lambda=\vert z\vert^{2}.
\end{equation}

This occupation number distribution can be compared with that of the CS of the harmonic oscillator with eigenvalue $z_{\rm CS}\in\mathbb{C}$, which is a Poisson distribution $P_{\tau}(n)=\exp(-\tau)\tau^{n}/n!$, with mean $\tau=\vert z_{\rm CS}\vert^{2}$. Figure~\ref{fig:occupation} shows that the function $P_{z}(\nu)$ coincides with the Poisson distribution $P_{\tau}(n)$ for large values of $\lambda$ such that $\tau=\lambda$.

\subsection{Mean energy value}
We compute the mean energy value for the Schr\"{o}dinger-type 2D coherent states by considering the Hamiltonian in Eq.~(\ref{ScrCS}),
\begin{equation}\label{Hz}
\frac{\langle H\rangle_{z}}{\hbar v'_{F}\sqrt{\omega_{\rm B}}}=\frac{2}{2\exp\left(\vert z\vert^{2}\right)-\exp\left(\vert z\beta\vert^{2}\right)}\sum_{\nu=0}^{\infty}\sum_{n=1}^{\nu}\frac{\vert z\vert^{2\nu}(1-\vert\beta\vert^{2})^{n}\vert\beta\vert^{2(\nu-n)}}{(\nu-n)!\,n!}\sqrt{n},
\end{equation}
and the mean value $\langle H^{2}\rangle_{z}$,
\begin{equation}\label{H2}
\frac{\langle H^{2}\rangle_{z}}{\hbar^{2}v_{F}^{'2}\omega_{\rm B}}=\frac{2}{2\exp\left(\vert z\vert^{2}\right)-\exp\left(\vert z\beta\vert^{2}\right)}\sum_{\nu=0}^{\infty}\sum_{n=1}^{\nu}\frac{\vert z\vert^{2\nu}(1-\vert\beta\vert^{2})^{n}\vert\beta\vert^{2(\nu-n)}}{(\nu-n)!\,n!}n,
\end{equation}
in order to estimate the relative value of the energy of the state $\vert\Psi_{z}\rangle$,
\begin{equation}
H_{\rm rel}=\frac{\langle H^{2}\rangle_{z}-\left(\langle H\rangle_{z}\right)^{2}}{\langle H\rangle_{z}}.
\end{equation}

\begin{figure}[h!]
	\centering
	\includegraphics[width=0.6\linewidth]{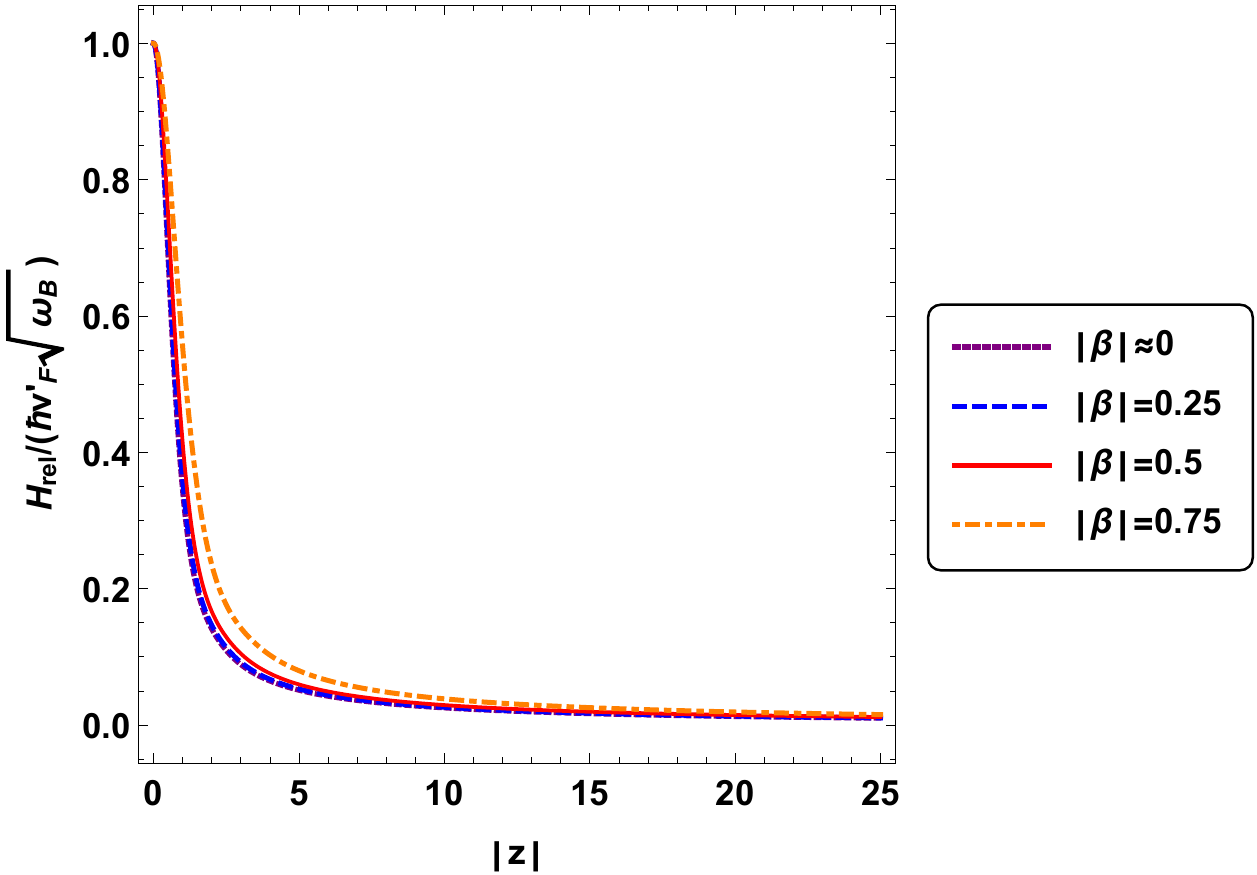}
	\caption{Dimensionless relative value of the energy $H_{\rm rel}$ as a function of $\vert z\vert$ for different values of $\vert\beta\vert$. As $\vert z\vert$ increases, $H_{\rm rel}\ll\hbar\,v'_{\rm F}\sqrt{\omega_{\rm B}}$.}
	\label{fig:energy}
\end{figure}

Although $H_{\rm rel}$ shall not depend on $\vert z\vert$ in a trivial way, according to Fig.~\ref{fig:energy}, as $\vert z\vert$ takes large values we have that
\begin{equation}\label{energy}
\frac{H_{\rm rel}}{\hbar\,v'_{\rm F}\sqrt{\omega_{\rm B}}}\ll1.
\end{equation}
Hence, the relative value of energy of $\vert\Psi_{z}\rangle$ is very well-defined, and an electron coherent state with $\vert z\vert\gg1$ can be considered as a good approximation to the classical limit of the charge carriers in graphene since the relative size of the quantum fluctuations vanish in the limit of $\vert z\vert\rightarrow\infty$.

\subsection{Time evolution}
Motivated by the work of the authors of Ref.~\cite{diazbetancur20}, we now study the time evolution of the $SU(2)$ and Schr\"{o}dinger-type 2D coherent
states in the $xy$-plane by applying the time-evolution unitary operator $U(t)=\exp\left(-iHt/\hbar\right)$ on the states $\vert\Psi_{\nu}\rangle$ and $\vert\Psi_{z}\rangle$ in Eqs.~(\ref{SU2}) and (\ref{ScrCS}), respectively. Thus, we get
\begin{subequations}\label{evolution}
	\begin{align}
	\nonumber U(t)\vert\Psi_{\nu}\rangle \quad \Longrightarrow\quad \rho_{\nu}(x,y,t)&=\mathcal{N}_{\nu}^{2}\Bigg(\sum_{m,n=1}^{\mu,\nu}\frac{\psi_{\mu-m,m-1}^{\ast}(x,y)\psi_{\nu-n,n-1}(x,y)}{\sqrt{(\mu-m)!(\nu-n)!m!n!}}\sqrt{\mu!\,\nu!}C_{\mu,\nu,m,n}(t)\\
	&\quad+\sum_{m,n=0}^{\mu,\nu}\frac{\psi_{\mu-m,m-1}^{\ast}(x,y)\psi_{\nu-n,n-1}(x,y)}{\sqrt{(\mu-m)!(\nu-n)!m!n!}}\sqrt{\mu!\,\nu!}C_{\mu,\nu,m,n}(t)\Bigg),\label{evolution1}\\
\nonumber U(t)\vert\Psi_{z}\rangle \quad \Longrightarrow \quad \rho_{z}(x,y,t)&=\mathcal{N}_{z}^{2}\Bigg(\sum_{\mu,\nu}^{\infty}\sum_{m,n=1}^{\mu,\nu}\frac{\psi_{\mu-m,m-1}^{\ast}(x,y)\psi_{\nu-n,n-1}(x,y)}{\sqrt{(\mu-m)!(\nu-n)!m!n!}}D_{\mu,\nu,m,n}(t)\\
&\quad+\sum_{\mu,\nu}^{\infty}\sum_{m,n=0}^{\mu,\nu}\frac{\psi_{\mu-m,m-1}^{\ast}(x,y)\psi_{\nu-n,n-1}(x,y)}{\sqrt{(\mu-m)!(\nu-n)!m!n!}}D_{\mu,\nu,m,n}(t)\Bigg), \label{evolution2}
\end{align}
\end{subequations}
where $D_{\mu,\nu,m,n}(t)=z^{\ast \mu}z^{\nu}C_{\mu,\nu,m,n}(t)$ and $C_{\mu,\nu,m,n}(t)=\alpha^{\ast m}\alpha^{n}\beta^{\ast \mu-m}\beta^{\nu-n}\exp\left(i(E_{m}-E_{n})t/\hbar\right)$. Setting $t=0$, we recover the expressions~(\ref{density}) and (\ref{densityz}), respectively.

\begin{figure}[h!]
	\centering
	\begin{tabular}{ccc}
		(a) $t'=0$ \qquad \qquad \quad & (b) $t'=5$ \qquad \qquad & (c) $t'=15$ \qquad \qquad \\
		\includegraphics[trim = 0mm 0mm 0mm 0mm, scale= 0.34, clip]{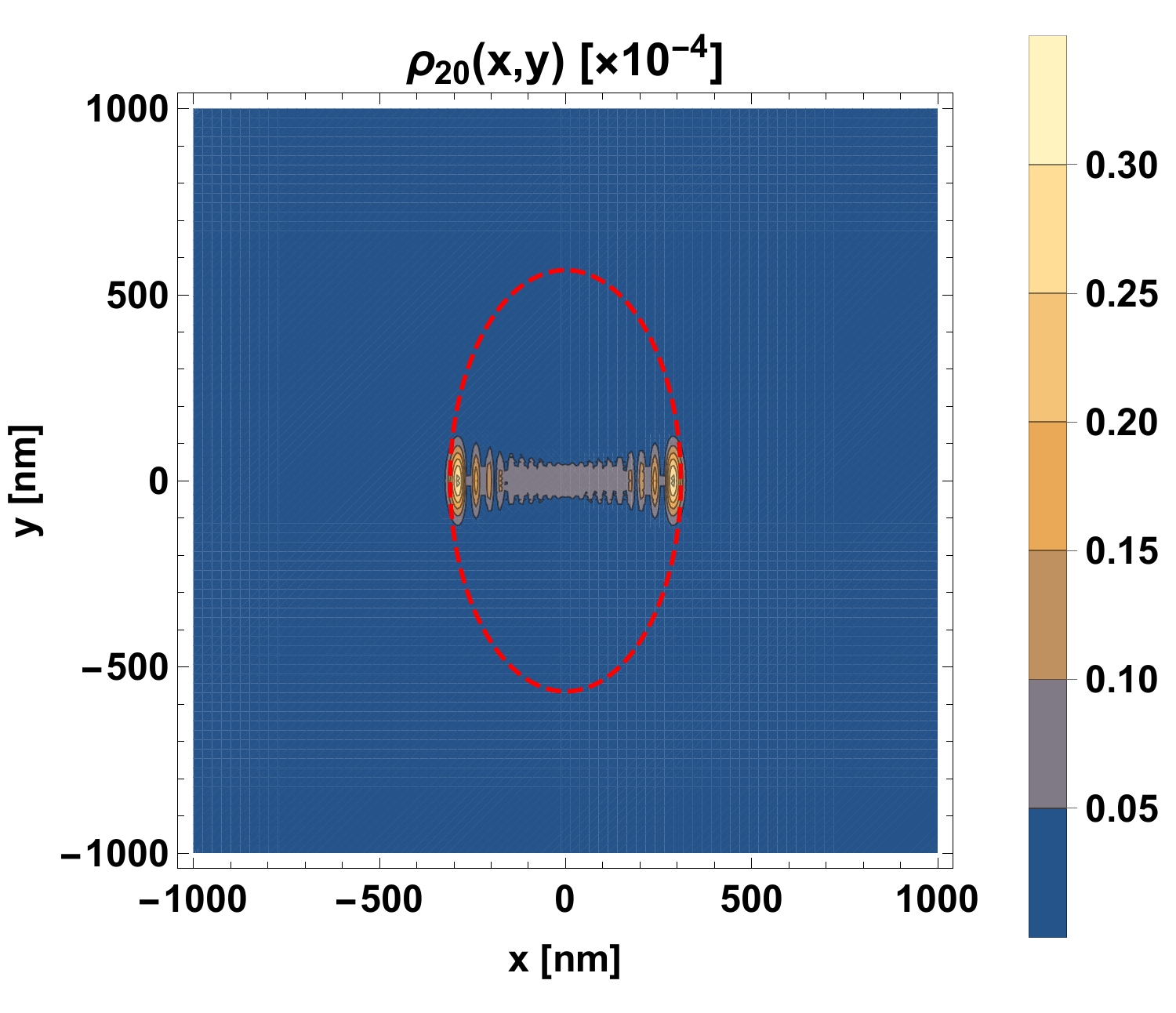} &
		\includegraphics[trim = 0mm 0mm 0mm 0mm, scale= 0.34, clip]{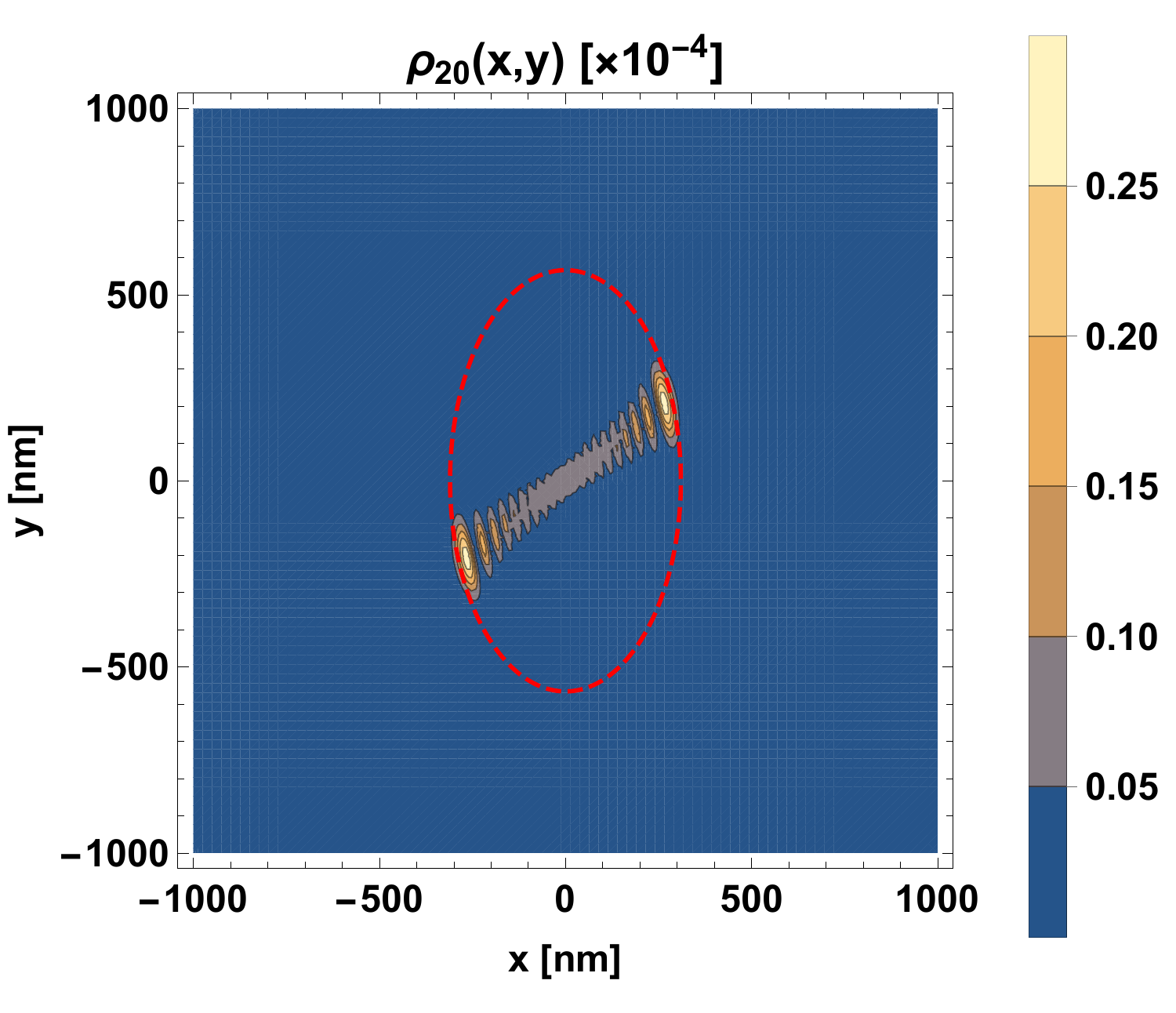}&
		\includegraphics[trim = 0mm 0mm 0mm 0mm, scale= 0.34, clip]{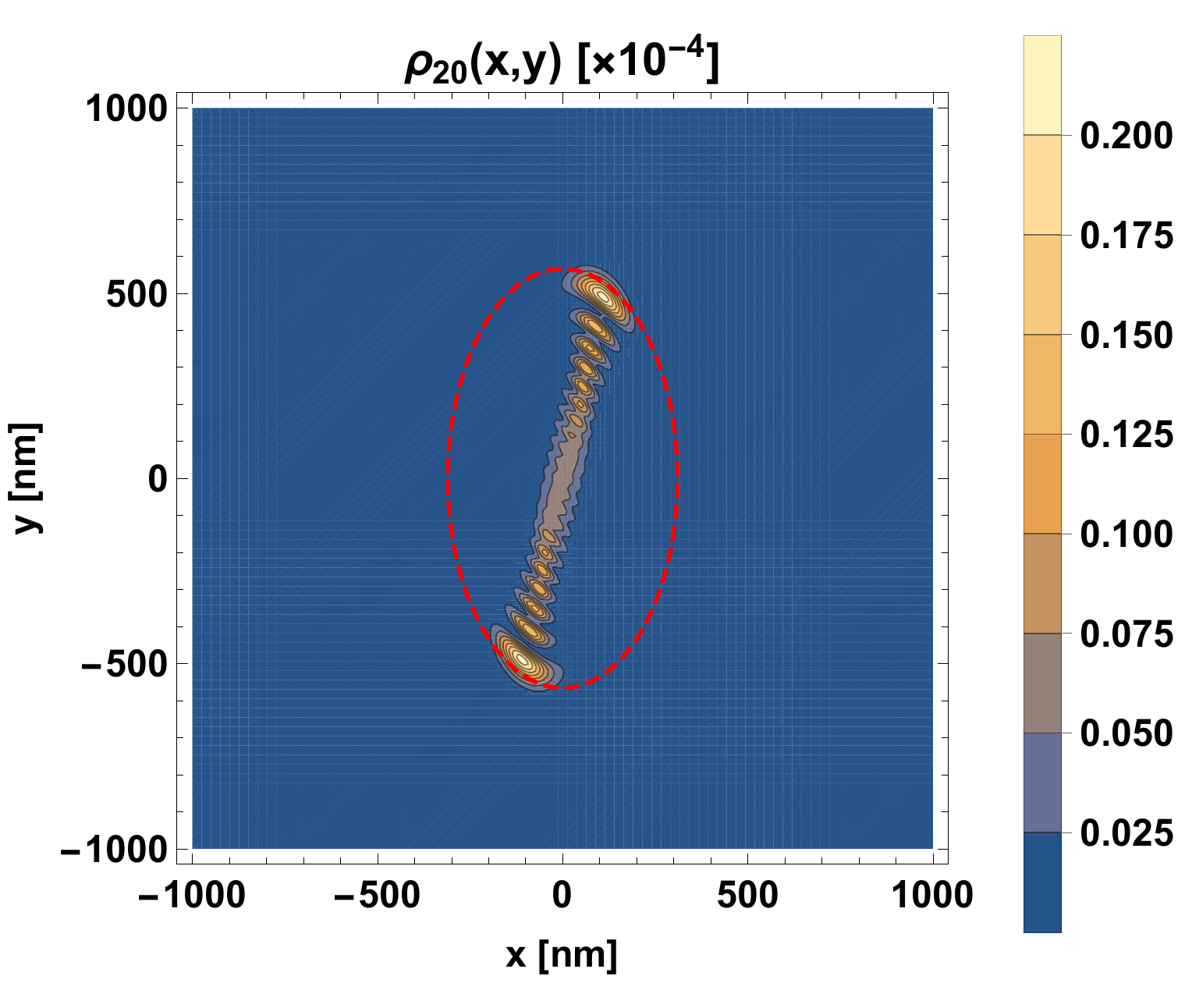}
	\end{tabular}
	\begin{tabular}{ccc}
		(d) $t'=30$ \qquad \qquad \quad & (e) $t'=60$ \qquad \qquad & (f) $t'=540$ \qquad \qquad \\
		\includegraphics[trim = 0mm 0mm 0mm 0mm, scale= 0.34, clip]{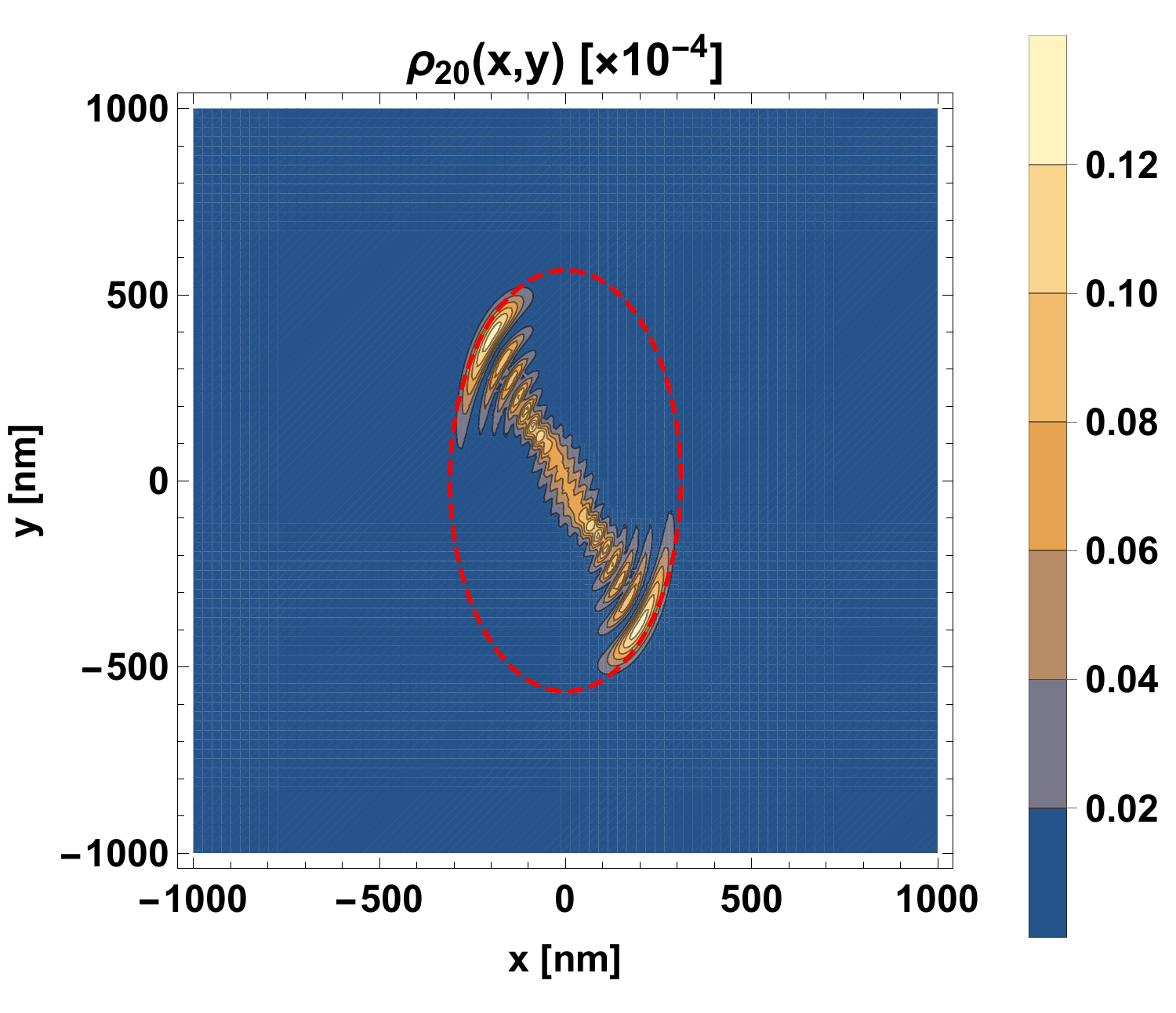} &
		\includegraphics[trim = 0mm 0mm 0mm 0mm, scale= 0.34, clip]{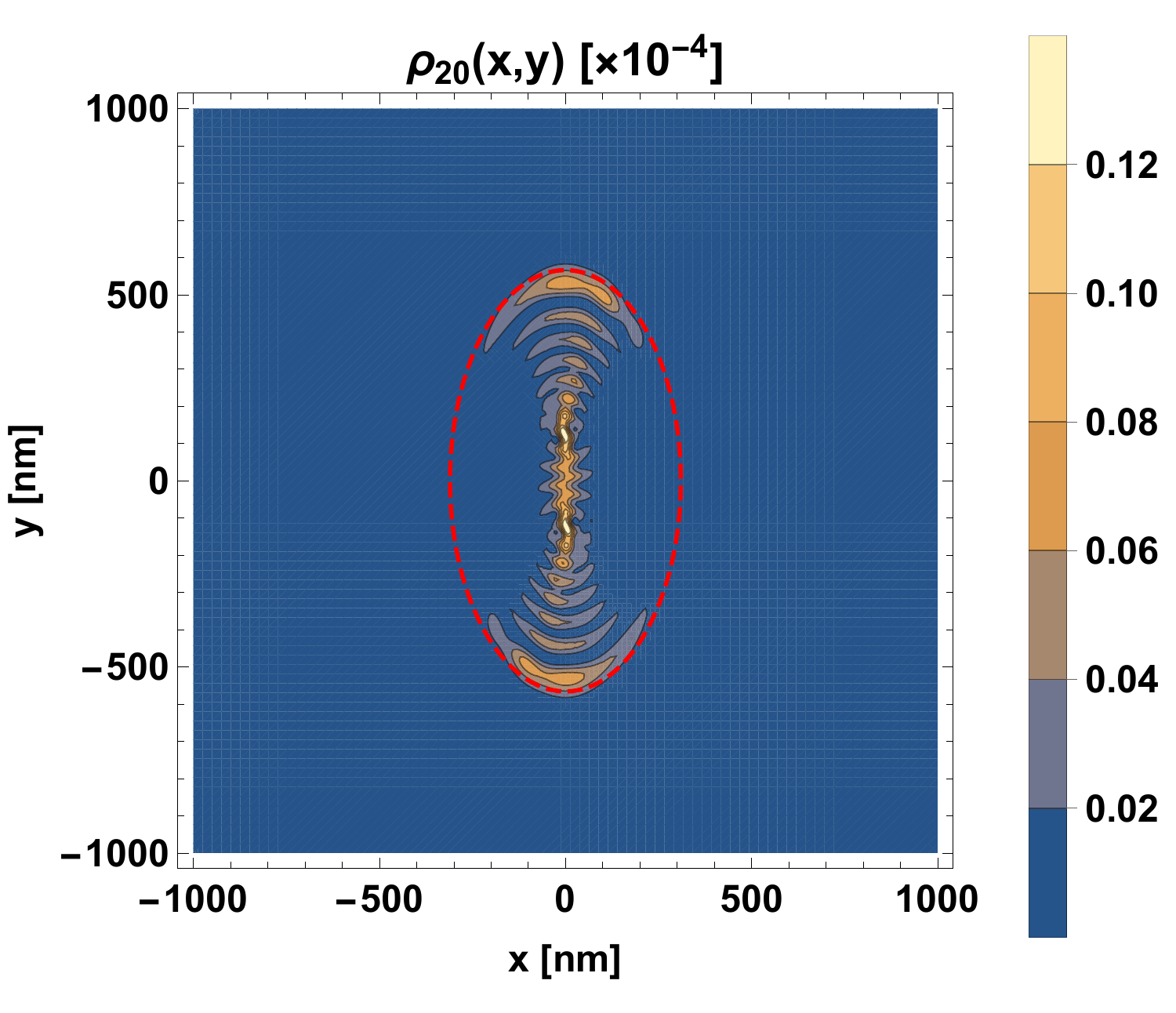}&
		\includegraphics[trim = 0mm 0mm 0mm 0mm, scale= 0.34, clip]{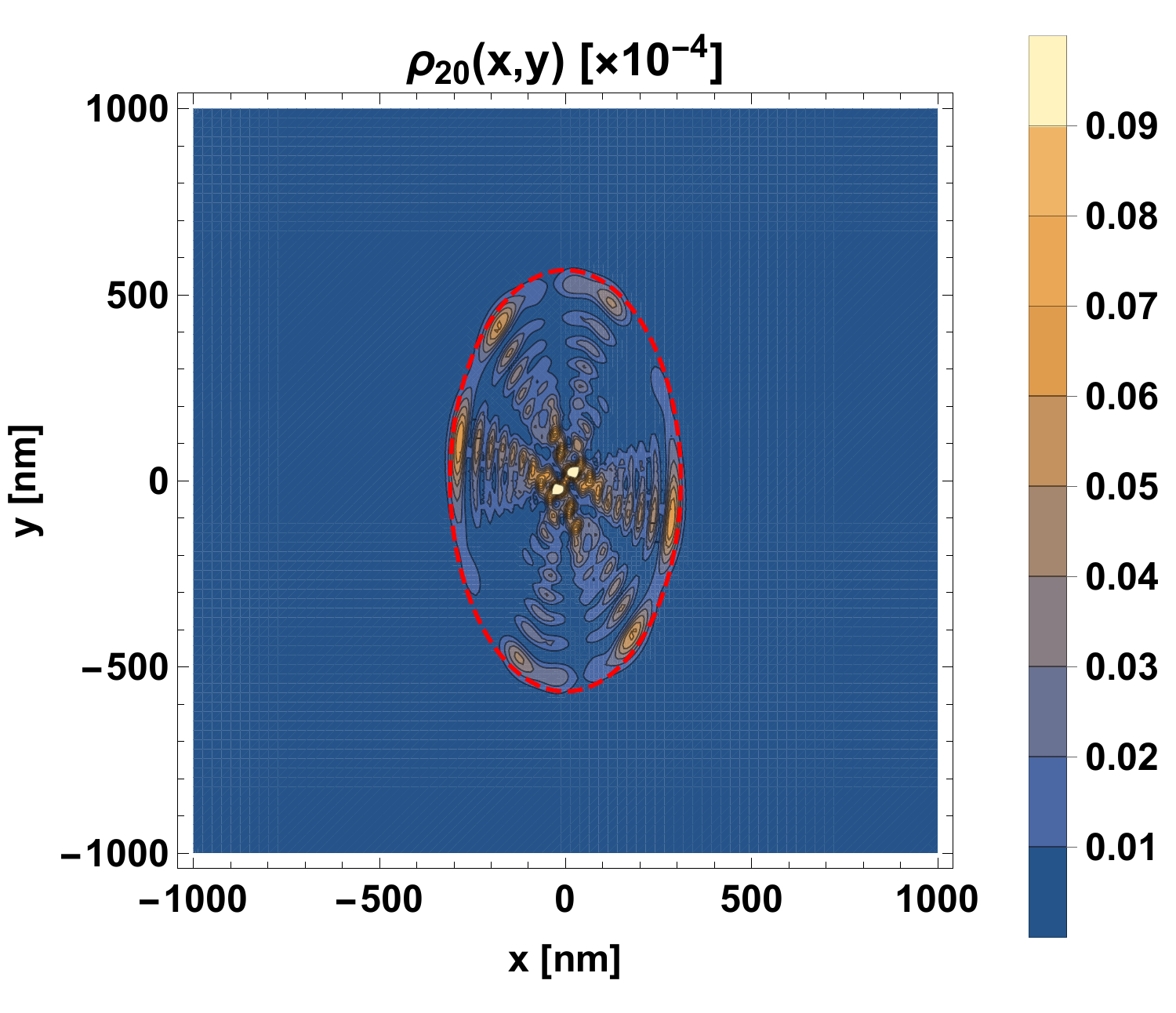}
	\end{tabular}
	\caption{\label{fig:density6}Time evolution of the probability density $\rho_{\nu}(x,y,t)$ in Eq.~(\ref{evolution1}) for different values of $t=t'/(v'_{\rm F}\sqrt{\omega_{\rm B}})$ and $\epsilon=20\%$ along the $\mathcal{Z}$ direction. The red dashed line shows the classical trajectory in~(\ref{path2}). $\nu=20$, $B_{0}=0.3$ T, and $\beta=1/\sqrt{2}$.}
\end{figure}

\begin{figure}[h!]
	\centering
	\begin{tabular}{ccc}
		(a) $t'=0$ \qquad \qquad \quad & (b) $t'=10$ \qquad \qquad & (c) $t'=20$ \qquad \qquad \\
		\includegraphics[trim = 0mm 0mm 0mm 0mm, scale= 0.34, clip]{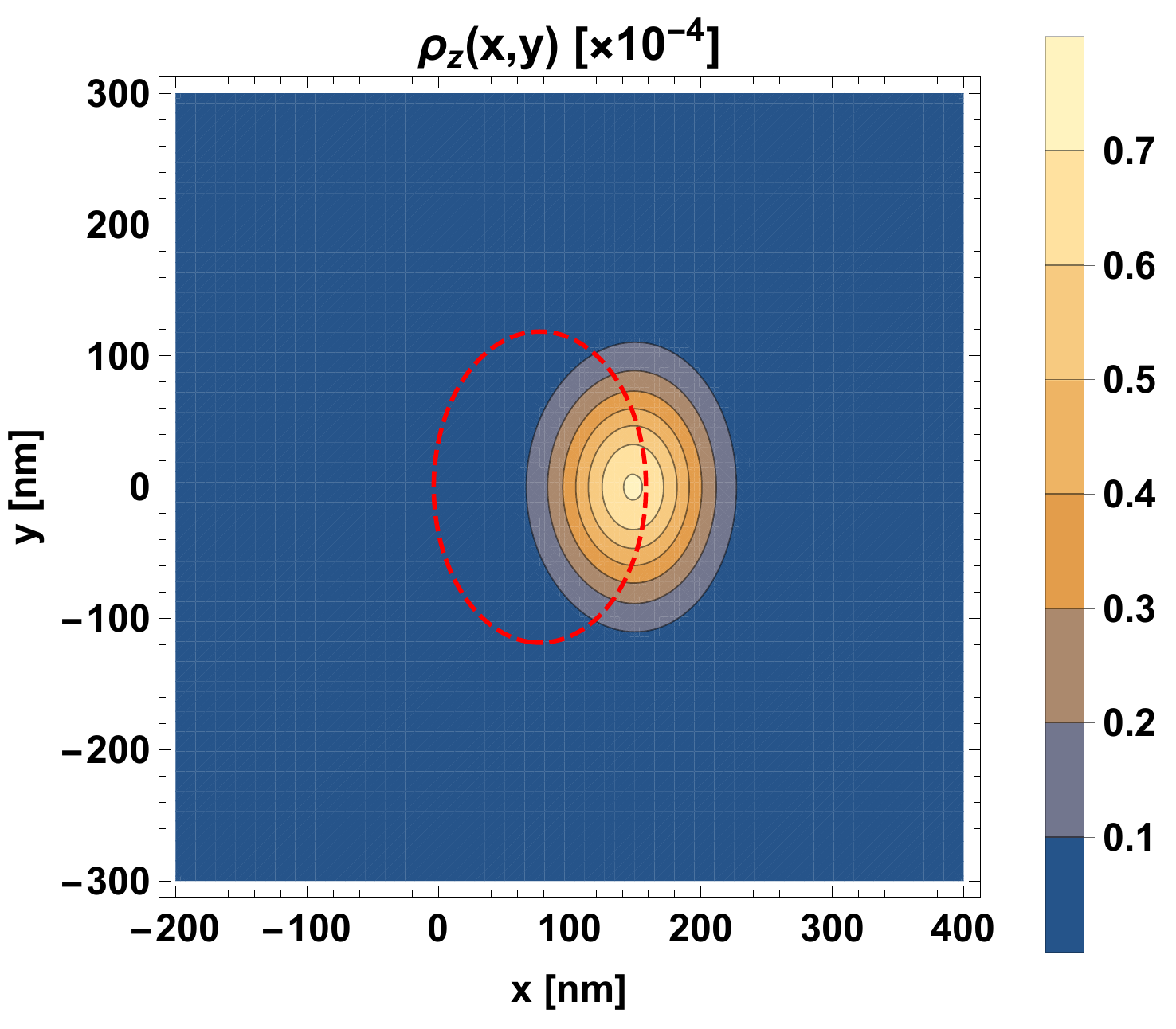} &
		\includegraphics[trim = 0mm 0mm 0mm 0mm, scale= 0.34, clip]{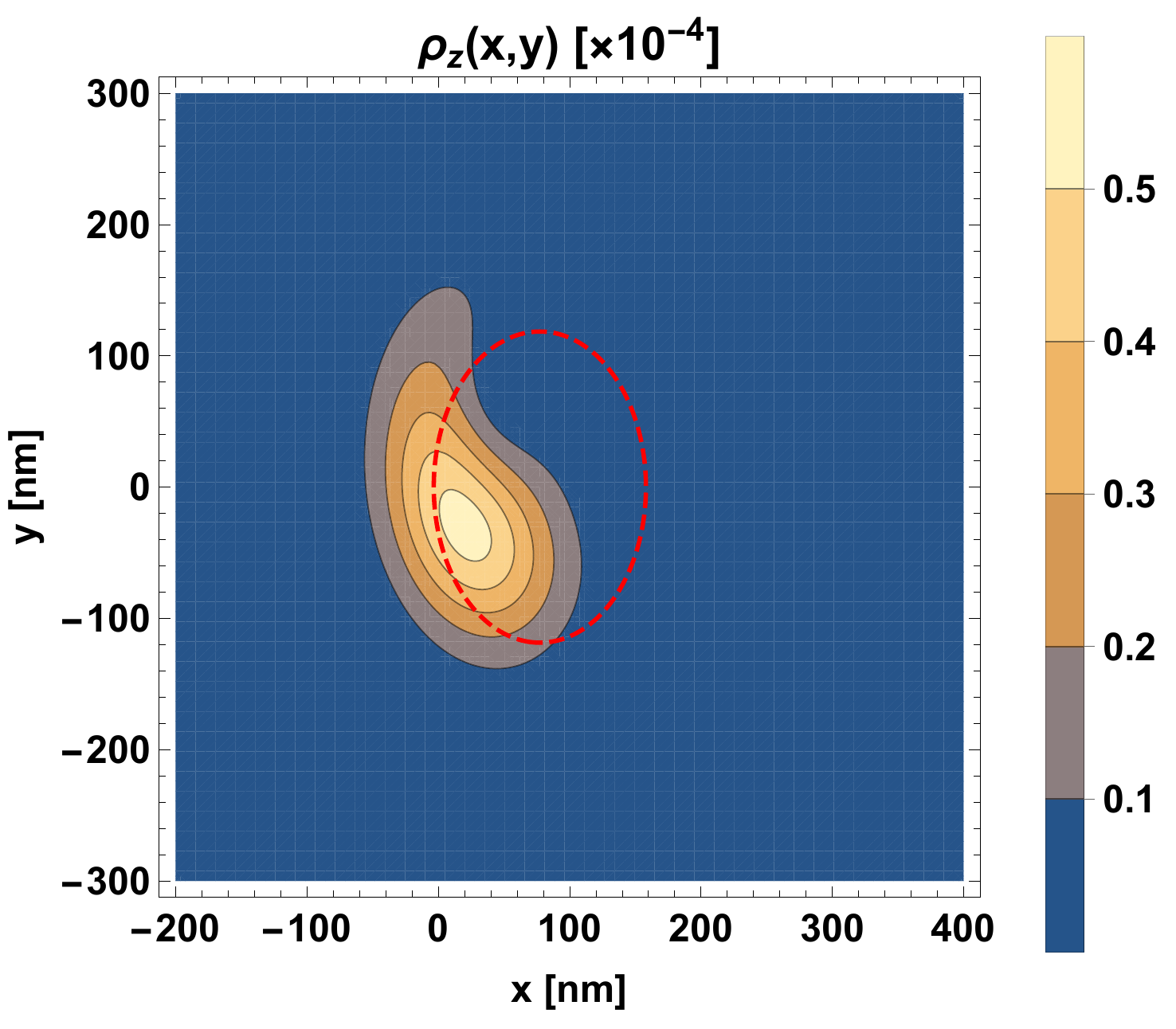}&
		\includegraphics[trim = 0mm 0mm 0mm 0mm, scale= 0.34, clip]{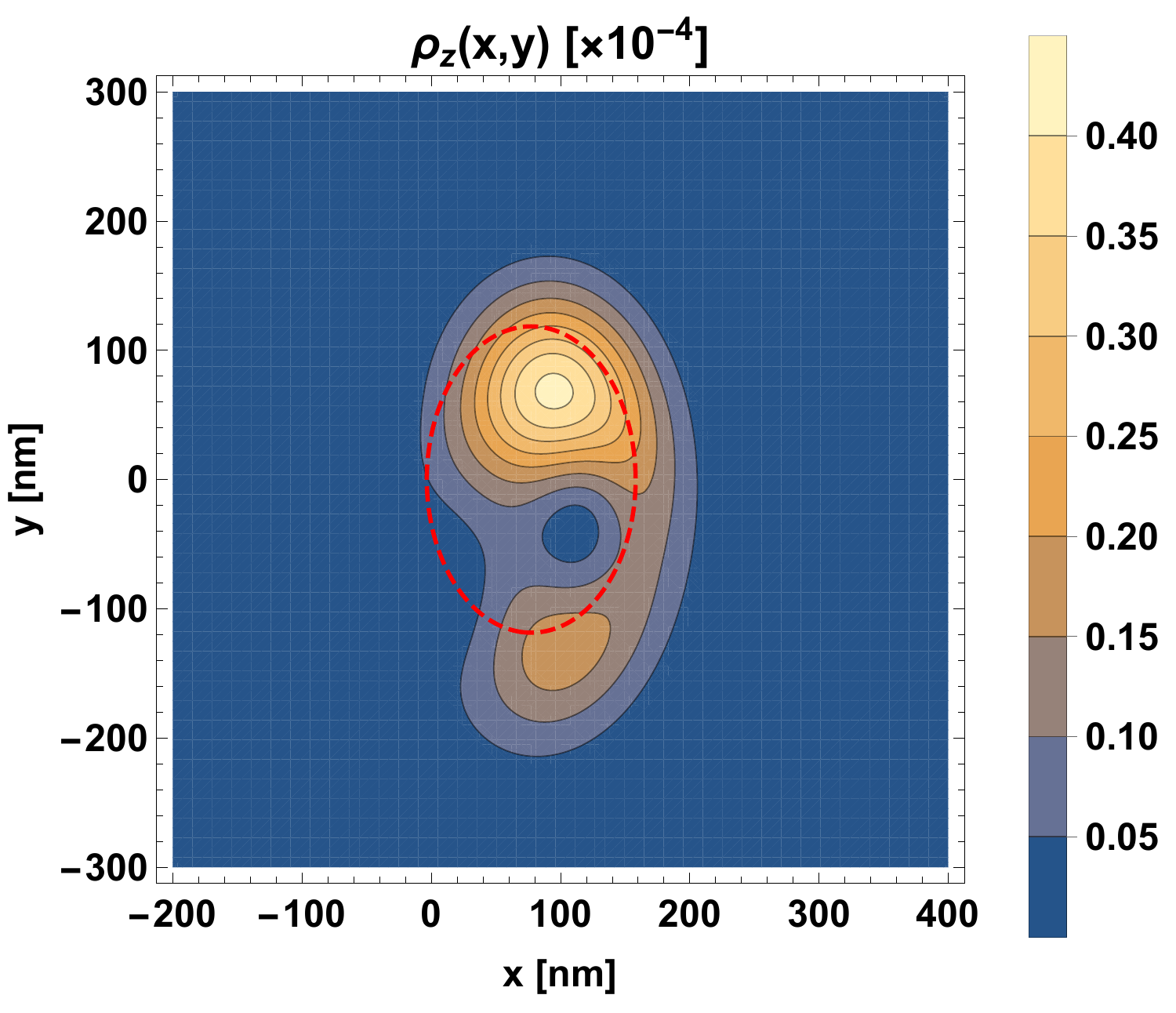}
	\end{tabular}
	\begin{tabular}{ccc}
	(d) $t'=60$ \qquad \qquad \quad & (e) $t'=540$ \qquad \qquad & (f) $t'=900$ \qquad \qquad \\
	\includegraphics[trim = 0mm 0mm 0mm 0mm, scale= 0.34, clip]{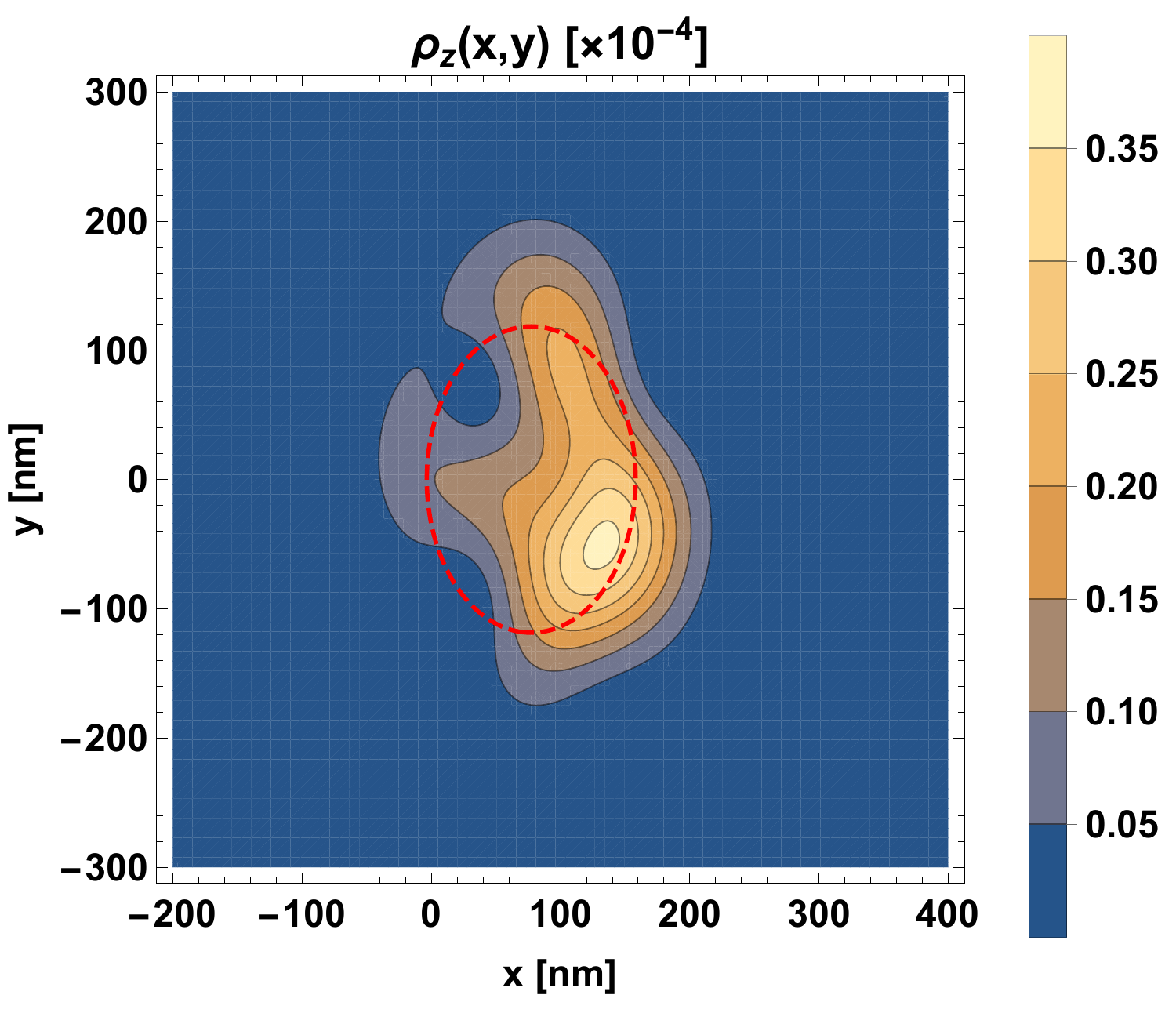} &
	\includegraphics[trim = 0mm 0mm 0mm 0mm, scale= 0.34, clip]{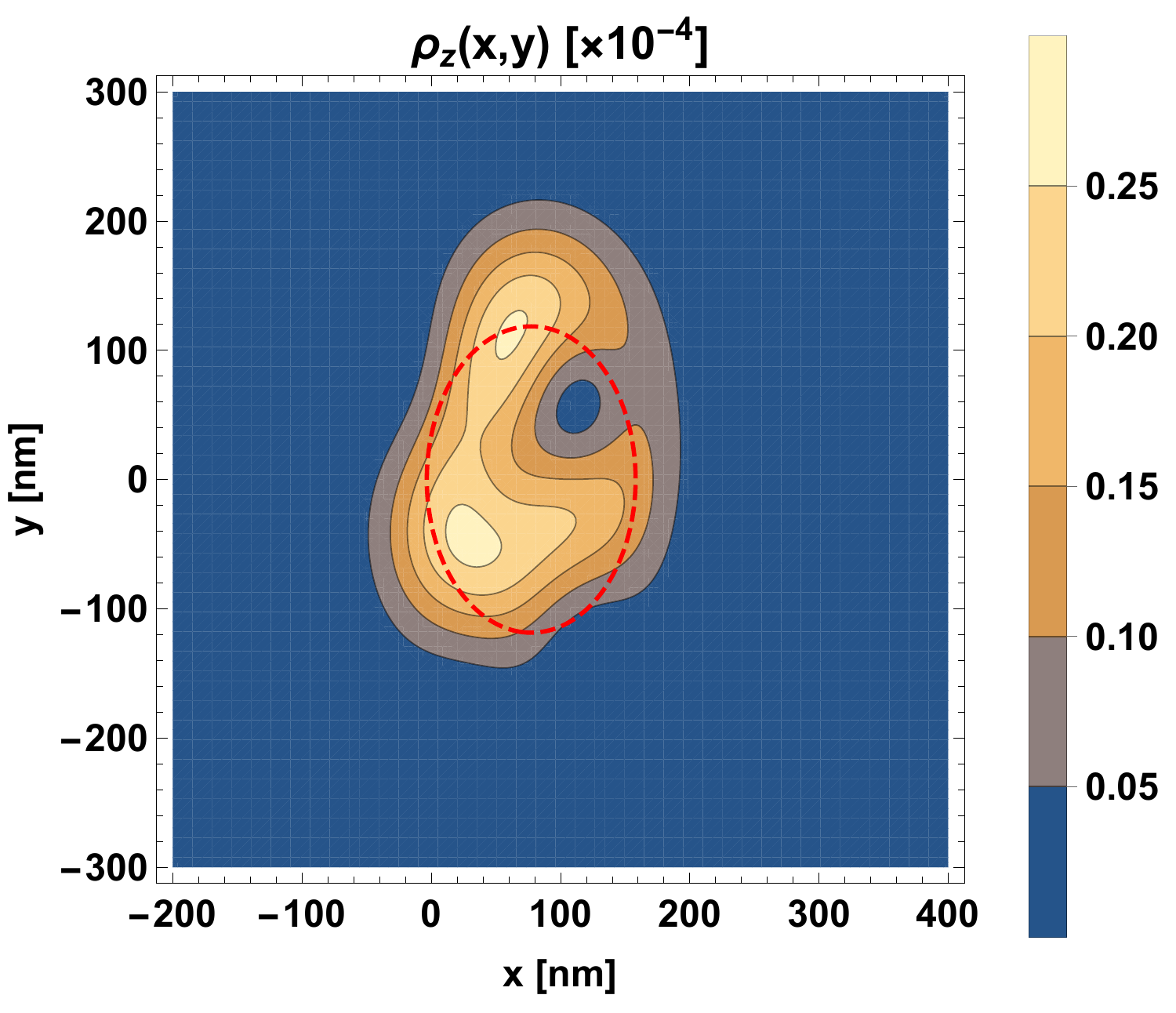}&
	\includegraphics[trim = 0mm 0mm 0mm 0mm, scale= 0.34, clip]{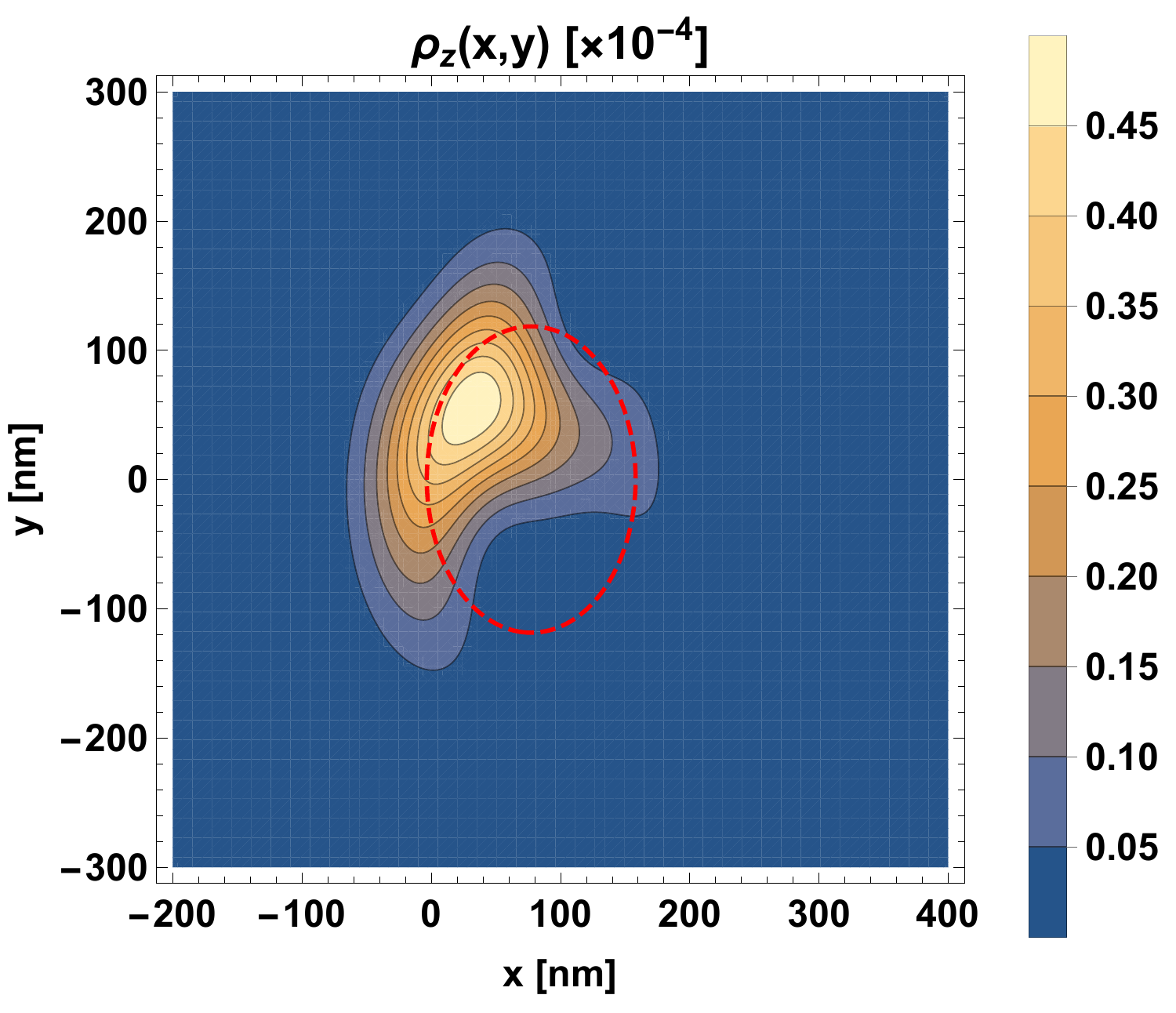}
\end{tabular}
	\caption{\label{fig:density7}Time evolution of the probability density $\rho_{z}(x,y,t)$ in Eq.~(\ref{evolution2}) for different values of $t=t'/(v'_{\rm F}\sqrt{\omega_{\rm B}})$ and $\epsilon=15\%$ along the $\mathcal{Z}$ direction. The red dashed line shows the classical trajectory in~(\ref{path2}). $z=2$, $B_{0}=0.3$ T, and $\beta=1/\sqrt{2}$.}
\end{figure}

Figures~\ref{fig:density6} and \ref{fig:density7} show the time evolution of the probability densities of electron coherent states. In particular, the shape of the function $\rho_{z}(x,y,t)$ changes as $t$ increases. Due to the non-equidistant LL in graphene, we do not have the same behavior for the CS of the harmonic oscillator, which are stable in time with a periodic time evolution. Instead, we can identify that for certain values of $t$, there are revivals, i.e., the probability function adopts a shape similar to what it was in $t=0$.

\begin{figure}[h!]
	\centering
	\begin{tabular}{ccc}
	(a)  \qquad \qquad \qquad \qquad \qquad \qquad \qquad   & (b)  \qquad \qquad \qquad \qquad \qquad \qquad \qquad & \\
		\includegraphics[trim = 0mm 0mm 0mm 0mm, scale= 0.65, clip]{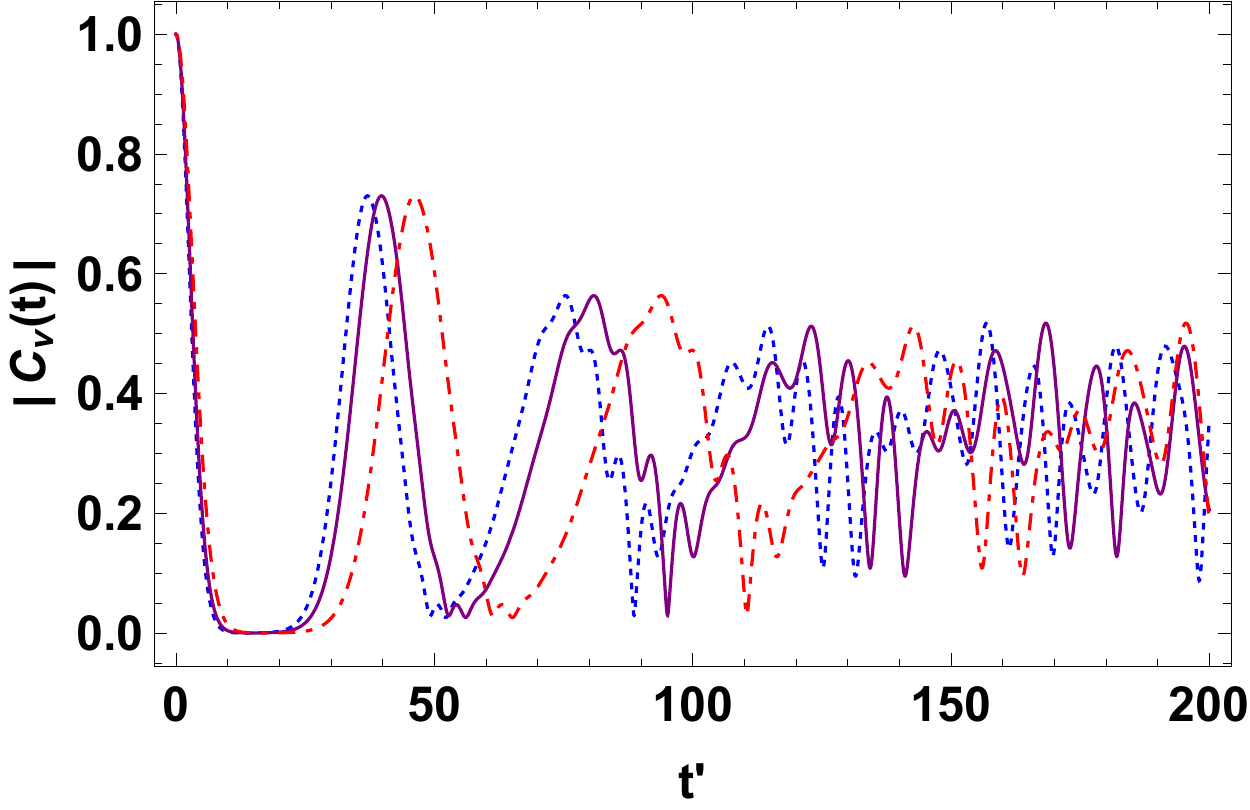} &
		\includegraphics[trim = 0mm 0mm 0mm 0mm, scale= 0.65, clip]{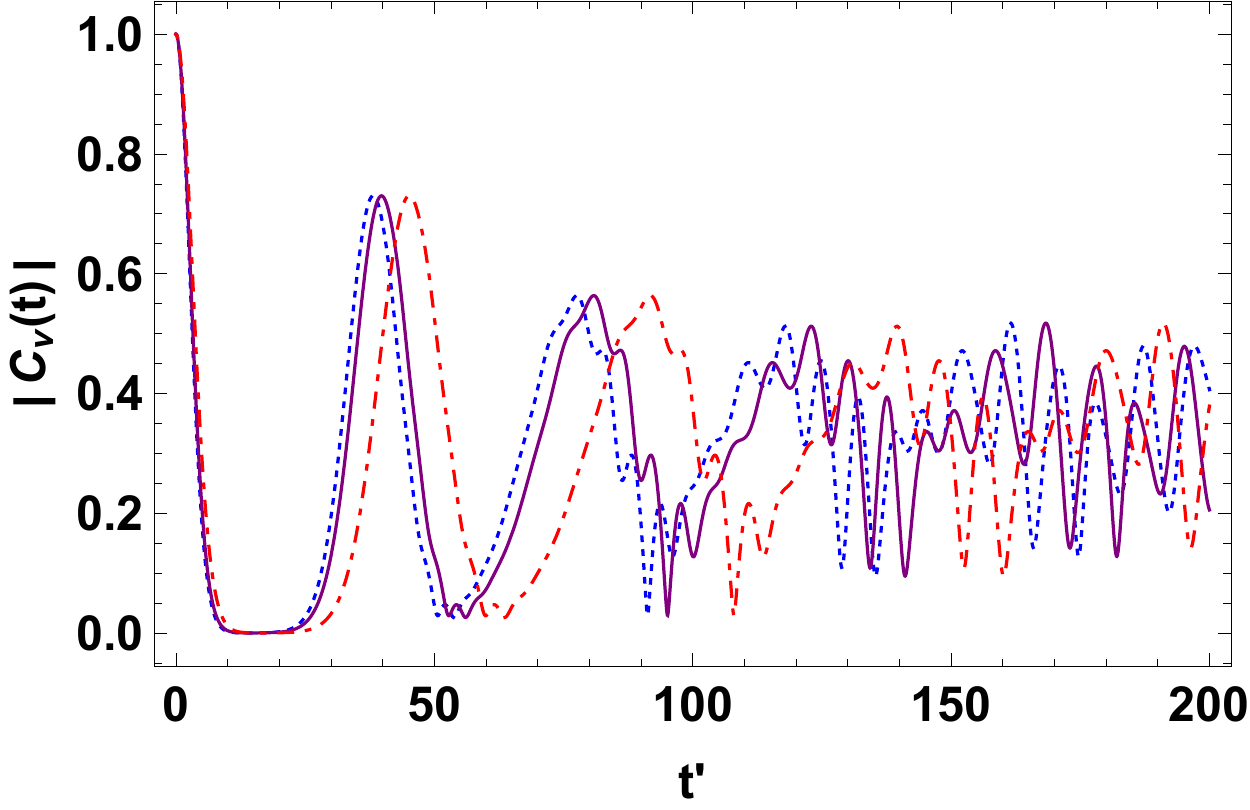}
	\end{tabular}
	\caption{\label{fig:autocollation1}Absolute value of auto-correlation as a function $\vert C_{\nu}(t)\vert$ as function of $t=t'/(v_{\rm F}\sqrt{\omega_{\rm B}})$ for different values of $\epsilon$ along (a) the $\mathcal{Z}$ direction and (b) the $\mathcal{A}$ direction: $\epsilon=-15\%$ (blue, \dotted), $\epsilon=0\%$ (purple, \full) and $\epsilon=15\%$ (red, \chain). $\nu=20$, $B_{0}=0.3$ T, and $\beta=1/\sqrt{2}$.}
\end{figure}

\begin{figure}[h!]
	\centering
	\begin{tabular}{ccc}
		(a) $z=2$ \qquad \qquad \qquad \qquad \qquad \qquad & (b) $z=10$ \qquad \qquad \qquad \qquad \qquad & \\
		\includegraphics[trim = 0mm 0mm 0mm 0mm, scale= 0.65, clip]{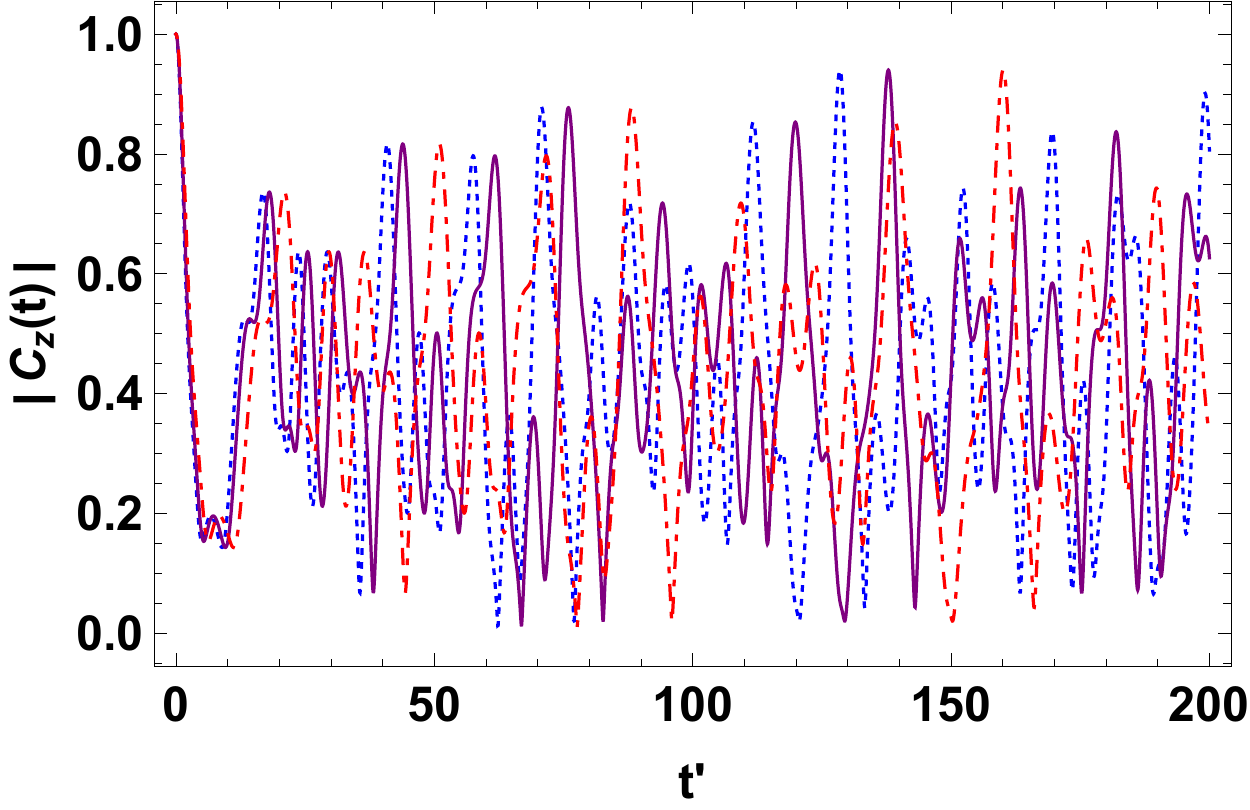} &
		\includegraphics[trim = 0mm 0mm 0mm 0mm, scale= 0.65, clip]{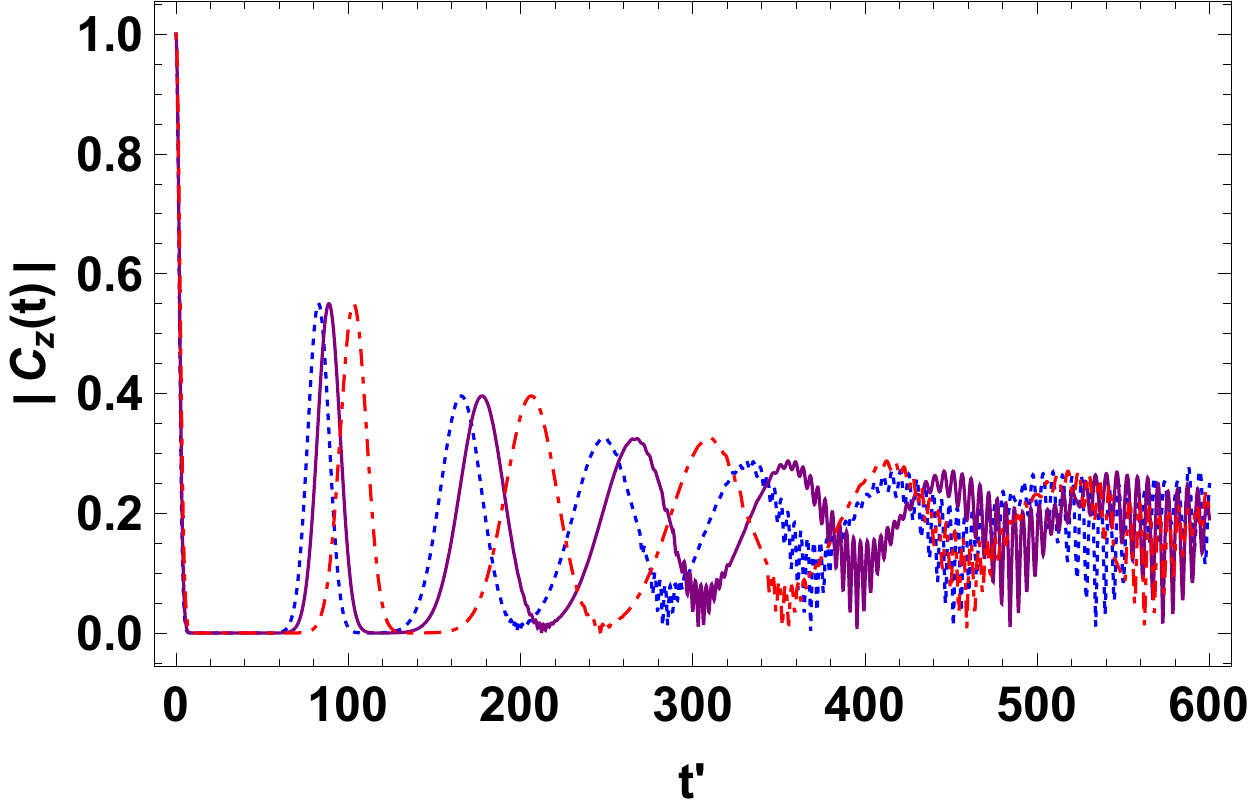}
	\end{tabular}
		\begin{tabular}{ccc}
		(c) {\color{white}$z=2$} \qquad \qquad \qquad \qquad \qquad \qquad & (d) {\color{white}$z=10$} \qquad \qquad \qquad \qquad \qquad & \\
		\includegraphics[trim = 0mm 0mm 0mm 0mm, scale= 0.65, clip]{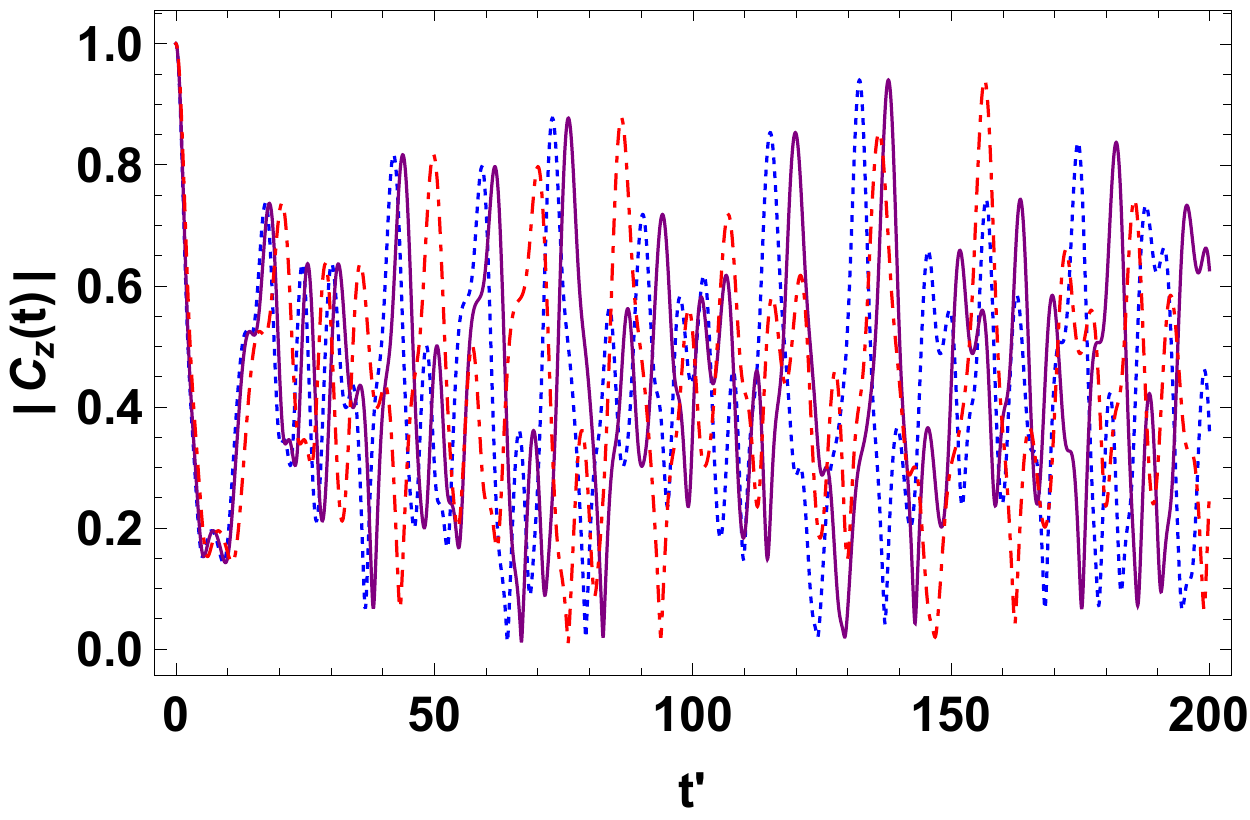} &
		\includegraphics[trim = 0mm 0mm 0mm 0mm, scale= 0.65, clip]{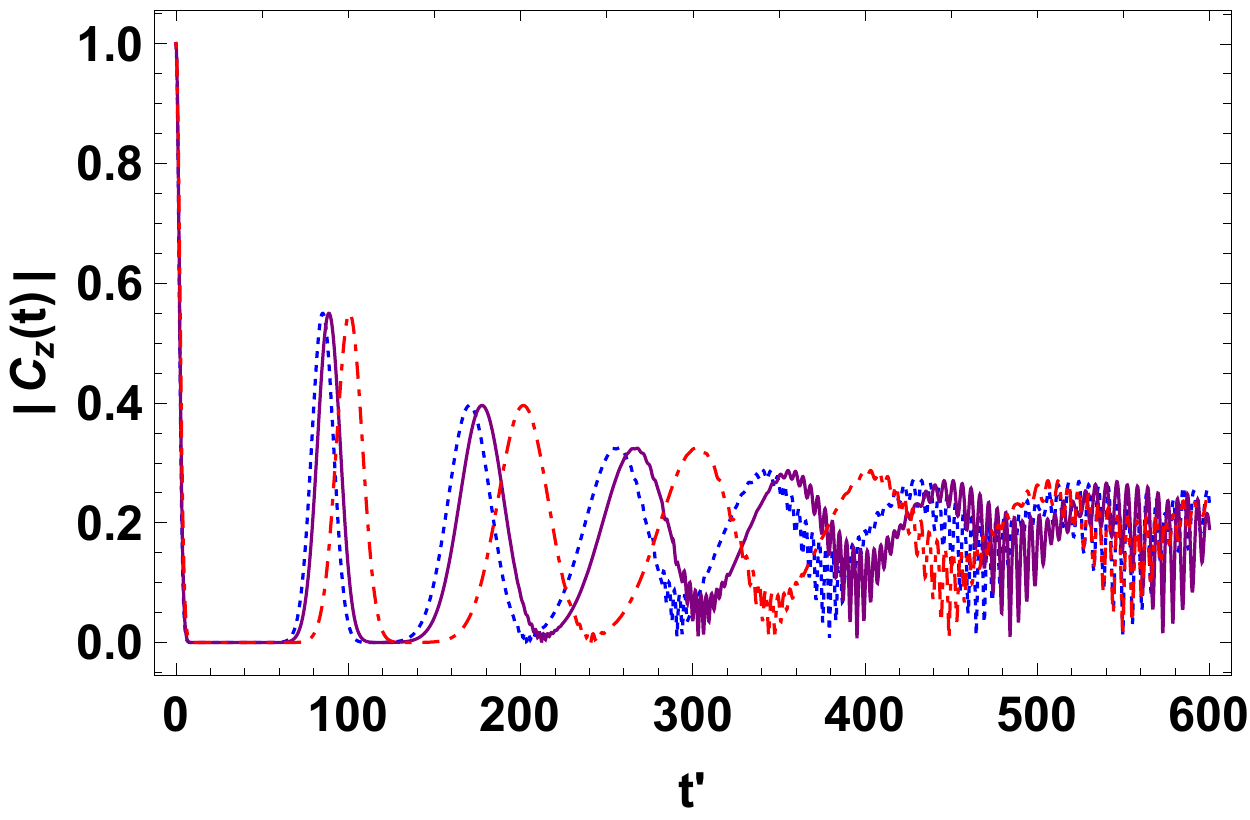}
	\end{tabular}
	\caption{\label{fig:autocollation2}Absolute value of auto-correlation as a function $\vert C_{z}(t)\vert$ as function of $t=t'/(v_{\rm F}\sqrt{\omega_{\rm B}})$ for different values of $\epsilon$ along (a, b) the $\mathcal{Z}$ direction and (c, d) the $\mathcal{A}$ direction: $\epsilon=-15\%$ (blue, \dotted), $\epsilon=0\%$ (purple, \full) and $\epsilon=15\%$ (red, \chain). $B_{0}=0.3$ T and $\beta=1/\sqrt{2}$.}
\end{figure}

Now, in order to investigate the values of $t$ for which the revivals~\cite{krueckl09} occurs and how they are affected by mechanical deformations, let us consider the time evolution of the auto-correlation function\cite{delande01,sergi04,sakata13,trushechkin17,dey18,alhambra20}, given by
\begin{equation}
C(t)=\langle \Psi(t=0)\vert\Psi(t)\rangle,
\end{equation}
which allows us to study the time evolution of the coherent states previously built inasmuch as it correlates the same state at two points in  time. Qualitatively, a time-correlation function describes, in general, how long a given property of a system persists until it is averaged out by microscopic motions  and interactions with its surroundings. Hence, we have
\begin{subequations}\label{correllation}
	\begin{align}
	C_{\nu}(t)&=\mathcal{N}_{\nu}^{2}\left(\vert\beta\vert^{2\nu}+2\sum_{n=1}^{\nu}\left(\begin{array}{c}
	\nu \\
	n
	\end{array}\right)\vert\alpha\vert^{2n}\vert\beta\vert^{2(\nu-n)}\exp\left(-iE_{n}t/\hbar\right)\right), \label{correllation1}\\
	C_{z}(t)&=\mathcal{N}_{z}^{2}\left(2\sum_{\nu=0}^{\infty}\sum_{n=0}^{\nu}\frac{\vert z\vert^{2\nu}\vert\alpha\vert^{2n}\vert\beta\vert^{2(\nu-n)}}{(\nu-n)!\,n!}\exp\left(-iE_{n}t/\hbar\right)-\exp\left(\vert z\beta\vert^{2}\right)\right). \label{correllation2}
	\end{align}
\end{subequations}

As we can see in Figs.~\ref{fig:autocollation1} and \ref{fig:autocollation2}, the time evolution of electron coherent states is affected by the tensile strain $\epsilon$ and the eigenvalue $z$. More precisely, when $\vert z\vert$ is close to zero, the auto-correlation function $C_{z}(t)$ oscillates rapidly in short-time intervals, while if $\vert z\vert$ increases, $C_{z}(t)$ first oscillates smoothly and afterward faster oscillations with a sinusoidal-like enveloping can be found at larger values of $t$. On the other hand, tensile and compression deformations also modify at what time the fast oscillations appear. Our conclusion is that positive (negative) deformations increase (decrease) the time interval in which one can find smooth oscillations. Therefore, although the probability density does not maintain its shape over time --due to the non-equidistant energy spectrum--  as $\vert z\vert$ increases, it is possible to retard abrupt changes in $\rho_{z}(x,y,t)$ by applying stress deformations to the material, which contracts the spacing of LLs. These results are according to Ehrenfest theorem that establishes that the propagation of a wave function is described for short times by classical equation of motion~\cite{ehrenfest27}.

Moreover, we can procedure as in~\cite{fernandezmartinez20} for finding a possible approximate period $\mathcal{T}$ for electron coherent states. Setting the eigenvalue $z$, we compute the mean energy value $\langle H\rangle_{z}$ in Eq.~(\ref{Hz}) and the energy interval in which it lies, i.e., $E_{j}<\langle H\rangle_{z}<E_{j+1}$. Thus, the approximate period is determined as
\begin{equation}
\mathcal{T}=\frac{2\pi\hbar}{E_{j+1}-E_{j}}=\frac{2\pi}{v'_{\rm F}\sqrt{\omega_{\rm B}}}\left(\sqrt{j+1}-\sqrt{j}\right)^{-1}.
\end{equation}
Due to the fact that energy spectrum depends on the strain parameter $\zeta$, the period $\mathcal{T}$ will be also affected. For instance, for Schr\"{o}dinger-type 2D coherent states with eigenvalues $\vert z\vert=2$ and $\vert z\vert=10$ (see Fig.~\ref{fig:autocollation2} with $\epsilon=0\%$), we have that $E_{1}<\langle H\rangle_{z=2}=0.027\,{\rm eV}<E_{2}$ and $E_{49}<\langle H\rangle_{z=10}=0.140\,{\rm eV}<E_{50}$, respectively. Thus, their respective quasi-periods turn out to be
\begin{equation}
\mathcal{T}_{z=2}\approx\frac{5\pi}{v'_{\rm F}\sqrt{\omega_{\rm B}}}, \quad \mathcal{T}_{z=10}\approx\frac{28\pi}{v'_{\rm F}\sqrt{\omega_{\rm B}}}.
\end{equation}

\section{Conclusions and final remarks}\label{sec6:conclusions}
The analysis of anisotropy effects on the charge carrier dynamics in anisotropic 2D Dirac materials through the coherent state formalism has been studied in~\cite{diaz2019coherent}. Motivated by this work, we addressed the construction of Schr\"{o}dinger-type 2D coherent states in strained graphene, in which the anisotropic Fermi velocity is induced by tensile and compression mechanical deformations. Hence, in order to describe the bidimensional effects of uniaxial deformations on dynamics of electrons in graphene under the interaction of a uniform homogeneous magnetic field and since the physical problem has a two-dimensional nature, in this work, the background magnetic field was defined through the symmetric gauge $\vec{A}=(\vec{B}\times\vec{r})/2$, while the applied mechanical deformations were encoded in two purely geometrical parameters $a$ and $b$, in Eq.~(\ref{ayb}).

The coherent state formulation was implemented building first states $\vert\Psi_{\nu}\rangle$ as linear combinations of the eigenstates $\vert\Psi_{m,n}\rangle$ labeled by the occupation number $\nu=m+n$, where the quantum number $n$ indicates the corresponding LL. The states $\vert\Psi_{\nu}\rangle$ satisfy, up to a constant, the orthogonality condition $\langle\Psi_{\mu}\vert\Psi_{\nu}\rangle=\delta_{\mu\nu}$ and the completeness relation in each Hilbert subspace $\mathcal{H}_{\nu}$ such that $\mathcal{H}=\bigoplus_{\nu=0}^{\infty}\mathcal{H}_{\nu}$. In addition, we defined a matrix annihilation operator $\mathbb{J}^{-}$, which allows us to obtain states in $\mathcal{H}_{\nu-1}$ by the action of $\mathbb{J}^{-}$ onto the states in $\mathcal{H}_{\nu}$.

Taking into account the latter, we were able to construct the Schr\"{o}dinger-type 2D coherent states as eigenstates of the annihilation operator $\mathbb{J}^{-}$ with complex eigenvalue $z$. Such states were constructed as a linear combination of the states $\vert\Psi_{\nu}\rangle$, and form a over-complete basis inasmuch as two different coherent states are not orthogonal. In addition, they obey a closure relation and satisfy an occupation number distribution with a Poissonian-like probability with mean $\vert z\vert^{2}$ as $\vert z\vert$ increases.

On the other hand, the probability densities of both coherent states exhibit anisotropy effects induced by the mechanical deformations since the tensile strain $\epsilon$ modifies the Fermi velocity along either the $\mathcal{Z}$ or $\mathcal{A}$ direction, so the function $\rho_{\nu}(x,y)$ and the classical trajectory are aligned to the $x$- or $y$-axis (see Table~\ref{tab:table1}). For tensile deformations along the $\mathcal{Z}$ and $\mathcal{A}$ directions, the effects on the probability densities are according to those shown in~\cite{diaz2019coherent} for anisotropic 2D Dirac materials.

Furthermore, the time evolution of the $SU(2)$ and Schr\"{o}dinger-type 2D coherent states was studied revealing interesting facts about how it is affected by the uniaxial strain applied. First, although the time evolution of these states is not stable --the shape of the probability densities changes in time-- due to the energy spectrum is not linear in $n$, we can observe quasi-periodicity on these states, which is identified as revivals in the probability density $\rho_{z}(x,y,t)$. Second, such a quasi-period $\mathcal{T}$, as well as the interval in which the auto-correlation function $C_{z}(t)$ shows smooth oscillations, can be modified according to the mechanical deformation applied to the material. In particular, tensile deformations along the $\mathcal{Z}$ direction make both quantities increase their values. Besides, as $\vert z\vert$ increases, the Schr\"{o}dinger-type 2D coherent states tend to be more stable in time.

Our findings show that the coherent state formulation is a good tool for describing the electron dynamics in graphene from a semi-classical model. In particular, an interesting result makes evident the effect of the anisotropy induced by mechanical deformations on the time evolution of these states. This fact suggests the possibility of studying the Ehrenfest time scale~\cite{delande01,silvestrov02,sergi04,maia08,dey18,Schubert_2012}, which is defined as the time scale at which mean values of propagated localized states, which allow to recover classical mechanics from quantum dynamics, become delocalized~\cite{Schubert_2012}. 
Therefore, the study of the time evolution of some physical quantities and the establishment of different time scales~\cite{trushechkin17}, namely, classical, collapse, and revivals, in condensed matter systems might yield to interesting results in electronic transport and confinement models where a space dependence Fermi velocity and pseudo-magnetic fields are considered.


\section*{Acknowledgments}
E.D.-B. acknowledges David J Fern\'andez for his careful reading and suggestions for improving this article, as well as L.M. Nieto and J. Negro for the valuable discussions that motivated the mathematical aspect of this work. This work was supported by SIP-IPN grant 20201196.

\appendix
\numberwithin{equation}{section}
\section{Eigenvalues and eigenstates}\label{appA}
	 In order to solve the problem of an electron in strained graphene lying on the $xy$-plane interacting with a homogeneous magnetic field aligned to the $z$-axis, we proceed as follows.
	 
	 We remove the anisotropy term from the kinetic part of the Hamiltonians $\mathcal{H}^{\pm}$ by using the elliptical coordinates
	 \begin{equation}
	 x=\zeta^{1/2}\rho\cos(\theta), \quad y=\zeta^{-1/2}\rho\sin(\theta)
	 \end{equation}
	 such that
	 \begin{equation}
	 \frac{x^{2}}{\zeta\rho^{2}}+\frac{y^{2}}{\zeta^{-1}\rho^{2}}=1.
	 \end{equation}
	Since now the problem has an implicit geometrical rotational-like symmetry around the $z$-axis, it is convenient to express the Hamiltonians $\mathcal{H}^\pm$ (\ref{hamiltonians}) in the coordinates $(\rho,\theta)$ as~\cite{f28,d31,dknn17}
\begin{equation}\label{9}
\mathcal{H}^\pm=\frac{1}{\omega_{\rm B}}\left[-\left(\partial_\rho^2+\frac{1}{\rho}\partial_\rho+\frac{1}{\rho^2}\partial_\theta^2\right)+\frac{i\omega_{\rm B}}{2}\partial_\theta+\frac{\omega_{\rm B}^2}{16}\rho^2\right]\mp\frac12.
\end{equation}

By introducing the dimensionless variable $\xi$, defined as follows:
\begin{equation}\label{10}
\xi=\frac{\sqrt{\omega_{\rm B}}}{2}\rho,
\end{equation}
the corresponding eiganvalue equations take the form
\begin{subequations}\label{11}
	\begin{align}
	\mathcal{H}^-\psi_1(\xi,\theta)&=A^-A^+\psi_1(\xi,\theta)=\frac{1}{4}\left[-\left(\partial_\xi^2+\frac{1}{\xi}\partial_\xi+\frac{1}{\xi^2}\partial_\theta^2\right)+2i\partial_\theta+\xi^2+2\right]\psi_1(\xi,\theta)=\mathcal{E}_1\psi_1(\xi,\theta), \label{11a} \\
	\mathcal{H}^+\psi_2(\xi,\theta)&=A^+A^-\psi_2(\xi,\theta)=\frac{1}{4}\left[-\left(\partial_\xi^2+\frac{1}{\xi}\partial_\xi+\frac{1}{\xi^2}\partial_\theta^2\right)+2i\partial_\theta+\xi^2-2\right]\psi_2(\xi,\theta)=\mathcal{E}_2\psi_2(\xi,\theta). \label{11b}
	\end{align}
\end{subequations}

\subsection{Algebraic treatment}\label{treatment}
Now, let us assume that the functions in Eq.~(\ref{11}) can be expressed by separated solutions \cite{dknn17}
\begin{equation}\label{13}
\psi_i(x,y)\rightarrow\psi_i(\xi,\theta)=\xi^{-1/2}\phi_i(\xi,\theta), \quad \phi_i(\xi,\theta)\equiv R_i(\xi)\Theta_i(\theta), \quad i=1,2,
\end{equation}
where $\Theta_i(\theta)$ is an eigenfunction of the $z$-component angular momentum-like operator $L_{z}=-i\partial_\theta$ with integer eigenvalues $m_{z}=0,\pm1,\pm2,\dots$, i.e.,
\begin{equation}\label{14}
\Theta_i(\theta)=\exp(i\,m_{z}\theta), \quad L_{z}\Theta_i(\theta)=m_{z}\Theta_i(\theta).
\end{equation}

Hence, each radial function $R_i(\xi)$ satisfies the following one-dimensional problems:
\begin{subequations}\label{15}
	\begin{align}
	\mathcal{H}_{m_{z}}^-R_1(\xi)&=\frac{1}{4}\left[-\frac{d^2}{d\xi^2}+\frac{m_{z}^2-1/4}{\xi^2}-2m_{z}+\xi^2+2\right]R_1(\xi)=\mathcal{E}_1R_1(\xi), \label{15a} \\
	\mathcal{H}_{m_{z}}^+R_2(\xi)&=\frac{1}{4}\left[-\frac{d^2}{d\xi^2}+\frac{m_{z}^2-1/4}{\xi^2}-2m_{z}+\xi^2-2\right]R_2(\xi)=\mathcal{E}_2R_2(\xi). \label{15b}
	\end{align}
\end{subequations}
where both Hamiltonians can be factorized in terms of two set of differential operators \cite{df96,kka12}
\begin{subequations}\label{17}
	\begin{align}
	a_{m_{z}}^\pm&=\frac12\left(\mp\frac{d}{d\xi}+\frac{m_{z}-1/2}{\xi}+\xi\right), \label{17a}\\
	b_{m_{z}}^\pm&=\frac12\left(\mp\frac{d}{d\xi}-\frac{m_{z}+1/2}{\xi}+\xi\right), \label{17b}
	\end{align}
\end{subequations}
as follows:
\begin{equation}\label{18}
\mathcal{H}_{m_{z}}^+=\frac{1}{2}(a_{m_{z}}^+a_{m_{z}}^-+b_{m_{z}}^+b_{m_{z}}^-+m_{z}), \quad \mathcal{H}_{m_{z}}^-=\mathcal{H}_{m_{z}}^++1.
\end{equation}
It is possible to translate the previous radial operators $a_{m_{z}}^\pm$, $b_{m_{z}}^\pm$ into another ``dressed'' ones $a^\pm, b^\pm$  in elliptical coordinates as follows~\cite{dknn17}:
\begin{subequations}\label{19}
	\begin{align}
	a^-&=\frac{\exp(-i\theta)}{2}\left(\partial_\xi-\frac{i\partial_\theta+1/2}{\xi}+\xi\right), \quad a^+=(a^-)^\dagger, \label{19a}\\
	b^-&=\frac{\exp(i\theta)}{2}\left(\partial_\xi-\frac{-i\partial_\theta+1/2}{\xi}+\xi\right), \quad b^+=(b^-)^\dagger, \label{19b}
	\end{align}
\end{subequations}
which satisfy the following commutation relations:
\begin{equation}\label{20}
[a^-,a^+]=\mathbf{1}, \quad [b^-,b^+]=\mathbf{1}, \quad [a^\pm,b^\pm]=\mathbf{0}, \quad [a^\pm,b^\mp]=\mathbf{0}.
\end{equation}

Therefore, Eq.~(\ref{18}) is written as
\begin{equation}\label{21}
\mathsf{H}^+=\frac12(a^+a^-+b^+b^-+\mathsf{L}_{z}), \quad \mathsf{H}^-=\mathsf{H}^++1,
\end{equation}
where the $z$-component angular momentum-like operator $L_{z}$, which is expressed as
\begin{equation}\label{22}
\mathsf{L}_{z}=a^+a^--b^+b^-,
\end{equation}
satisfies the commutation relations,
\begin{equation}\label{23}
[\mathsf{L}_{z},a^\pm]=\pm a^\pm, \quad [\mathsf{L}_{z},b^\pm]=\mp b^\pm,
\end{equation}
This implies that the operators $b^+$ and $b^-$, acting on an eigenstate of $\mathsf{L}_{z}$, 
decreases or increases, respectively, the eigenvalue $m_{z}$ in an unity; meanwhile, the operators $a^\pm$ have the contrary effect.

\section{Eigenfunctions}\label{appB}
Defining now the corresponding number and angular momentum operators as
\begin{equation}\label{24}
\mathsf{H}^+=N_a\equiv a^+a^-, \quad N_b\equiv b^+b^-, \quad \mathsf{L}_{z}\equiv N_a-N_b,
\end{equation}
the eigenstates of the Hamiltonians $\mathcal{H}^\pm$ can be labeled by two positive integers $m,n=0,1,2,\dots$, corresponding to the number operators $N_a$ and $N_b$:
\begin{equation}\label{25}
\phi_1(\xi,\theta)\equiv\phi_{m_1,n}(\xi,\theta), \quad \phi_2(\xi,\theta)\equiv\phi_{m_2,n-1}(\xi,\theta),
\end{equation}
where the condition $m_1=m_2\equiv m$ must be fulfilled in order to still have eigenstates of the DW Hamiltonian $H$. Therefore, the operators $N_a$, $N_b$ and $\mathsf{L}_{z}$ act onto the states $\phi_{m,n}(\xi,\theta)$ as
\begin{equation}\label{26}
N_a\,\phi_{m,n}=n\,\phi_{m,n}, \quad N_b\,\phi_{m,n}=m\,\phi_{m,n}, \quad \mathsf{L}_{z}\,\phi_{m,n}=(n-m)\,\phi_{m,n}, 
\end{equation}
where the last expression implies that the states $\phi_{m,n}(\xi,\theta)$ are also eigenstates of the operator $\mathsf{L}_{z}$ with eigenvalue $m_{z}\equiv n-m$.



On the other hand, the wave functions of excited states $\phi_{m,n}(\xi,\theta)$ can be built from the successive action of the creation operators $a^+$ and $b^+$ on the fundamental state $\phi_{0,0}(\xi,\theta)$,
\begin{equation}\label{27}
\phi_{m,n}(\xi,\theta)=\frac{(b^+)^m(a^+)^{n}}{\sqrt{m!\,n!}}\phi_{0,0}(\xi,\theta), \quad m,\,n=0,1,2,\dots
\end{equation}
where the ground state wave function $\phi_{0,0}(\xi,\theta)$ is determinated by the conditions
\begin{equation}\label{groundstate}
a^-\phi_{0,0}(\xi,\theta)=b^-\phi_{0,0}(\xi,\theta)=0 \quad \Longrightarrow \quad \phi_{0,0}(\xi,\theta)=K_0\,\xi^{1/2}\exp\left(-\frac{1}{2}\xi^2\right),
\end{equation}
with $K_0$ being a normalization constant. 




In order to obtain the explicit form of the wave functions of the excited states $\psi_{m,n}(x,y)$, one can assume that the radial function
$R_1(\xi)$, in Eq.~(\ref{15a}), can be written as
\begin{equation}\label{29}
R_{1}(\xi)=K_{mn}\,\xi^{1/2+\vert m-n\vert}\exp\left(-\frac{1}{2}\xi^2\right)f_{mn}(\xi),
\end{equation}
where $K_{mn}$ are the normalization constants and $f_{mn}(\xi)$ are functions to determine. After the variable change $z=\xi^2$, we obtain the following differential equations:
\begin{subequations}\label{30}
	\begin{align}
	&z\frac{d^2f_{mn}(z)}{dz^2}+(1+m-n-z)\frac{df_{mn}(z)}{dz}+nf_{mn}(z)=0, \quad m>n, \label{30a} \\
	&z\frac{d^2\bar{f}_{mn}(z)}{dz^2}+(1+n-m-z)\frac{d\bar{f}_{mn}(z)}{dz}+m\bar{f}_{mn}(z)=0, \quad n>m. \label{30b}
	\end{align}
\end{subequations}
in which each one has as a solution the associated Laguerre polynomials $L_k^{\alpha}(x)$. Therefore, it follows that the eigenvalues $\mathcal{E}\rightarrow\mathcal{E}_n$ of the Hamiltonians $\mathcal{H}^\pm$ are related as in Eq.~(\ref{energies}), while the normalized eigenfunctions of the Hamiltonian $\mathcal{H}^+$ are given as in Eq.~(\ref{34}).


\section{Ladder operators}\label{appC}
Coming back to the initial DW problem, the ladder operators in Eq.~(\ref{ladderoperators}) can be expressed in dimensionless polar coordinates $(\xi,\theta)$ as follows:
\begin{subequations}\label{ladderA}
	\begin{align}		                               A^+=\xi^{-1/2}a^+\xi^{1/2}&=\frac{\exp(i\theta)}{2}\left(-\partial_\xi+\frac{-i\partial_\theta}{\xi}+\xi\right), \label{36a}\\
	A^-=\xi^{-1/2}a^-\xi^{1/2}&=\frac{\exp(-i\theta)}{2}\left(\partial_\xi+\frac{-i\partial_\theta}{\xi}+\xi\right), \label{36b}
	\end{align}
\end{subequations}
while through the operators $b^\pm$, one can build another pair of differential operators, given by
\begin{subequations}\label{ladderB}
	\begin{align}
	B^+=\xi^{-1/2}b^+\xi^{1/2}&=\frac{\exp(-i\theta)}{2}\left(-\partial_\xi+\frac{i\partial_\theta}{\xi}+\xi\right), \label{35a}\\
	B^-=\xi^{-1/2}b^-\xi^{1/2}&=\frac{\exp(i\theta)}{2}\left(\partial_\xi+\frac{i\partial_\theta}{\xi}+\xi\right), \label{35b}
	\end{align}
\end{subequations}
which satisfy the commutation relations,
\begin{equation}\label{37}
[B^-,B^+]=\mathbf{1}, \quad [A^\pm,B^\pm]=\mathbf{0}, \quad [A^\pm,B^\mp]=\mathbf{0},
\end{equation}
and their action onto the states $\vert\psi_{m,n}\rangle$ is
\begin{subequations}\label{38}
	\begin{align}
	A^-\vert\psi_{m,n}\rangle=\sqrt{n}\,\vert\psi_{m,n-1}\rangle, &\quad A^+\vert\psi_{m,n}\rangle=\sqrt{n+1}\,\vert\psi_{m,n+1}\rangle, \label{38a}\\
	B^-\vert\psi_{m,n}\rangle=\sqrt{m}\,\vert\psi_{m-1,n}\rangle, &\quad B^+\vert\psi_{m,n}\rangle=\sqrt{m+1}\,\vert\psi_{m+1,n}\rangle. \label{38b}
	\end{align}
\end{subequations}

Likewise, the operators in Eqs.~(\ref{21}) and (\ref{22}) are also transformed as
\begin{equation}
\mathcal{H}^{\pm}=\xi^{-1/2}\mathsf{H}^{\pm}\,\xi^{1/2}, \quad L_{z}=\xi^{-1/2}\mathsf{L}_{z}\,\xi^{1/2}=N-M,
\end{equation}
where $N=A^{+}A^{-}$ and $M=B^{+}B^{-}$ are the transformations of $N_{a}$ and $N_{b}$, respectively.

\section{Matrix operators $\mathbb{A}^{\pm}$ and $\mathbb{B}^{\pm}$}\label{appD}

The action of the matrix operators defined in Eq.~(\ref{annihop1}) onto the eigenstates $\vert\Psi_{m,n}\rangle$ is given by:
	\begin{subequations}\label{relations}
		\begin{align}
		\mathbb{A}^{-}\vert\Psi_{m,n}\rangle&=\sqrt{n}\vert\Psi_{m,n-1}\rangle, \quad m,\,n=0,1,2,\dots, \label{relationsa}\\ \mathbb{A}^{+}\vert\Psi_{m,n}\rangle&=\sqrt{n+1}\vert\Psi_{m,n+1}\rangle, \quad m+1,\,n=1,2,3,\dots, \label{relationsb}\\
		\mathbb{B}^{-}\vert\Psi_{m,n}\rangle&=\sqrt{m}\vert\Psi_{m-1,n}\rangle, \quad m,\,n=0,1,2,\dots, \label{relationsc}\\ \mathbb{B}^{+}\vert\Psi_{m,n}\rangle&=\sqrt{m+1}\vert\Psi_{m+1,n}\rangle, \quad m,\,n=0,1,2,\dots. \label{relationsd}
		\end{align}
	\end{subequations}
such that they satisfy the commutation relations,
\begin{subequations}\label{relations2}
	\begin{align}
[\mathbb{B}^-,\mathbb{B}^+]=\mathbb{I}, \quad [\mathbb{A}^{-},\mathbb{B}^\pm]=\mathbf{0},& \quad m,\,n=0,1,2,\dots, \label{relations2a} \\
[\mathbb{A}^{+},\mathbb{B}^\pm]=\mathbf{0},& \quad m+1,\,n=1,2,3,\dots. \label{relations2b}
\end{align}
\end{subequations}
	
	According to Eqs.~(\ref{relationsb}) and (\ref{relations2b}), the operator $\mathbb{A}^{+}$ cannot be considered as a creation operator for whole Hilbert space $\mathcal{H}$ since $\vert\Psi_{m,1}\rangle\neq\mathbb{A}^{+}\vert\Psi_{m,0}\rangle$. As a consequence, the commutation relation $[\mathbb{A}^{-},\mathbb{A}^{+}]=\mathbb{I}$ only fulfills for $n\geq2$ and for any $m$. However, we can define a different matrix operator that works as a creation operator even it is not the adjoint operator of $\mathbb{A}^{-}$. Such operator, which is defined as
	\begin{equation}
	\tilde{\mathbb{A}}^{+}=\left(\begin{array}{c c}
	A^{+}\frac{\sqrt{N+2}}{\sqrt{N+1}} & -i\sqrt{N+1} \\
	i(A^{+})^{2}\frac{1}{\sqrt{N+1}} & A^{+}
	\end{array}\right),
	\end{equation}
	and whose action onto the states $\vert\Psi_{m,n}\rangle$ reads as
	\begin{equation}
	\tilde{\mathbb{A}}^{+}\vert\Psi_{m,n}\rangle=2^{(1-\delta_{0n})}\sqrt{n+1}\vert\Psi_{m,n+1}\rangle, \quad n=0,1,2,\dots,
	\end{equation}
	can be linked to the so-called $\mathcal{D}$ pseudo-bosonic operators~\cite{trifonov09,bagarello13,bagarello132,bagarello15,bagarello17}, for which the canonical commutation relation, $[c,c^{\dagger}]=1$, is modified as $[a,b]=1$, where $b\neq a^{\dagger}$. It is straightforward to verify that
\begin{subequations}
		\begin{align}
	[\tilde{\mathbb{A}}^{+},\mathbb{B}^{\pm}]&=\mathbf{0}, \quad n,\,m=0,1,2,\cdots, \\
	[\mathbb{A}^{-},\tilde{\mathbb{A}}^{+}]\vert\Psi_{m,n}\rangle&=c(n)\vert\Psi_{m,n}\rangle, \quad c(n)=\begin{cases}
	1, & n=0,\\
	3, & n=1,\\
	2, & n>1.
	\end{cases}
	\end{align}
\end{subequations}

Thus, we are able to obtain excited states from the fundamental one $\vert\Psi_{0,0}\rangle$ as follows:
\begin{equation}\label{eigenstate}
\vert\Psi_{m,n}\rangle=\frac{2^{(1-\delta_{0n}-n)}}{\sqrt{n!}}(\tilde{\mathbb{A}}^+)^{n}\vert\Psi_{m,0}\rangle=\frac{2^{(1-\delta_{0n}-n)}}{\sqrt{m!\,n!}}(\mathbb{B}^+)^{m}(\tilde{\mathbb{A}}^+)^{n}\vert\Psi_{0,0}\rangle, \quad m,n=0,1,2,\dots.
\end{equation}


\bibliographystyle{ieeetr}
\bibliography{biblio}

\end{document}